\documentclass[nofootinbib,aps,11pt]{revtex4-1}
\usepackage{graphicx}
\usepackage{hyperref}
\usepackage{cancel}
\usepackage{amssymb}
\usepackage{textcomp}
\usepackage{amsmath}
\usepackage{bm}
\usepackage{times}
\usepackage{epsfig}
\usepackage{color}
\newcommand{\fb}{\textrm{fb}}

\newcommand{\GeV}{{\rm GeV}}

\newcommand{\TeV}{{\rm TeV}}

\begin{document}
\title{\LARGE LHC phenomenology of the type II seesaw mechanism: Observability of neutral scalars in the nondegenerate case}
\bigskip
\author{Zhi-Long Han~$^{1}$}
\email{hanzhilong@mail.nankai.edu.cn}
\author{Ran Ding~$^{2}$}
\email{dingran@mail.nankai.edu.cn}
\author{Yi Liao~$^{3,2,1}$}
\email{liaoy@nankai.edu.cn}
\affiliation{
$^1$~School of Physics, Nankai University, Tianjin 300071, China
\\
$^2$ Center for High Energy Physics, Peking University, Beijing 100871, China
\\
$^3$ State Key Laboratory of Theoretical Physics, Institute of Theoretical Physics,
Chinese Academy of Sciences, Beijing 100190, China}
\date{\today}

\begin{abstract}
This is a sequel to our previous work on LHC phenomenology of the type II seesaw model in the nondegenerate case. In this work, we further study the pair and associated production of the neutral scalars $H^0/A^0$. We restrict ourselves to the so-called negative scenario characterized by the mass order $M_{H^{\pm\pm}}>M_{H^\pm}>M_{H^0/A^0}$, in which the $H^0/A^0$ production receives significant enhancement from cascade decays of the charged scalars $H^{\pm\pm},~H^\pm$. We consider three important signal channels---$b\bar{b}\gamma\gamma$, $b\bar{b}\tau^+\tau^-$, $b\bar{b}\ell^+\ell^-\cancel{E}_T$---and perform detailed simulations. We find that at the 14 TeV LHC with an integrated luminosity of $3000~\fb^{-1}$, a $5\sigma$ mass reach of $151$, $150$, and $180~\GeV$, respectively, is possible in the three channels from the pure Drell-Yan $H^0A^0$ production, while the cascade-decay-enhanced $H^0/A^0$ production can push the mass limit further to $164$, $177$, and $200~\GeV$. The neutral scalars in the negative scenario are thus accessible at LHC run II.
\end{abstract}

\maketitle

%%%%%%%%%%%%%%%%%%%%%%%
\section{Introduction}
%%%%%%%%%%%%%%%%%%%%%%%

In a previous paper~\cite{Han:2015hba}, we presented a comprehensive analysis on the LHC signatures of the type II seesaw model of neutrino masses in the nondegenerate case of the triplet scalars. In this companion paper, another important signature---the pair and associated production of the neutral scalars--is explored in great detail. This is correlated to the pair production of the standard model (SM) Higgs boson, $h$, which has attracted lots of theoretical and experimental interest~\cite{Aad:2013wqa,Chatrchyan:2013lba} since its discovery~\cite{Aad:2012tfa,Chatrchyan:2012ufa}, because the pair production can be used to gain information on the electroweak symmetry breaking sector~\cite{Plehn:1996wb}. Since any new ingredients in the scalar sector can potentially alter the production and decay properties of the Higgs boson, a thorough examination of the properties offers a diagnostic tool to physics effects beyond the SM. The Higgs boson pair production has been well studied for collider phenomenology in the framework of the SM and beyond~\cite{Plehn:1996wb,Dawson:1998py,Djouadi:1999rca,Baur:2002qd,Asakawa:2010xj,
Dolan:2012rv,Papaefstathiou:2012qe,Goertz:2013kp,Gupta:2013zza,Barr:2013tda,
deFlorian:2013jea,Dolan:2013rja,Barger:2013jfa,Englert:2014uqa,Liu:2014rva,deLima:2014dta,
Barr:2014sga}, and extensively studied in various new physics models~\cite{Dolan:2012ac,Arhrib:2009hc,Craig:2013hca,Hespel:2014sla,Kribs:2012kz,
Cao:2013si,Nhung:2013lpa,Ellwanger:2013ova,Bhattacherjee:2014bca,Christensen:2012si,
Wu:2015nba,Cao:2014kya,Han:2013sga,Gouzevitch:2013qca,No:2013wsa,Grober:2010yv,
Gillioz:2012se,Liu:2013woa,Arhrib:2008pw,Heng:2013cya,Dawson:2012mk,
Chen:2014xwa,Dib:2005re,Yang:2014gca,Chen:2014ask}, as well as in the effective field theory approach of anomalous couplings~\cite{Contino:2012xk,Nishiwaki:2013cma,Liu:2014rba,Dawson:2015oha} and effective operators~\cite{Azatov:2015oxa,Goertz:2014qta,Pierce:2006dh,Kang:2015nga,He:2015spf}.

The pair production of the SM Higgs boson proceeds dominantly through the gluon fusion process~\cite{Plehn:1996wb,Djouadi:1999rca}, and has a cross section at the $14~\TeV$ LHC (LHC14) of about $18~\fb$ at leading order~\cite{Plehn:1996wb}.
\footnote{This number is modified to $33~\fb$ at next-to-leading order~\cite{Dawson:1998py} and to $40~\fb$ at next-to-next-to-leading order~\cite{deFlorian:2013jea}.} It can be utilized to measure the Higgs trilinear coupling. A series of studies have surveyed its observability in the $b\bar{b}\gamma\gamma$, $b\bar{b}\tau^+\tau^-$, $b\bar{b}W^+W^-$, $b\bar{b}b\bar{b}$, and $WW^*WW^*$ signal channels~\cite{Baglio:2012np,Dolan:2012rv,Gouzevitch:2013qca,Papaefstathiou:2012qe,
Goertz:2013kp,deLima:2014dta,Barr:2014sga}. For the theoretical and experimental status of the Higgs trilinear coupling and pair production at the LHC, see Refs.~\cite{Baglio:2012np,Dawson:2013bba}. In summary, at the $14~\TeV$ LHC with an integrated luminosity of $3000~\fb^{-1}$ (LHC14@3000), the trilinear coupling could be measured at an accuracy of $\sim 40\%$~\cite{Barger:2013jfa}, and thus leaves potential space for new physics.

As we pointed out in Ref.~\cite{Han:2015hba}, in the negative scenario of the type II seesaw model where the doubly charged scalars $H^{\pm\pm}$ are the heaviest and the neutral ones $H^0/A^0$ the lightest, i.e., $M_{H^{\pm\pm}}>M_{H^\pm}>M_{H^0/A^0}$, the associated $H^0A^0$ production gives the same signals as the SM Higgs pair production while enjoying a larger cross section. The leading production channel is the Drell-Yan process $pp\to Z^*\to H^0A^0$, with a typical cross section $20$-$500~\fb$ in the mass region {$130$-$300~\GeV$}. Additionally, there exists a sizable enhancement from the cascade decays of the heavier charged scalars, which also gives some indirect evidence for these particles. The purpose of this paper is to examine the importance of the $H^0A^0$ production with an emphasis on the contribution from cascade decays and to explore their observability.

The paper is organized as follows. In Sec.~\ref{decay}, we summarize the relevant part of the type II seesaw and explore the decay properties of $H^0,~A^0$ in the negative scenario. Sections \ref{Eh} and \ref{signal} contain our systematical analysis of the impact of cascade decays on the $H^0/A^0$ production in the three signal channels, $b\bar{b}\gamma\gamma$, $b\bar{b}\tau^+\tau^-$, and $b\bar{b}\ell^+\ell^-\cancel{E}_T$. We discuss the observability of the signals and estimate the required integrated luminosity for a certain mass reach and significance. Discussions and conclusions are presented in Sec.~\ref{Dis}. In most cases, we will follow the notations and conventions in Ref.~\cite{Han:2015hba}.

\section{Decay Properties of Neutral Scalars in the Negative Scenario}
\label{decay}

The type II seesaw and its various experimental constraints have been reviewed in our previous work \cite{Han:2015hba}. Here we recall the most relevant content that is necessary for our study of the decay properties of the scalars in this section and of their detection at the LHC in later sections.

The type II seesaw model introduces an extra scalar triplet $\Delta$ of hypercharge two~\cite{typeII} on top of the SM Higgs doublet $\Phi$ of hypercharge unity. Writing $\Delta$ in matrix form, the most general scalar potential is
\begin{eqnarray}
\label{Vpotential}
V(\Phi,\Delta)&=&
m^2\Phi^\dagger\Phi+M^2\text{Tr}(\Delta^\dagger\Delta)+\lambda_1(\Phi^\dagger\Phi)^2
+\lambda_2\left(\text{Tr}(\Delta^\dagger\Delta)\right)^2
+\lambda_3\text{Tr}(\Delta^\dagger\Delta)^2\notag\\
&&+\lambda_4(\Phi^\dagger\Phi)\text{Tr}(\Delta^\dagger\Delta)
+\lambda_5\Phi^\dagger\Delta\Delta^\dagger\Phi+\left(\mu \Phi^T i\tau^2\Delta^\dagger \Phi+\text{H.c.}\right).
\end{eqnarray}
As in the SM, $m^2 < 0$ is assumed to trigger spontaneous symmetry breaking, while $M^2 > 0$ sets the mass scale of the new scalars. The vacuum expectation value (vev) $v$ of $\Phi$ then induces via the $\mu$ term a vev $v_\Delta$ for $\Delta$. The components of equal charge (and also of identical $CP$ in the case of neutral components) in $\Delta$ and $\Phi$ then mix into physical scalars $H^\pm$; $A^0$; $H^0,~h$ and would-be Goldstone bosons $G^{\pm;0}$, with the mixing angles specified by (see, for instance, Refs.~\cite{Arhrib:2011uy,Aoki:2012jj})
\begin{align}
\tan \theta_+ = \frac{\sqrt{2} v_{\Delta}}{v},~
\tan \alpha = \frac{2 v_{\Delta}}{v},~
\tan 2\theta_0 = \frac{2v_{\Delta}}{v} \frac{v^2(\lambda_4+\lambda_5)-2M_{\Delta}^2}
{2v^2\lambda_1-M_{\Delta}^2-v_\Delta^2(\lambda_2+\lambda_3)},
\label{mixangles}
\end{align}
where an auxiliary parameter is introduced for convenience,
\begin{align}
M_\Delta^2=\frac{v^2\mu}{\sqrt{2}v_\Delta}.
\end{align}
To a good approximation, the SM-like Higgs boson $h$ has the mass $M_h \approx\sqrt{2\lambda_1}v$, the new neutral scalars $H^0,~A^0$ have an equal mass $M_{H^0}\approx M_{A^0} \approx M_{\Delta}$, and the new scalars of various charges are equidistant in squared masses:
\begin{equation}
M^2_{H^{\pm\pm}}-M^2_{H^{\pm}}\approx M^2_{H^{\pm}}-M^2_{H^0/A^0}\approx -\frac{1}{4}\lambda_5v^2.
\label{massrelation}
\end{equation}
There are thus two scenarios of spectra, positive or negative, according to the sign of $\lambda_5$. For convenience, we define $\Delta M\equiv M_{H^\pm}-M_{H^0/A^0}$.

\begin{figure}[!htbp]
\begin{center}
\includegraphics[width=0.45\linewidth]{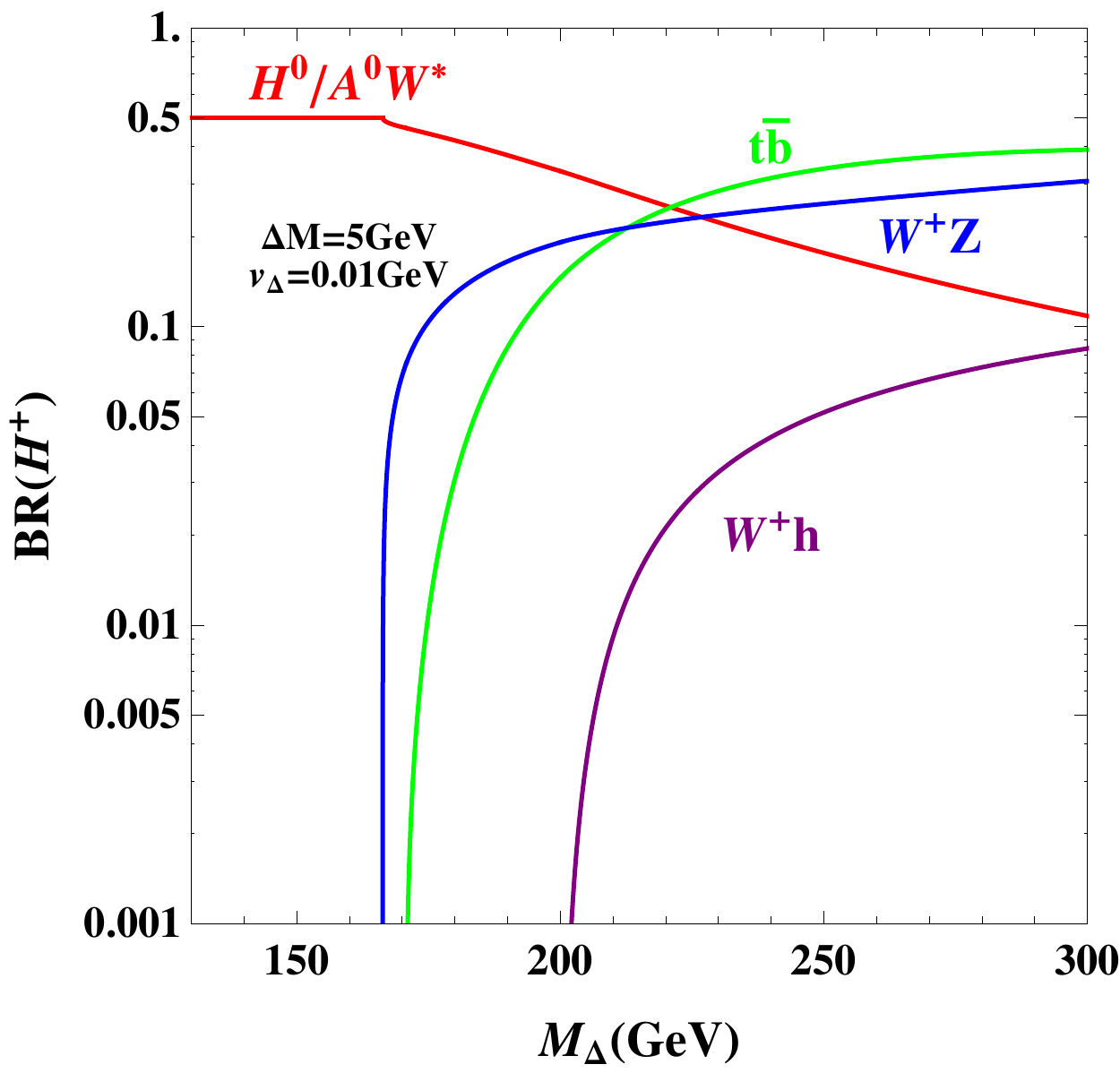}
\includegraphics[width=0.45\linewidth]{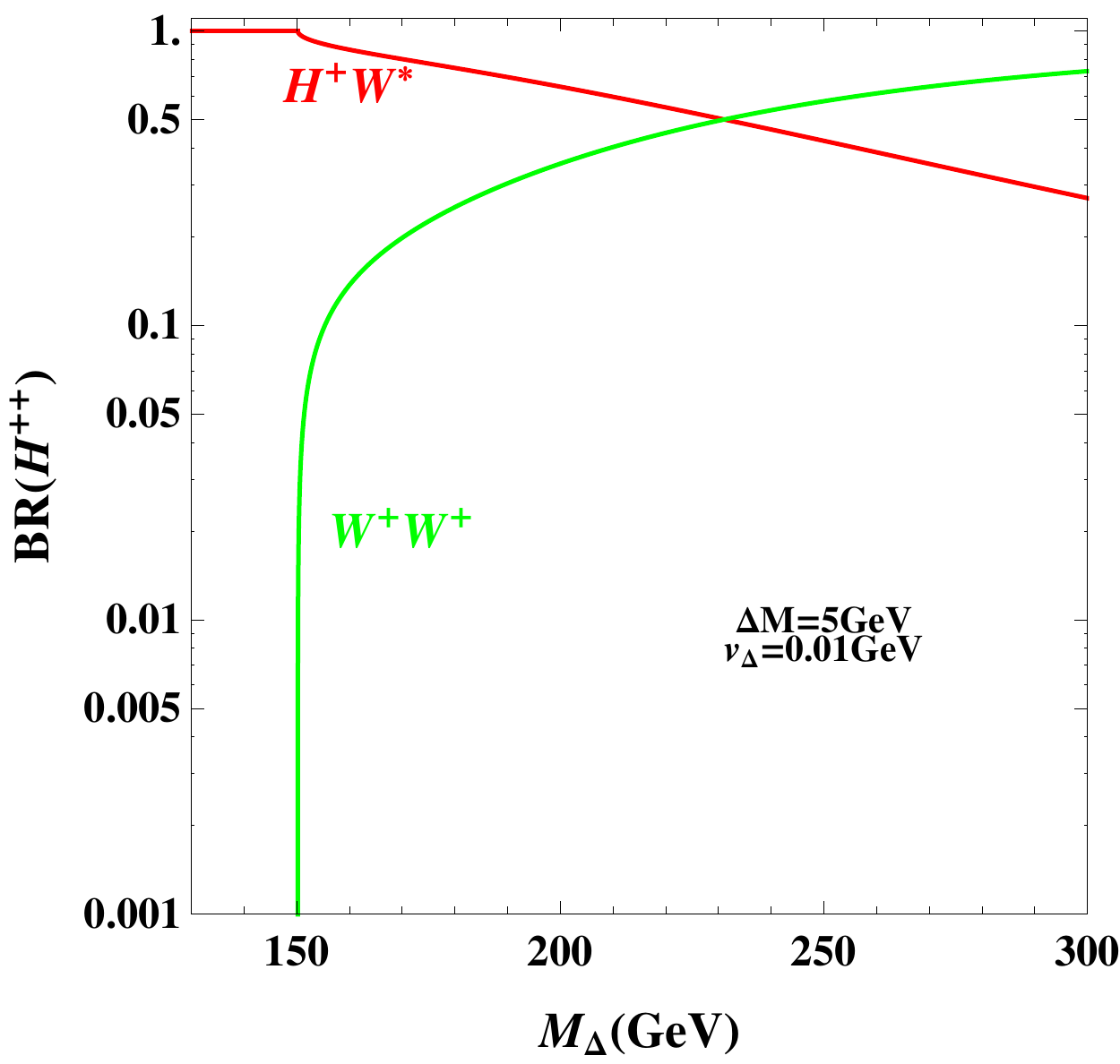}
\includegraphics[width=0.45\linewidth]{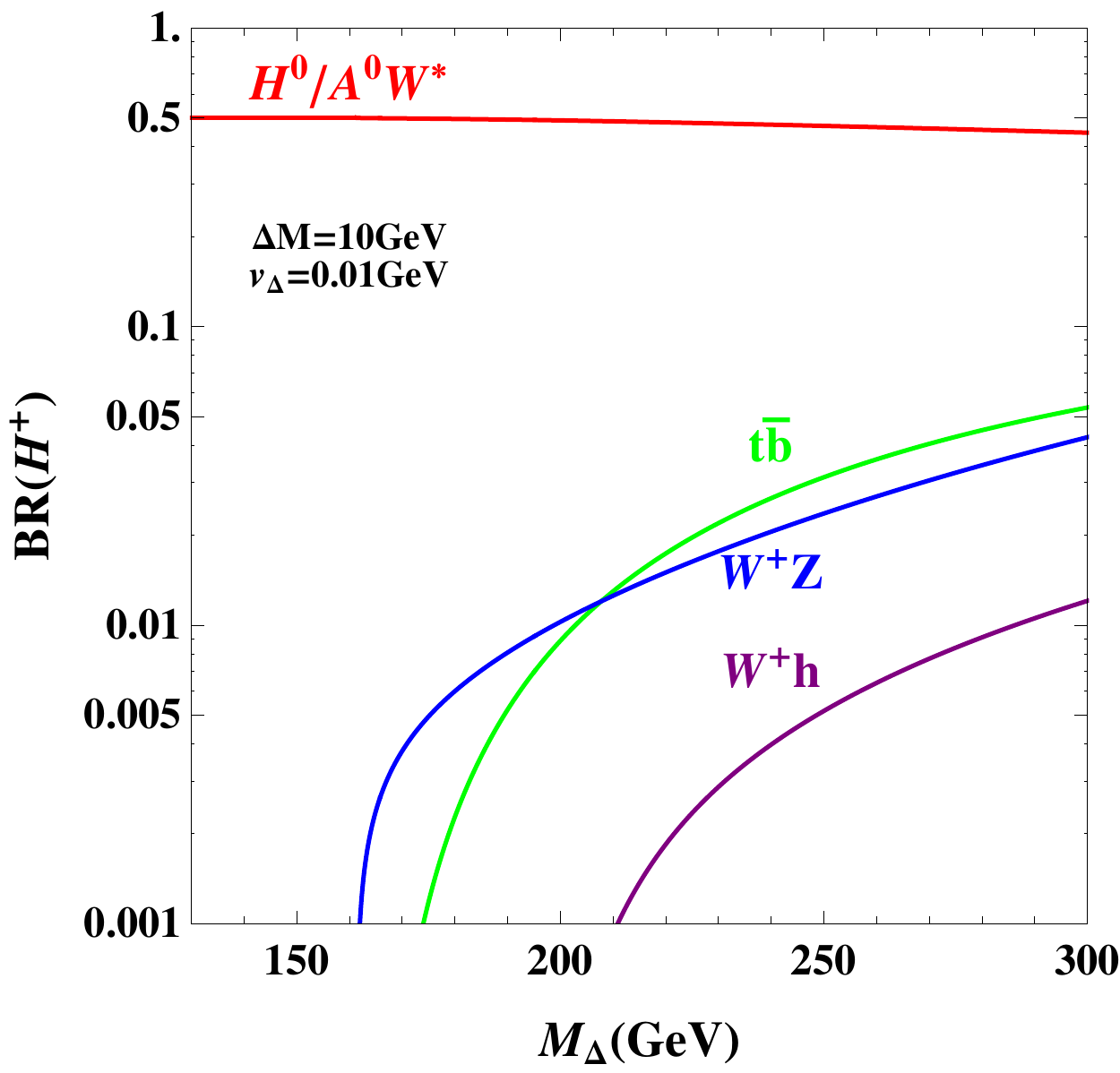}
\includegraphics[width=0.45\linewidth]{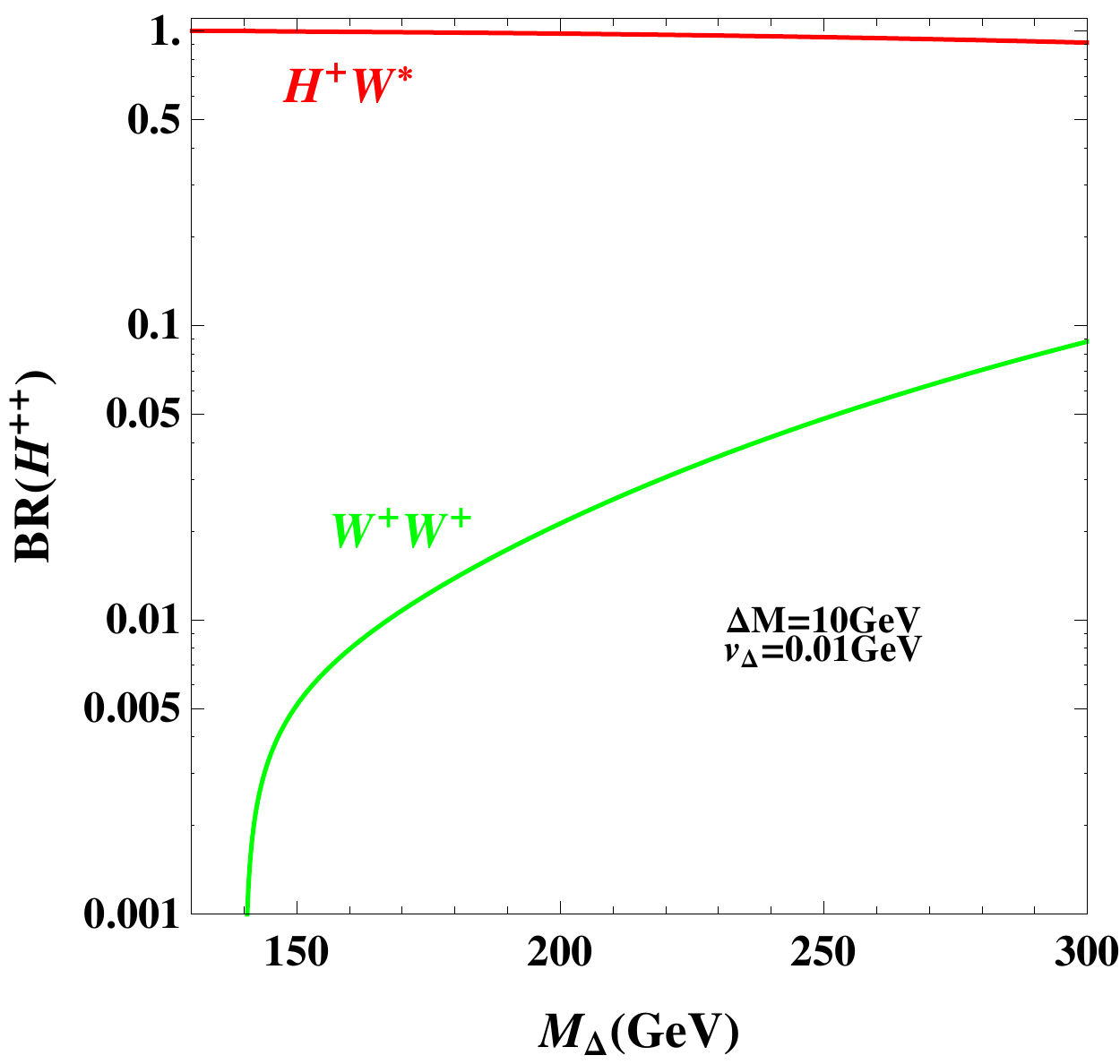}
\includegraphics[width=0.45\linewidth]{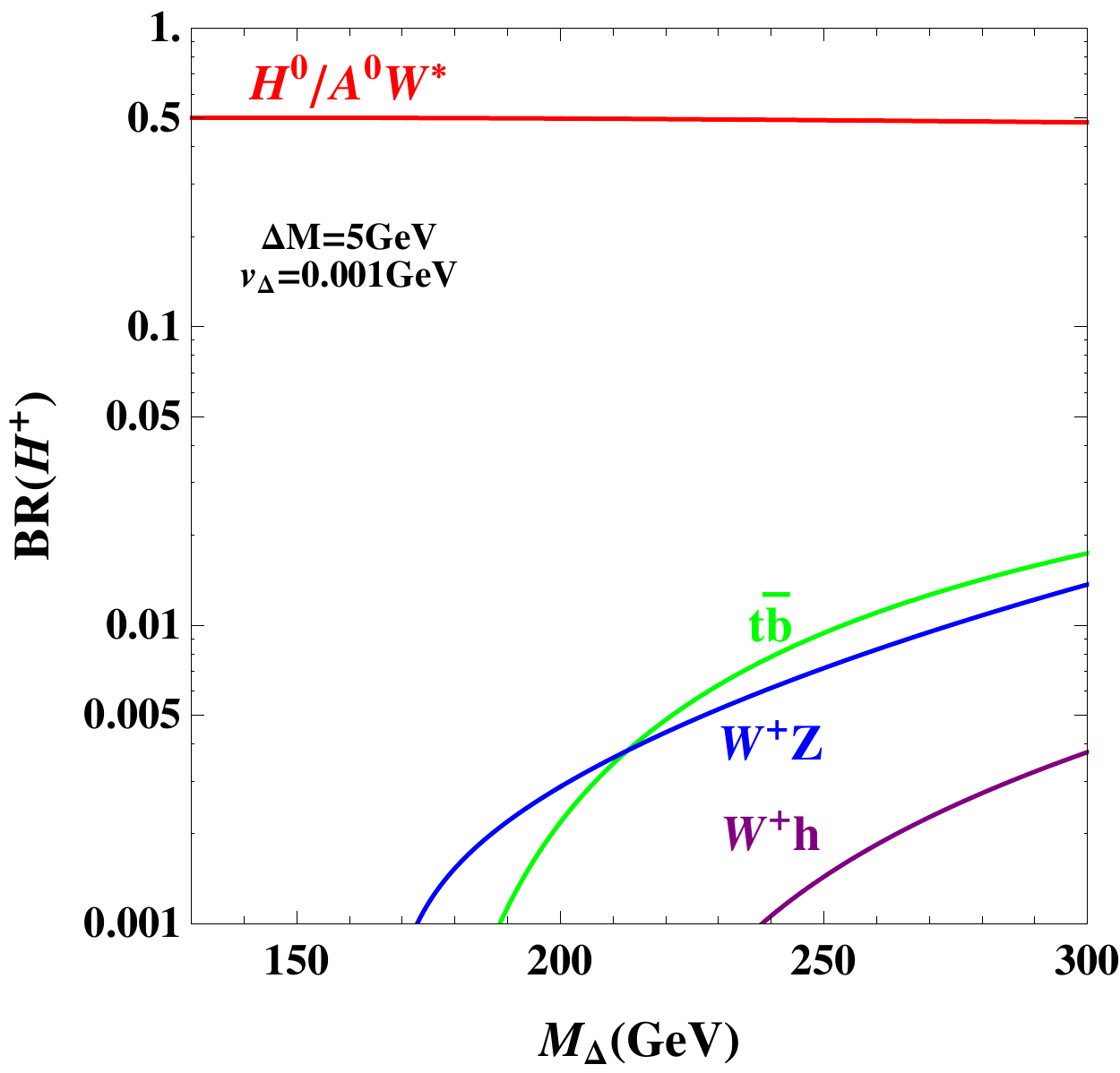}
\includegraphics[width=0.45\linewidth]{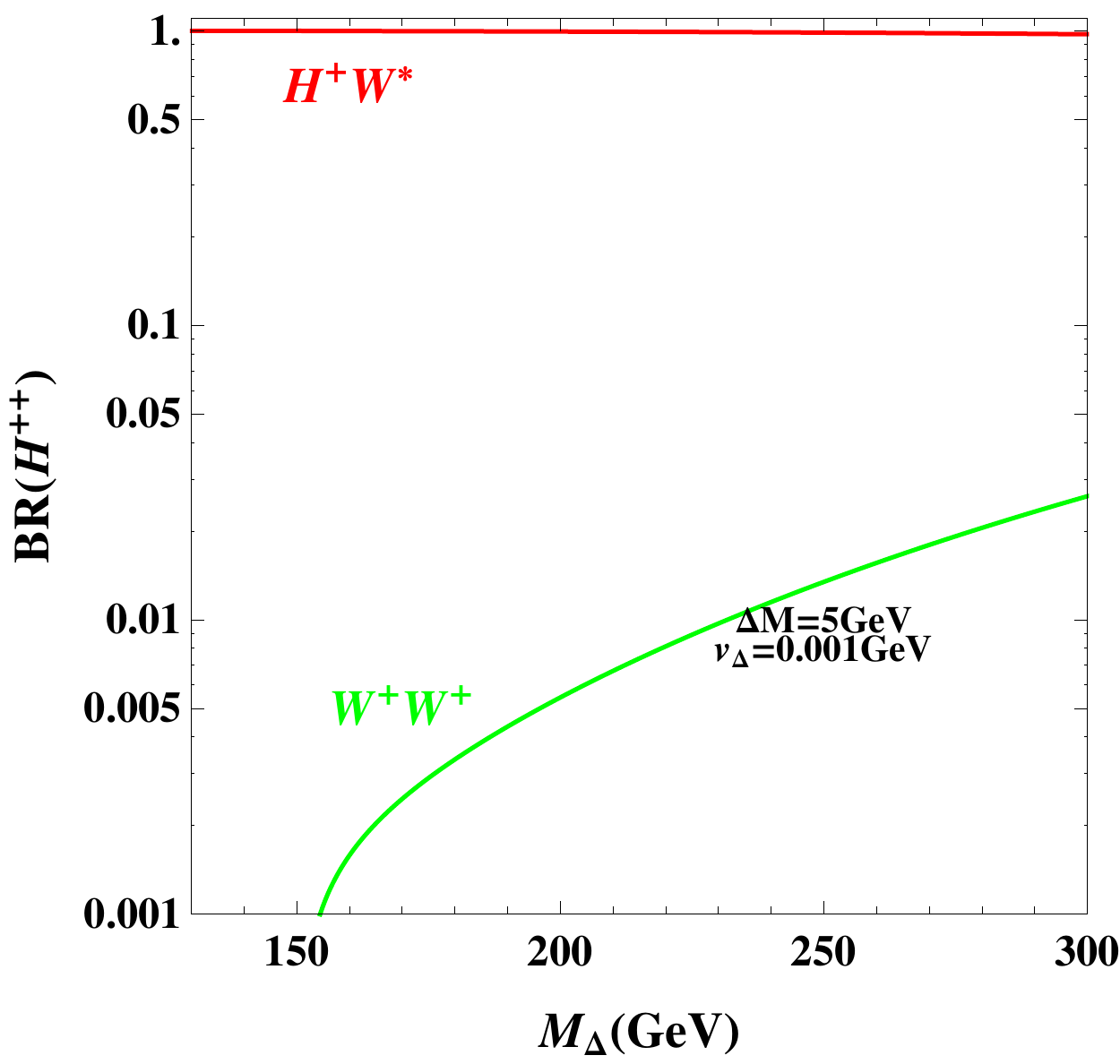}
\end{center}
\caption{Branching ratios of $H^{\pm}$ and $H^{\pm\pm}$ versus $M_{\Delta}$ at some benchmark points of $\Delta M$ and $v_{\Delta}$: $(\Delta M,v_\Delta)=(5,0.01),~(10,0.01),~(5,0.001)~\GeV$, from the upper to the lower panels. \label{brhp}}
\end{figure}

In the rest of this section, we discuss the decay properties of the new scalars in the negative scenario with an emphasis on $H^0$ and $A^0$. The explicit expressions for the relevant decay widths can be found in Refs.~\cite{Djouadi:2005gj,Aoki:2011pz,Chabab:2014ara}. It has been shown that $H^0/A^0$ decays dominantly into neutrinos for $v_{\Delta}<10^{-4}~\GeV$~\cite{Perez:2008ha}, resulting in totally invisible final states. We will restrict ourselves to $v_{\Delta}\gg 10^{-4}~\GeV$ in this work, where $H^0/A^0$ dominantly decays into visible particles. Before we detail their decay properties, we give a brief account of the cascade decays of the charged scalars. The branching ratios of the cascade decays are controlled by the three parameters, $v_{\Delta}$, $\Delta M$, and $M_{\Delta}$. The cascade decays dominate in the moderate region of $v_{\Delta}$ and for $\Delta M$ not too small, where a minimum value of $\Delta M\sim2~\GeV$ appears around $v_{\Delta}\sim10^{-4}~\GeV$~\cite{Perez:2008ha, Aoki:2011pz, Han:2015hba,Melfo:2011nx}. In Fig.~\ref{brhp}, the branching ratios of $H^{\pm}$ and $H^{\pm\pm}$ are shown as a function of $M_{\Delta}$ at some benchmark points of $v_{\Delta}$ and $\Delta M$. Basically speaking, in the mass region $M_\Delta=130$-$300~\GeV$, the cascade decays are dominant for a relatively large mass splitting $\Delta M$ (as shown in the middle panel of Fig.~\ref{brhp}) or a relatively small $v_{\Delta}$ (in the lower panel).

\subsection{$H^0$ decays}

At tree level, $H^0$ can decay to $f\bar{f}~(f=q,l)$, $\nu\nu$, $W^{+}W^{-}$, $ZZ$, and $hh$. It can also decay to $gg$, $\gamma\gamma$, and $Z\gamma$ through radiative effects. Similarly, $A^0 \to f\bar{f}$, $\nu\nu$, $Zh$ at tree level, and it has the same decay modes as $H^0$ at the loop level. Since we have chosen $v_{\Delta}\gg 10^{-4}~\GeV$, the neutrino mode can be safely neglected for both $H^0$ and $A^0$. Previous work usually concentrated on the decoupling region where the neutral scalars $H^0/A^0$ are much heaver than the light $CP$-even Higgs $h$ and the scalar self-couplings $\lambda_i$ are taken to be zero for simplicity~\cite{Perez:2008ha}. In this case, the mixing angle $\theta_0\approx\alpha$, and the $H^0W^+W^-$ coupling [being proportional to $\sin(\alpha-\theta_0)$] tends to vanish. As a consequence, the $W$-pair mode is absent and the dominant channels are $H^0 \to hh$, $ZZ$ for a heavy $H^0$. In contrast, we take into account the effect of scalar self-interactions and focus on the nondecoupling regime, i.e., $H^0/A^0$ are not much heavier than $h$.

For illustration, we choose the benchmark values $v_{\Delta}=10^{-3}~\GeV$, $\Delta M=5~\GeV$; then, $\lambda_5$ is determined by Eq.~(\ref{massrelation}) upon specifying $M_\Delta$. \footnote{As pointed out in Ref.~\cite{Aoki:2011pz}, varying $v_{\Delta}$ in the range $10^{-3}$-$1~\GeV$ would not change the branching ratios significantly.} To investigate the effect of the scalar self-interactions, we note the following features in the decays of $H^0$.
1) The decay widths of $H^0 \to f\bar{f},~gg$ differ from those of $h$ only by a factor of $\sin^2\theta_0$, which leads to similar behavior for $H^0$ and $h$.
2) The only free parameter for the mixing between $H^0$ and $h$ is $\lambda_4$, because [as shown in Eq.~(\ref{mixangles})] the impact of $\lambda_{2,3}$ is suppressed by a small $v_{\Delta}$ and a relatively large mass difference between $M_{\Delta}$ and $M_h$ while $\lambda_1$ is fixed by $M_h$.
3) $\lambda_4$ enters the $H^0W^+W^-$ and $H^0ZZ$ couplings and thus affects the decays $H^0\to W^+W^-,~ZZ$.
4) The $H^0hh$ coupling simplifies for $v_{\Delta}\ll v$ such that the only free parameter in the decay $H^0\to hh$ is again $\lambda_4$. As a consequence of these features, we shall choose $\lambda_4$ as a free parameter and vary it in the range $[-1.0,1.0]$, and fix the couplings $\lambda_2=\lambda_3=0.1$ which are involved in loop-induced decays.

\begin{figure}[!htbp]
\begin{center}
\includegraphics[width=0.43\linewidth]{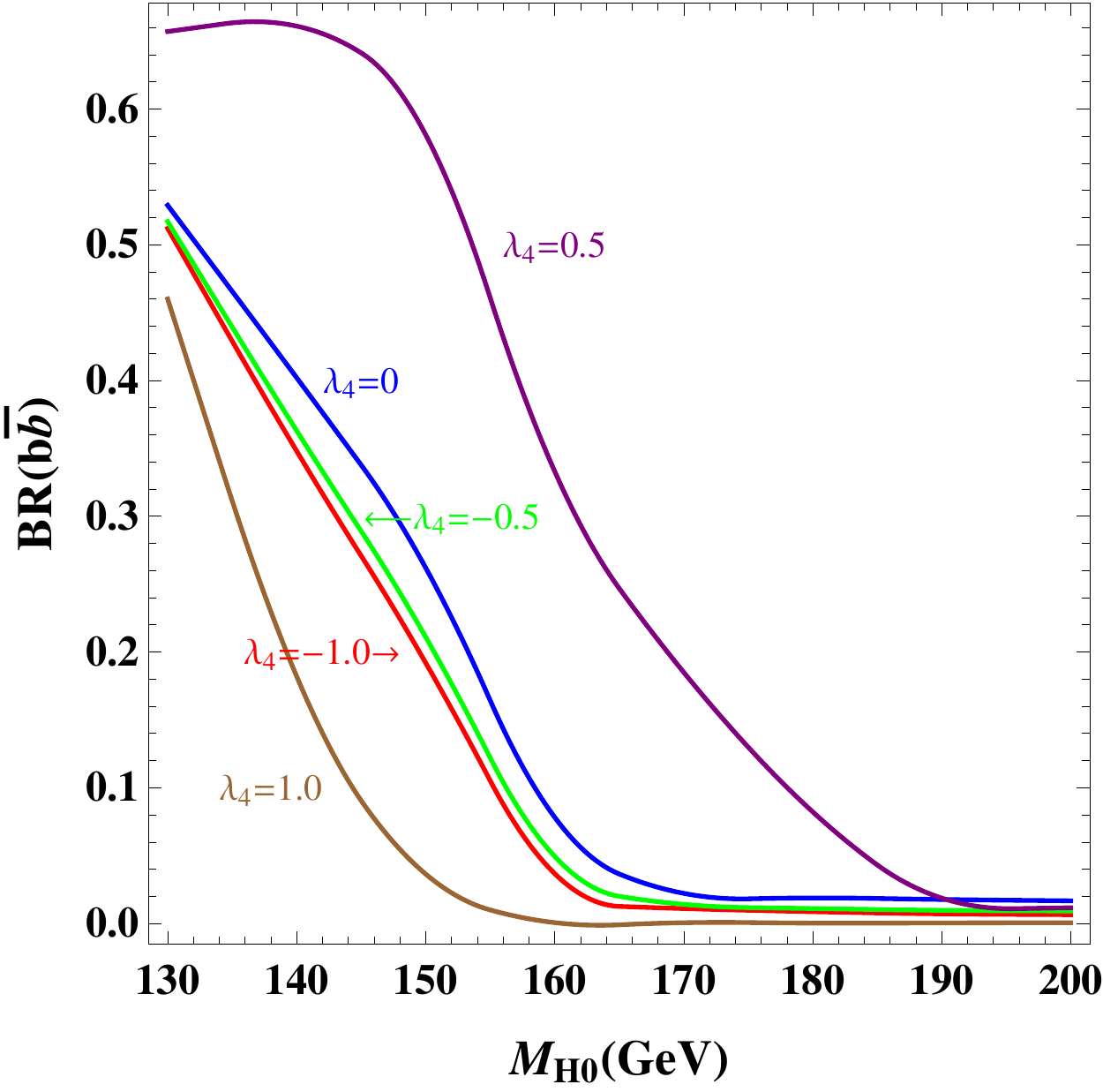}
\includegraphics[width=0.45\linewidth]{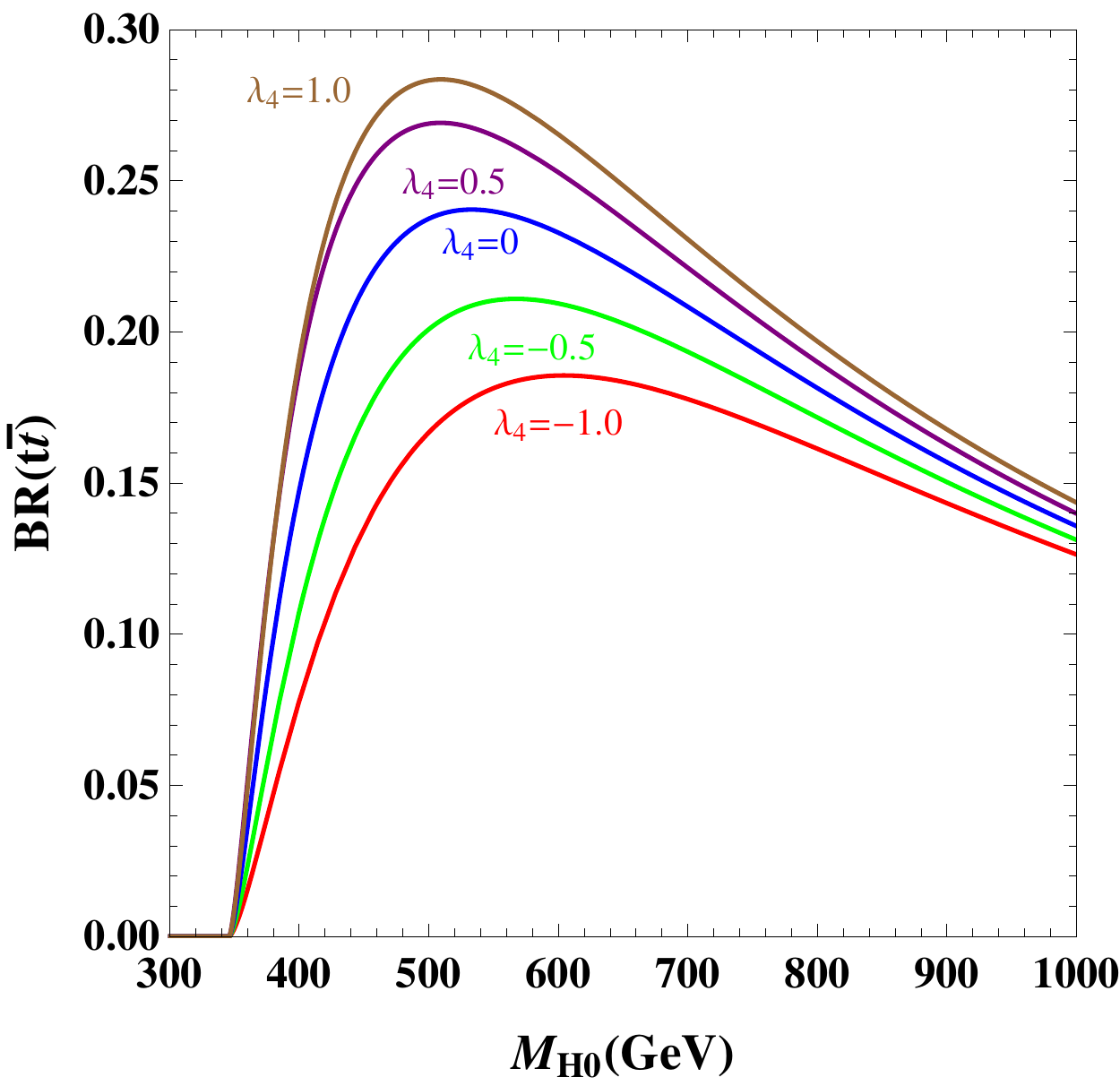}
\end{center}
\caption{Branching ratios of $H^0\to b\bar{b}$ and $H^0\to t\bar{t}$ as a function of $M_{H^0}$ for various values of $\lambda_4$.
\label{brh0tt}}
\end{figure}

We first examine the branching ratios of $H^0\to f\bar{f}$. BR($H^0\to b\bar{b}$) and BR($H^0\to t\bar{t}$) are plotted in Fig.~\ref{brh0tt} for different mass regions of $H^0$.
\footnote{The influence of $\lambda_4$ for light fermions $b,c,\tau,\mu$ and gluons is similar, so we only present BR($H^0\to b\bar{b}$) in Fig.~\ref{brh0tt}.}
It is clear that the variation of BR($H^0\to b\bar{b}$) is more dramatic for $\lambda_4>0$. The maximum of BR($H^0\to b\bar{b}$) appears at $\lambda_4\approx0.5$. Obviously, BR($H^0\to b\bar{b}$) is a nonmonotonic function of $\lambda_4$, while BR($H^0\to t\bar{t}$) monotonically increases with $\lambda_4$. As will be discussed later, this different behavior in the two mass regions is due mainly to a zero in the $H^0ZZ$ coupling.

\begin{figure}[!htbp]
\begin{center}
\includegraphics[width=0.45\linewidth]{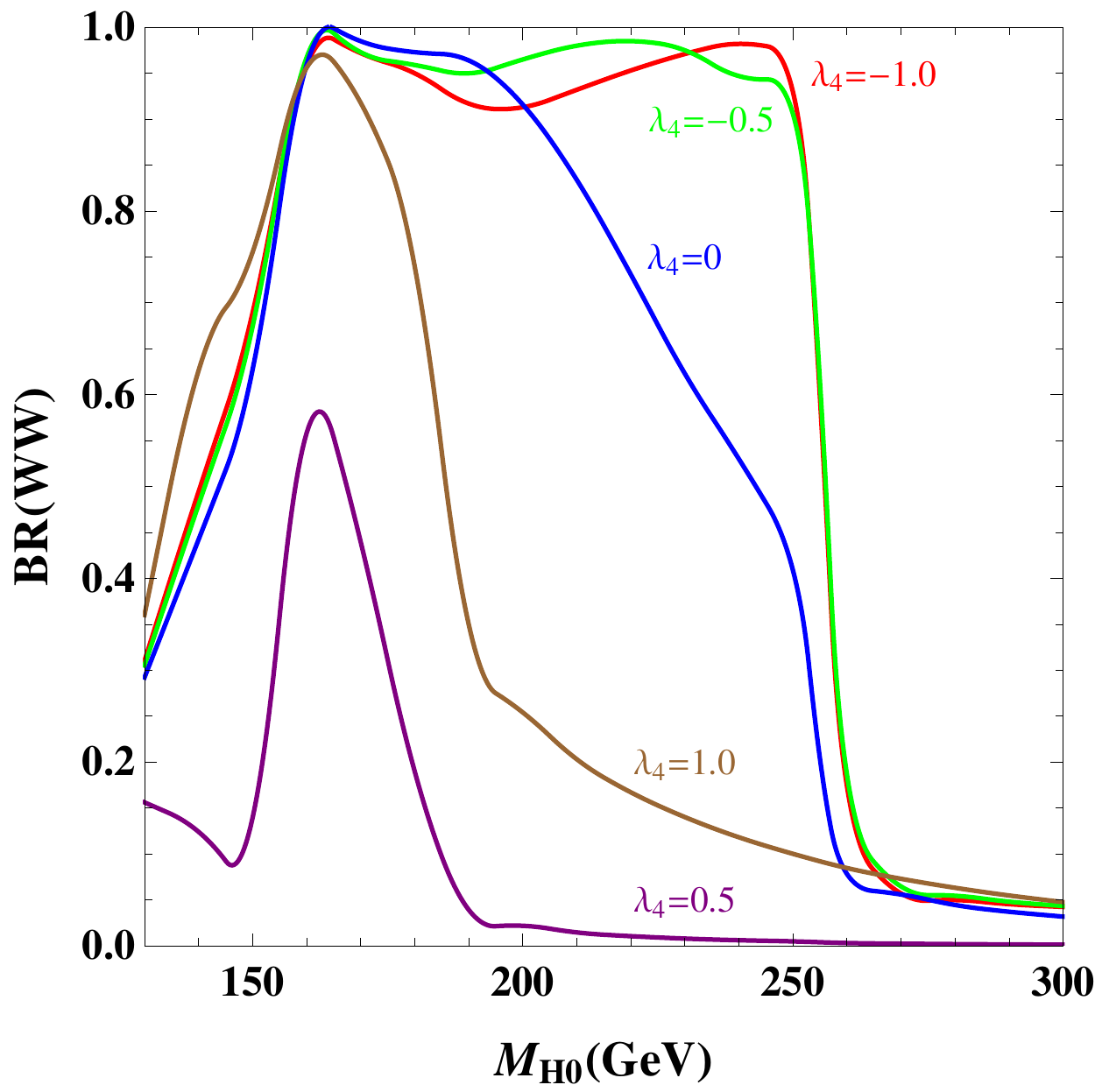}
\includegraphics[width=0.45\linewidth]{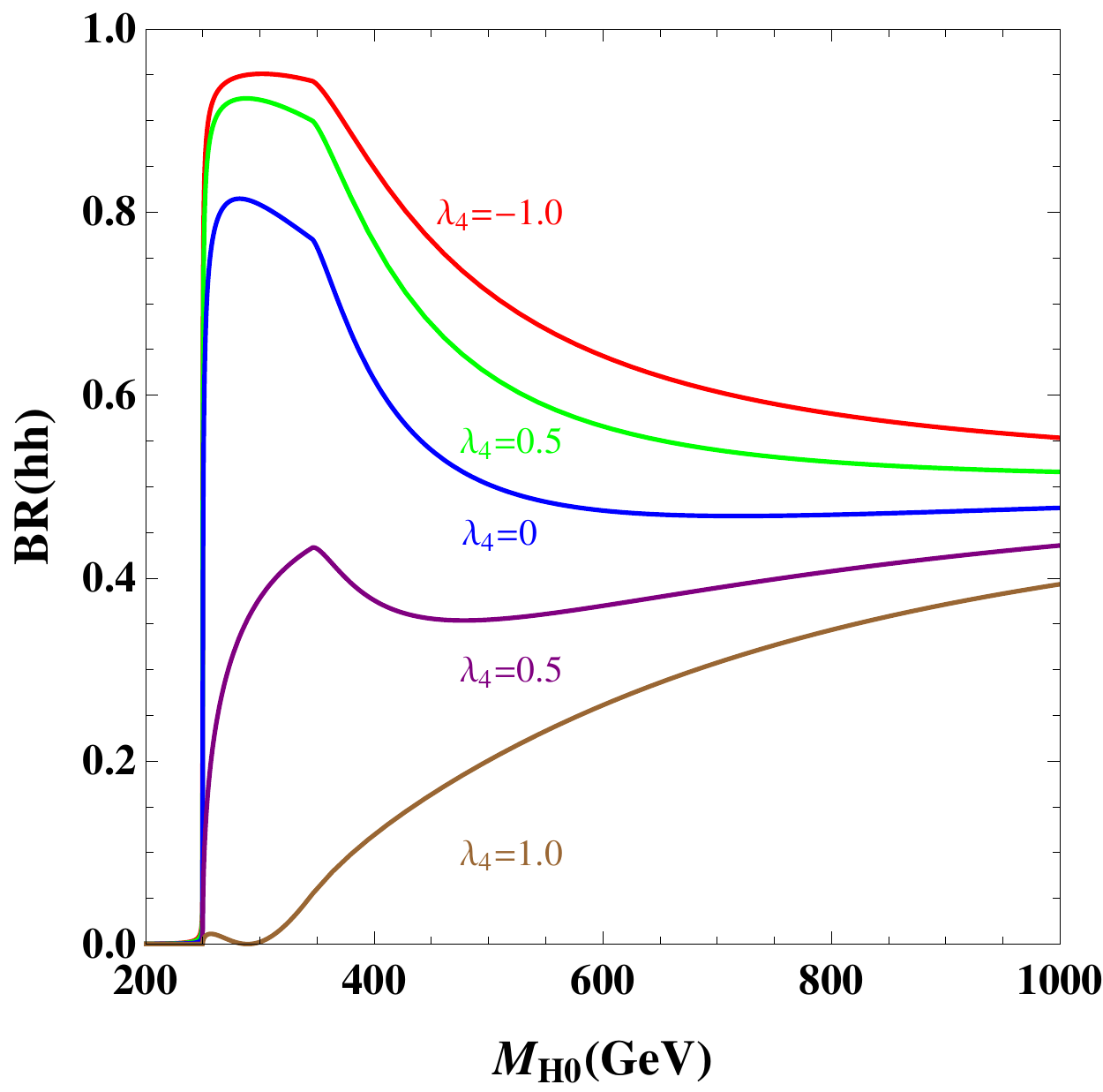}
\includegraphics[width=0.45\linewidth]{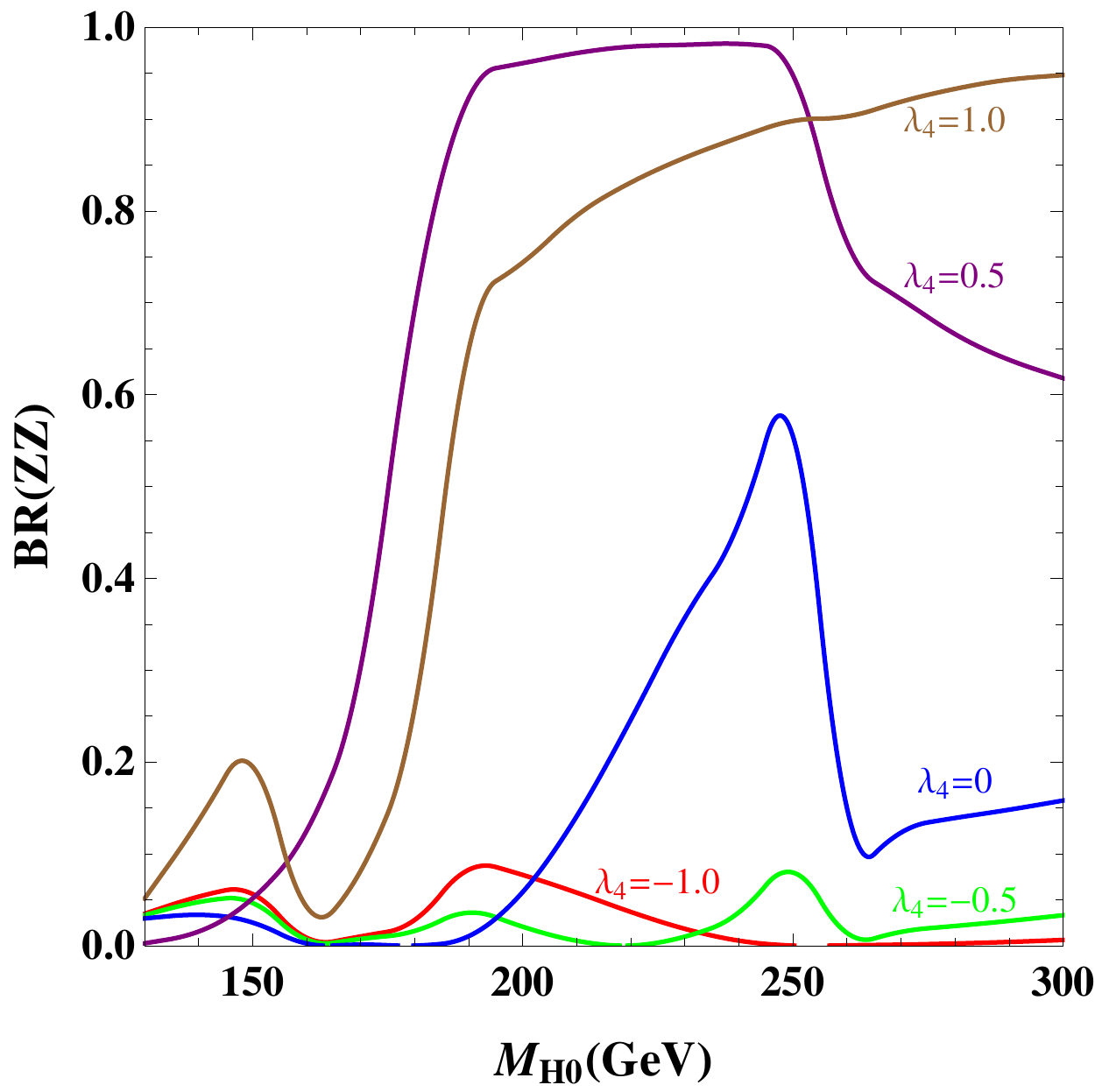}
\includegraphics[width=0.45\linewidth]{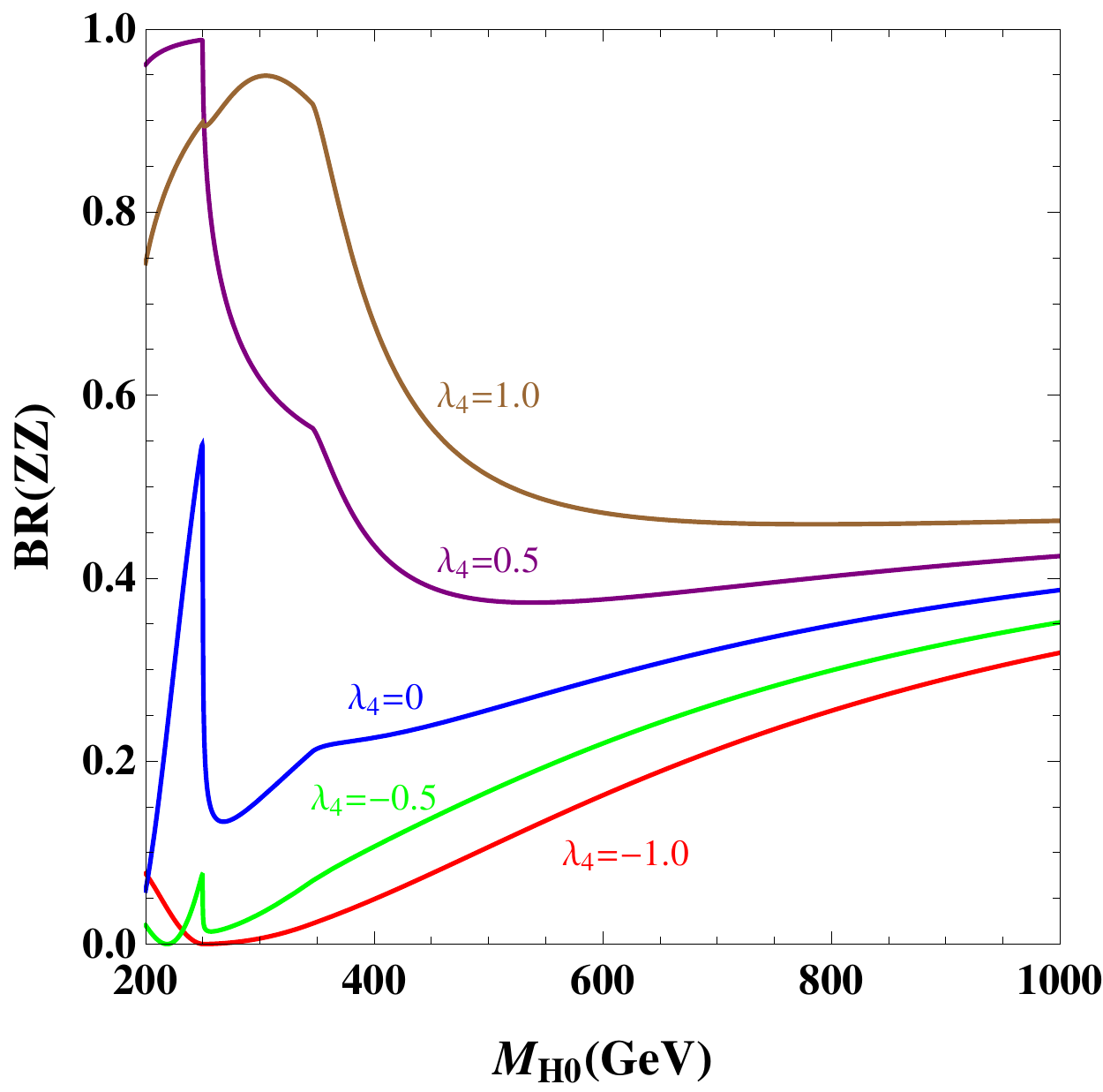}
\end{center}
\caption{Left: Branching ratios of $H^0\to W^+W^-,~ZZ$ as a function of $M_{H^0}$ in the mass region $130$-$300~\GeV$. Right: Branching ratios of $H^0\to hh,~ZZ$ as a function of $M_{H^0}$ in the mass region $200$-$1000~\GeV$.
\label{brh0WW}}
\end{figure}

Now we study the bosonic decays $H^0\to W^+W^-,~ZZ,~hh$. In the left panel of Fig.~\ref{brh0WW}, we present the branching ratios of $H^0\to W^+W^-,~ZZ$ in the mass region $130$-$300~\GeV$. For most values of $\lambda_4$, BR($H^0\to W^+W^-$) increases with $M_{H^0}$ when $M_{H^0}<2M_{W}$, and varying $\lambda_4$ for $\lambda_4>0$ changes it considerably. $\lambda_4$ has a strong impact on BR($H^0\to W^+W^-$) in the mass region $2M_{Z}<M_{H^0}<2M_{h}$ where the decay channel dominates
overwhelmingly for $\lambda_4<0$ but becomes negligible for $\lambda_4$ approaching about $0.5$. However, once the $H^0\to hh$ channel is opened, $H^0\to W^+W^-$ is suppressed significantly independent of $\lambda_4$. The decay $H^0\to ZZ$ cannot dominate when $M_{H^0}<2M_{W}$. In the mass region $2M_{Z}<M_{H^0}<2M_{h}$, it is complementary with the $W^+W^-$ channel, so their behavior is just opposite. More interestingly, there is a zero point for the $H^0ZZ$ coupling, which is proportional to $(v\sin\theta_0-4v_{\Delta}\cos\theta_0)$. According to Eq. (\ref{mixangles}), one obtains the corresponding $M_{\Delta}$ at the zero:
\begin{equation}
M_{\Delta}^0(ZZ)=\sqrt{2M_h^2-\frac{1}{2}(\lambda_4+\lambda_5)v^2}.
\end{equation}
Note that the above relation only holds for $\lambda_4+\lambda_5<2M_h^2/v^2\approx0.5$, since we are working in the scenario where $M_{\Delta}>M_h$. The existence of the zero coupling explains the presence of the nodes in BR($H^0\to ZZ$) for $\lambda_4\leq0$.

\begin{figure}[!htbp]
\begin{center}
\includegraphics[width=0.45\linewidth]{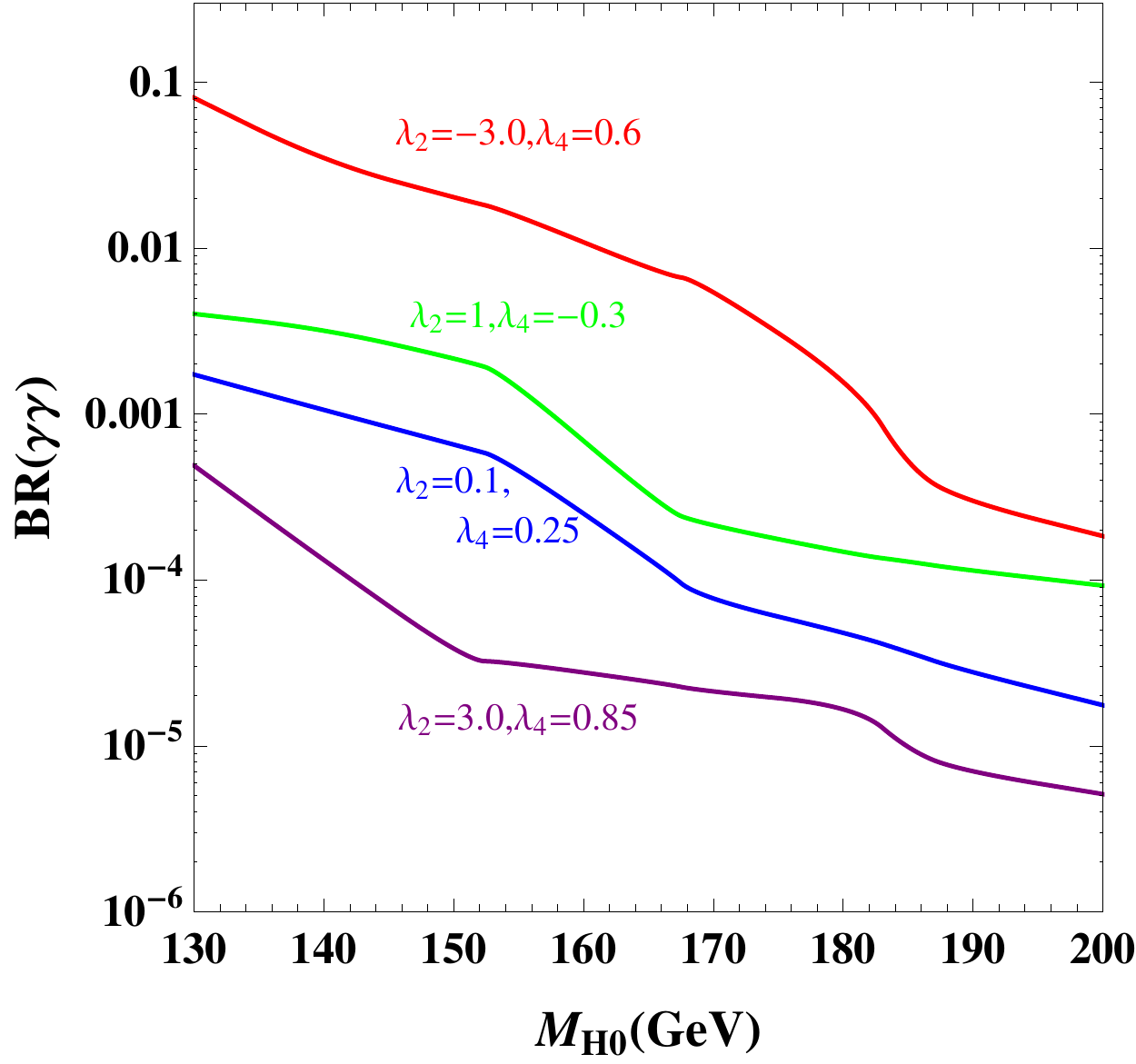}
\includegraphics[width=0.44\linewidth]{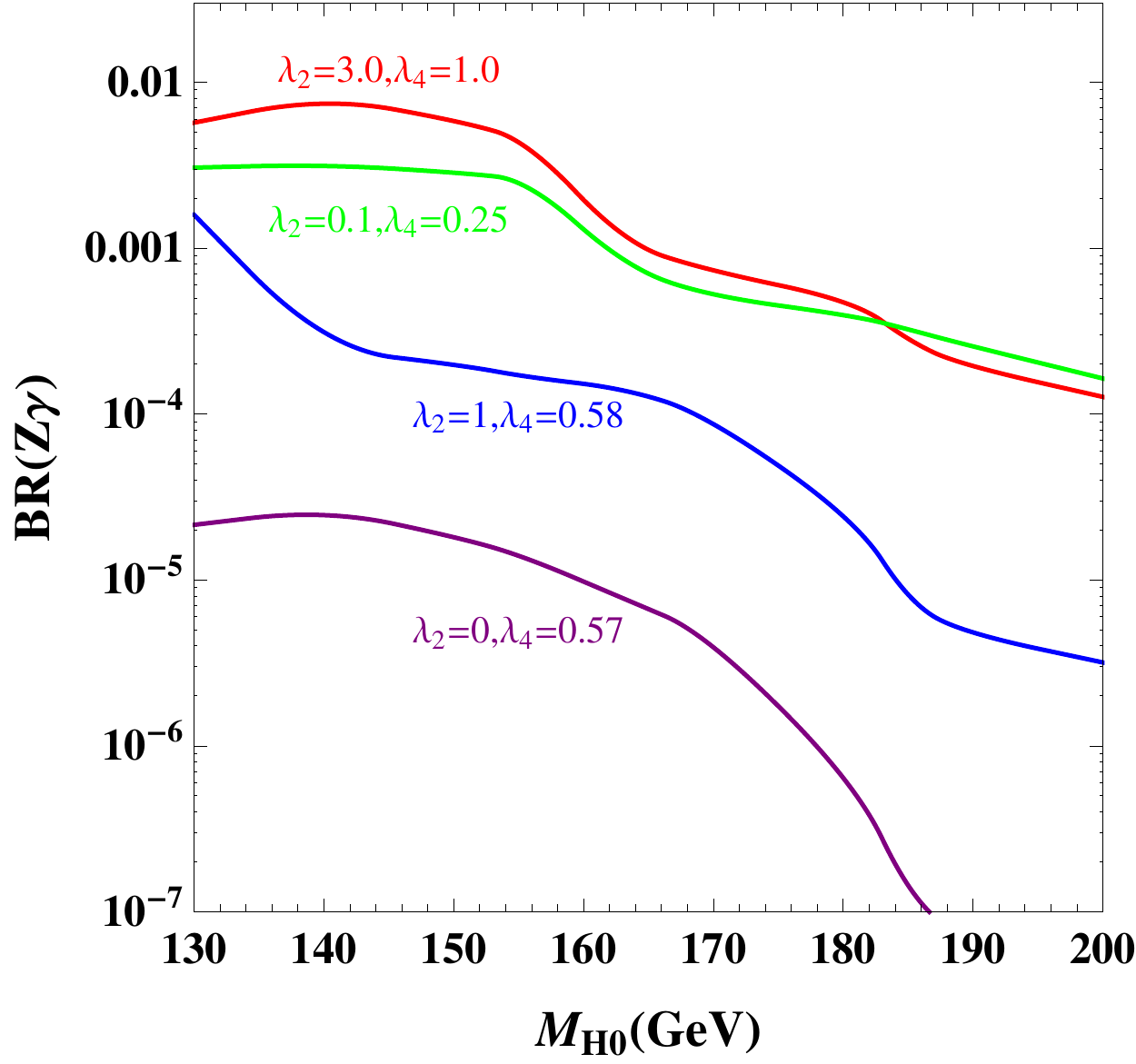}
\end{center}
\caption{Branching ratios of $H^0\to \gamma\gamma,~Z\gamma$ as a function of $M_{H^0}$ for various sets of $\lambda_{2,4}$ values.
\label{brh0AA}}
\end{figure}

In the right panel of Fig. \ref{brh0WW}, BR($H^0\to hh,ZZ$) are shown in the mass region $200$-$1000~\GeV$. When $M_{H^0}>2M_{h}$, the dependence on $\lambda_4$ is simple: a larger $\lambda_4$ corresponds to a smaller BR($H^0\to hh$) and a larger BR($H^0\to ZZ$). It is clear that $\lambda_4$ has a more significant impact in the mass region $200\sim350~\GeV$, and varying $\lambda_4$ could change BR($H^0\to ZZ$) from $0$ to $0.9$. Once $M_{H^0}$ exceeds $2M_t$, the evolution of Br($H^0\to hh,ZZ$) becomes smooth with the increase of $M_{H^0}$. There also exists a zero point for the $H^0hh$ coupling, which can be obtained as for the $ZZ$ channel:
\begin{equation}
M_{\Delta}^0(hh)=\sqrt{2(\lambda_4+\lambda_5)v^2-2M_h^2}~,
\end{equation}
which is valid for $\lambda_4+\lambda_5>3M_h^2/2v^2\approx0.375$.

Finally, we investigate the loop-induced decays, $H^0\to \gamma\gamma,~Z\gamma$. In addition to the usual contributions from the top quark and $W$ boson, the new charged scalars $H^{\pm}$ and $H^{\pm\pm}$ also contribute to the decays. These new terms involve the $H^0H^+H^-$ and $H^0H^{++}H^{--}$ couplings, which are proportional to
\begin{eqnarray}
H^0H^+H^-&:& [(2\lambda_2+2\lambda_3-\lambda_5)\sin\alpha\cos\theta_0-(2\lambda_4+\lambda_5)\cos\alpha\sin\theta_0],
\nonumber
\\
H^0H^{++}H^{--}&:&(\lambda_2\sin\alpha\cos\theta_0-\lambda_4\cos\alpha\sin\theta_0).
\end{eqnarray}
One therefore has to consider the scalar self-couplings $\lambda_{2,3}$. For simplicity, we set $\lambda_2=\lambda_3$ and vary them from $-3.0$ to $3.0$. In Fig. \ref{brh0AA}, we display BR($H^0\to \gamma\gamma$) and BR($H^0\to Z\gamma$) versus $M_{H^0}$ for some typical sets of $\lambda_{2,4}$ values. The evolution of both branching ratios crosses 3 orders of magnitude in this parameter region. The resulting enhancement compared with $h\to\gamma\gamma$ in the SM looks significant: the maximal enhancement can be achieved at the level of $9\%$ for the $H^0\to \gamma\gamma$ channel at $M_{H^0}=130~\GeV$, and of $0.7\%$ for the $H^0\to Z\gamma$ channel at $M_{H^0}\approx140~\GeV$.

\subsection{$A^0$ decays}

\begin{figure}[!htbp]
\begin{center}
\includegraphics[width=0.43\linewidth]{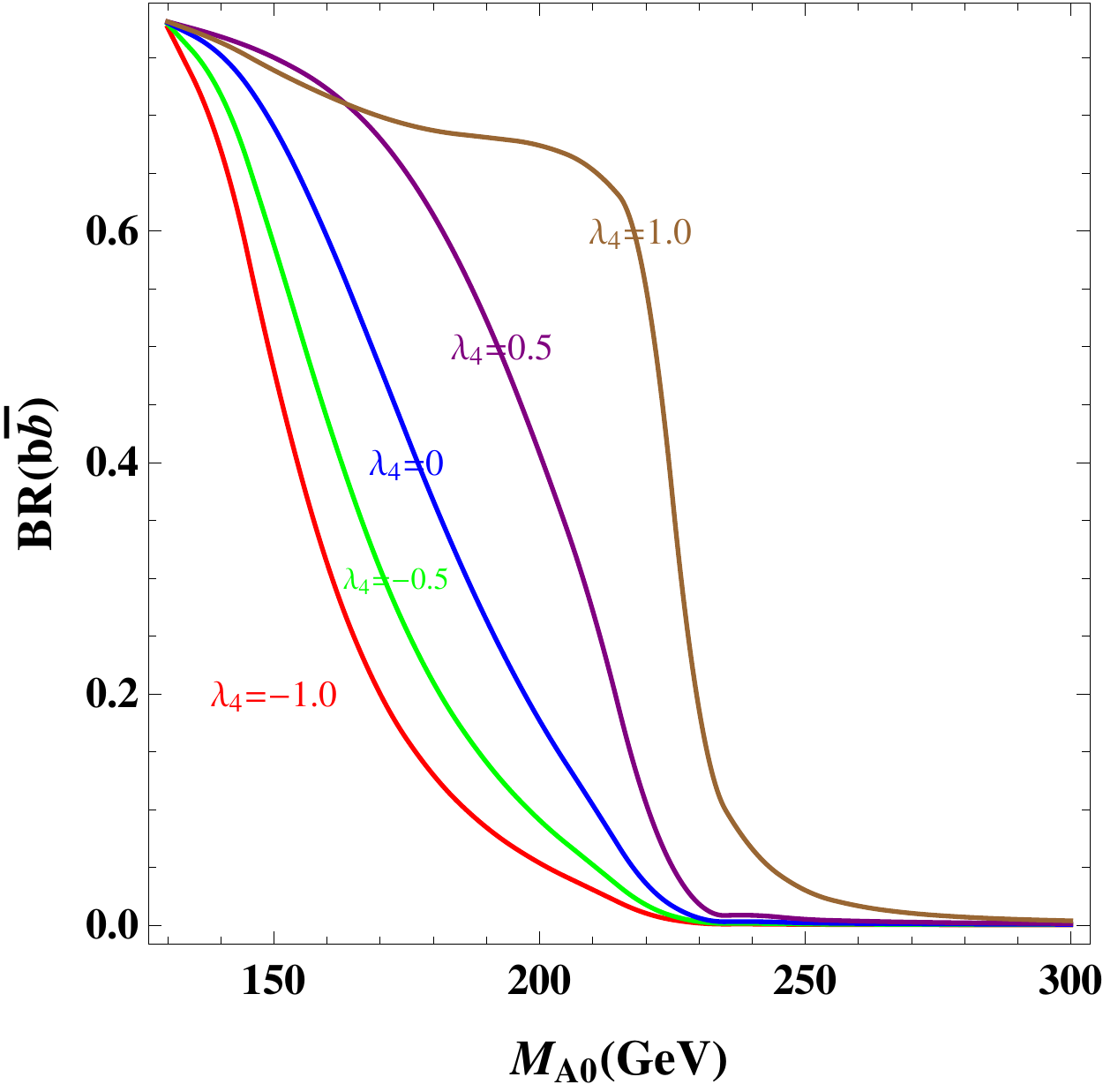}
\includegraphics[width=0.45\linewidth]{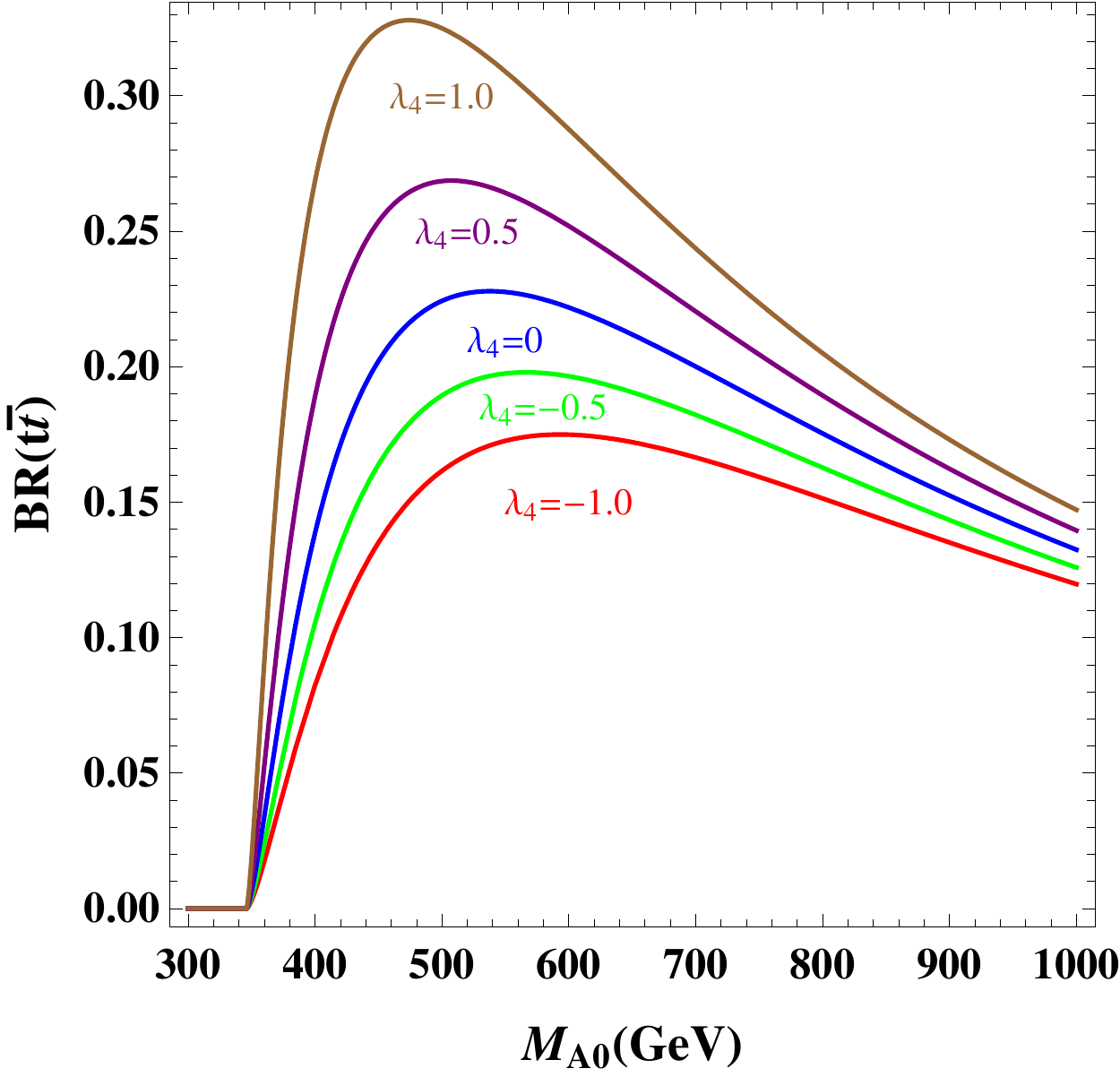}
\end{center}
\caption{Branching ratios of $A^0\to b\bar{b},~t\bar{t}$ as a function of $M_{A^0}$ for various values of $\lambda_4$.
\label{bra0tt}}
\end{figure}

Similar to $H^0$, the decay widths of $A^0 \to f\bar{f},~gg$ differ from those of $h$ by a factor of $\sin^2\alpha$ with $\alpha$ being given in Eq. (\ref{mixangles}). Moreover, the only vertex which involves $\lambda_i$ is the $A^0Zh$ coupling proportional to $(\cos\theta_0\sin\alpha-2\sin\theta_0\cos\alpha)$. As a consequence, one can only choose $\lambda_4$ as a free parameter to illustrate the influence of scalar interactions. In this section, we also vary $\lambda_4$ from $-1.0$ to $1.0$ and take the same benchmark values for $v_{\Delta}$ and $\Delta M$ as for the $H^0$ decays.

In the left panel of Fig. \ref{bra0tt}, we present BR($A^0\to b\bar{b}$) as a function of $M_{A^0}$.~\footnote{As before, the influence of $\lambda_4$ on the $A^0\to f \bar{f},~gg$ channels is similar to the $b\bar{b}$ mode.} For a fixed value of $\lambda_4$, BR($A^0\to b\bar{b}$) decreases as $M_{A^0}$ increases. The dependence of BR($A^0\to b\bar{b}$) on $\lambda_4$ is simple: The larger $\lambda_4$ is, the larger BR($A^0\to b\bar{b}$) is. And BR($A^0\to b\bar{b}$) can be dominant with $\lambda_4=1.0$ as long as $A^0\to Zh$ is not fully opened. The right panel of Fig. \ref{bra0tt} shows BR($A^0\to t\bar{t}$), which is very similar to BR($H^0\to t\bar{t}$).

\begin{figure}[!htbp]
\begin{center}
\includegraphics[width=0.44\linewidth]{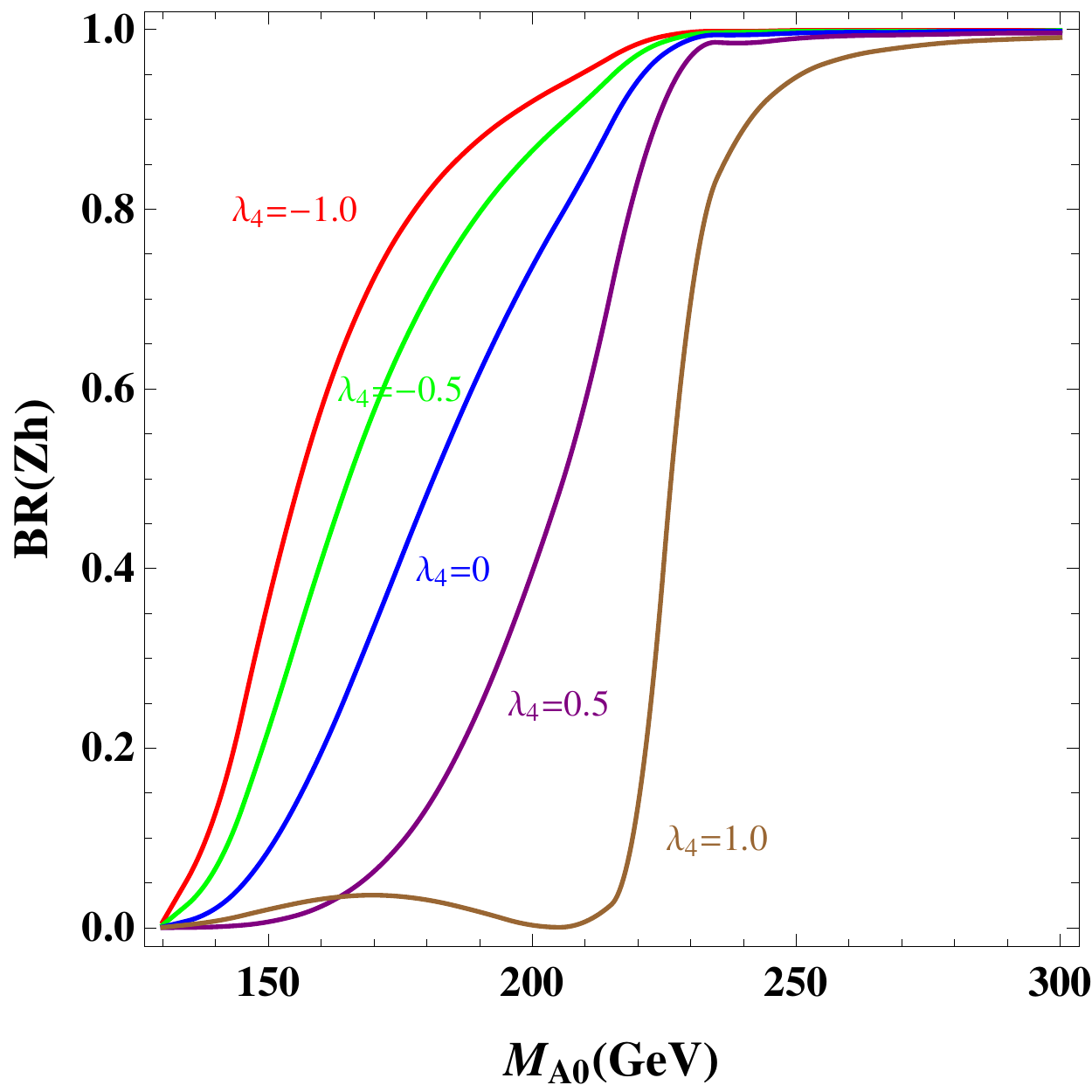}
\includegraphics[width=0.45\linewidth]{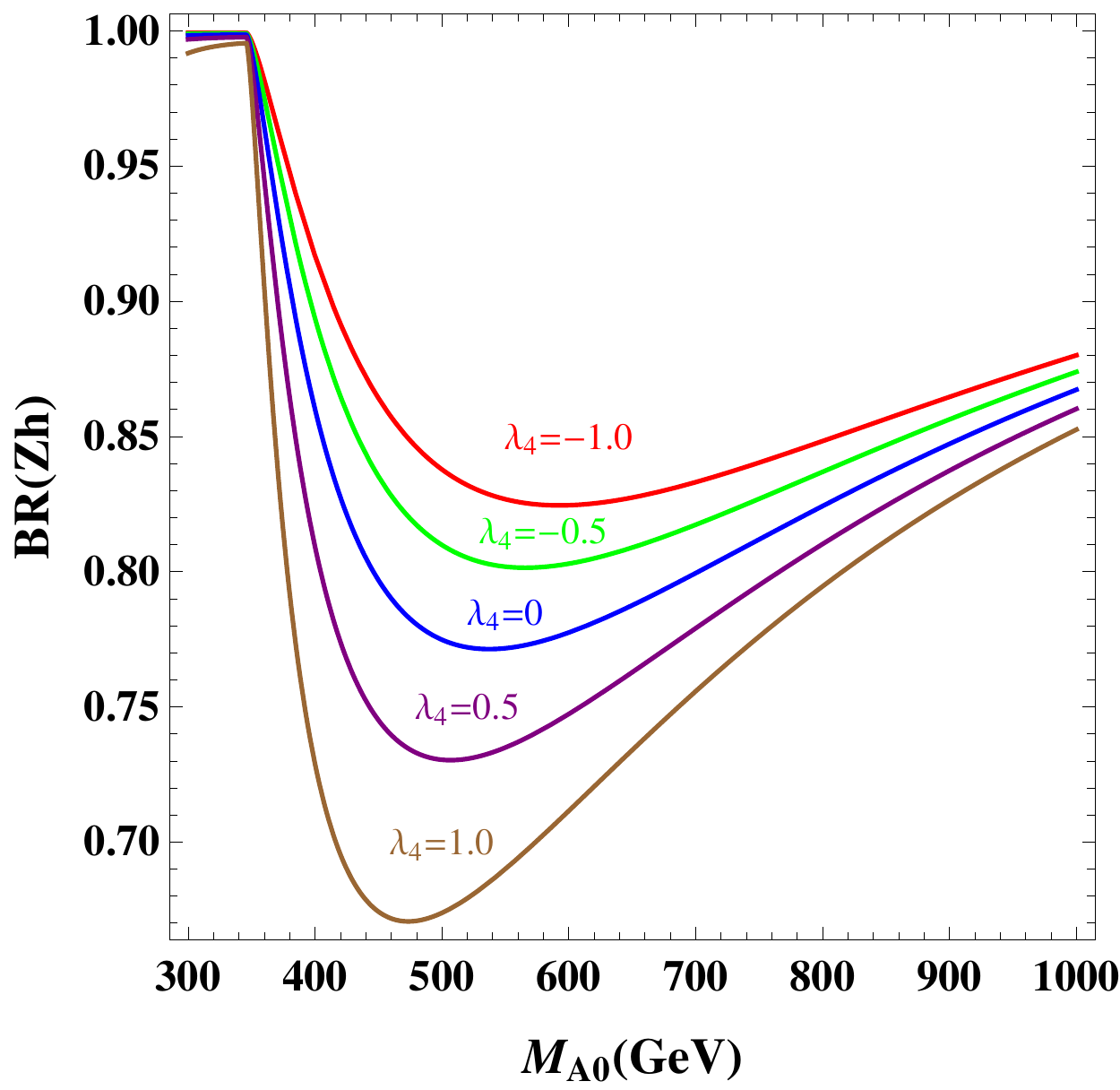}
\end{center}
\caption{Branching ratios of $A^0\to Zh$ as a function of $M_{A^0}$ for various values of $\lambda_4$..
\label{bra0zh}}
\end{figure}

\begin{figure}[!htbp]
\begin{center}
\includegraphics[width=0.43\linewidth]{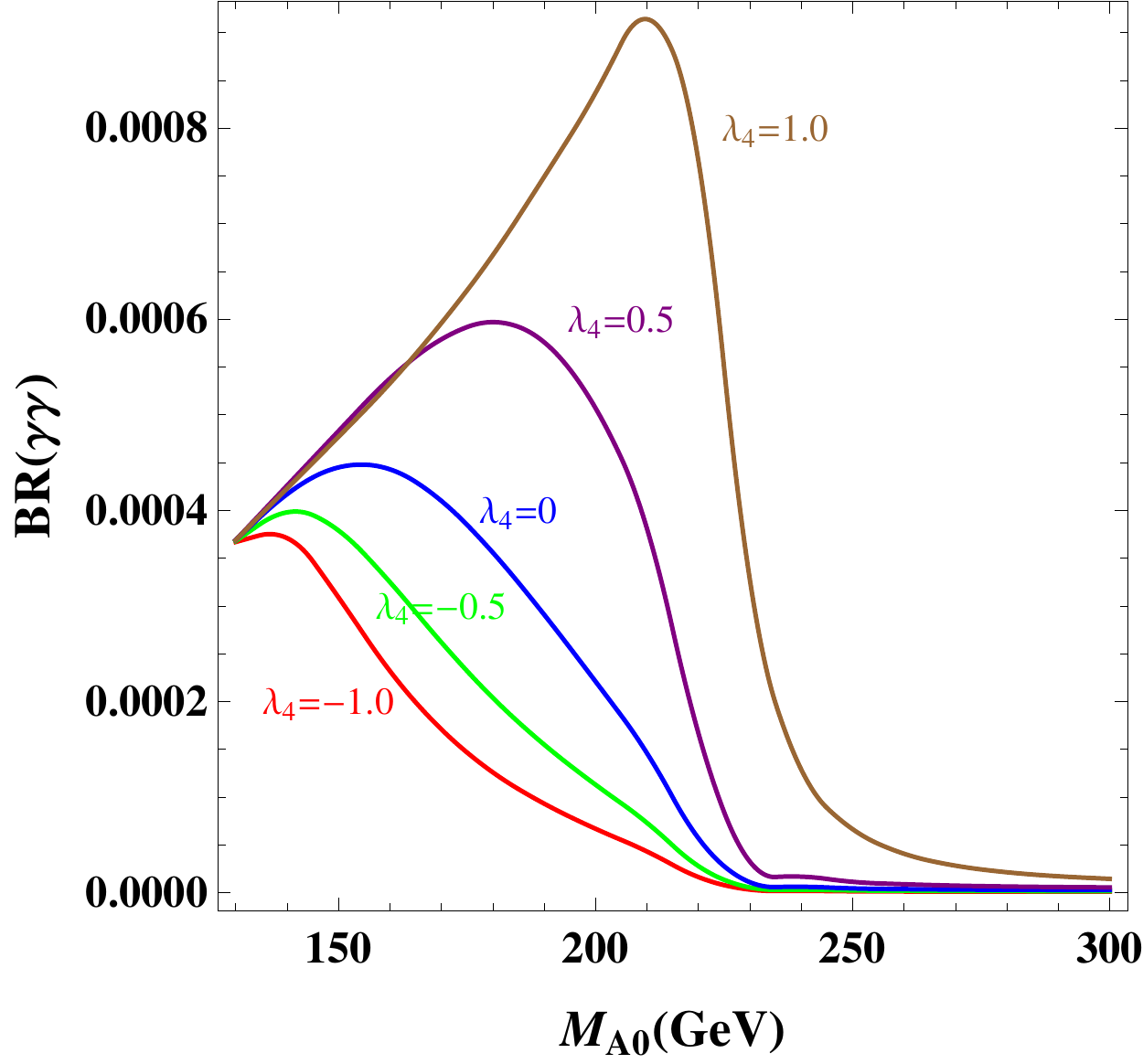}
\includegraphics[width=0.45\linewidth]{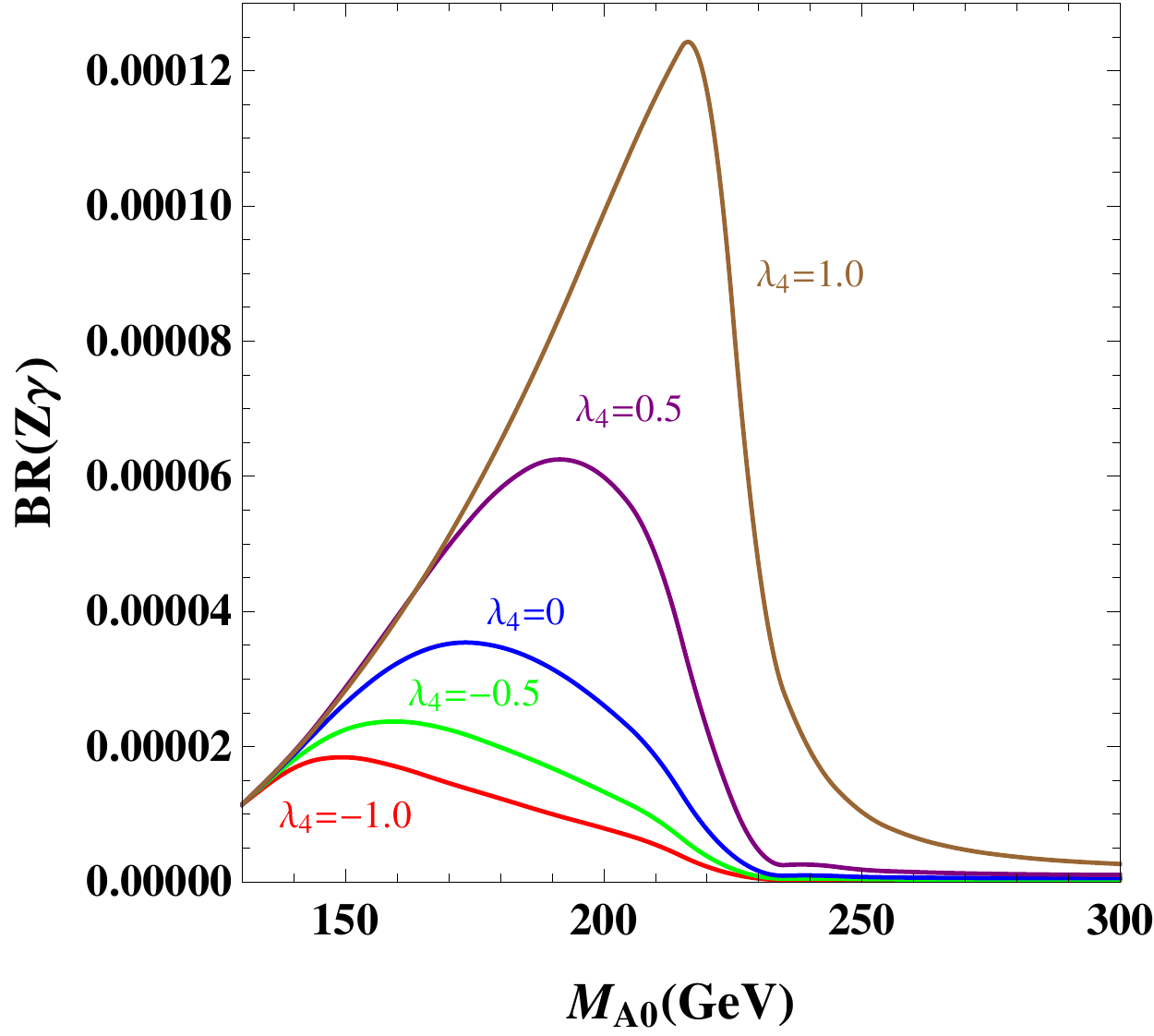}
\end{center}
\caption{Branching ratios of $A^0\to \gamma\gamma$ and $A^0\to Z\gamma$ as a function of $M_{A^0}$ for various values of $\lambda_4$.
\label{bra0aa}}
\end{figure}

We then study the most important decay $A^0\to Zh$. In Fig. \ref{bra0zh}, we present BR($A^0\to Zh$) as a function of $M_{A^0}$ in the low-mass region ($130$-$300~\GeV$) and  high-mass region ($300$-$1000~\GeV$), respectively. The evolution of BR($A^0\to Zh$) with $M_{A^0}$ and $\lambda_4$ is just opposite to that of $A^0\to b\bar{b}~(t\bar t)$ in the low- (high-) mass region. The variation of BR($A^0\to Zh$) with $\lambda_4$ is dramatic below the $Zh$ threshold. In particular, near the $Zh$ threshold BR($A^0\to Zh)\sim 1.0$ for $\lambda_4=-1.0$, while BR($A^0\to Zh$) tends to vanish for $\lambda_4=1.0$, which corresponds to the zero point of the $A^0Zh$ coupling:
\begin{equation}
M_{\Delta}^0(Zh)=\sqrt{(\lambda_4+\lambda_5)v^2-M_h^2},
\end{equation}
with $\lambda_4+\lambda_5>2M_h^2/v^2\approx0.5$. BR($A^0\to Zh$) is totally dominant in the mass region between the $Zh$ and $t\bar t$ thresholds, and becomes comparable to BR($A^0\to t\bar{t}$) when $M_{A^0}>2M_t$.

At last, we study the one-loop-induced decays, $A^0\to \gamma\gamma,~Z\gamma$. These two channels can only be induced by the top quark in the loop since the $A^0W^+W^-$, $A^0H^+H^-$, and $A^0H^{++}H^{--}$ couplings are absent in the $CP$-conserving case. In Fig. \ref{bra0aa}, both BR($A^0\to \gamma\gamma$) and BR($A^0\to Z\gamma$) are displayed. For $M_{A^0}$ below the $Zh$ threshold, the variation in $\lambda_4$ of BR($A^0\to \gamma\gamma$) increases as $M_{A^0}$ increases. BR($A^0\to\gamma\gamma$) could reach $9\times10^{-4}$ for $M_{A^0}\approx 210~\GeV$ and $\lambda_4=1.0$, which is much smaller than the maximum of BR($H^0\to\gamma\gamma$). The variation in $\lambda_4$ of BR($A^0\to Z\gamma$) is slightly steeper, with a maximum of $1.2\times10^{-4}$ at $M_{A^0}\approx 215~\GeV$ and $\lambda_4=1.0$.

\begin{figure}[!htbp]
\begin{center}
\includegraphics[width=0.45\linewidth]{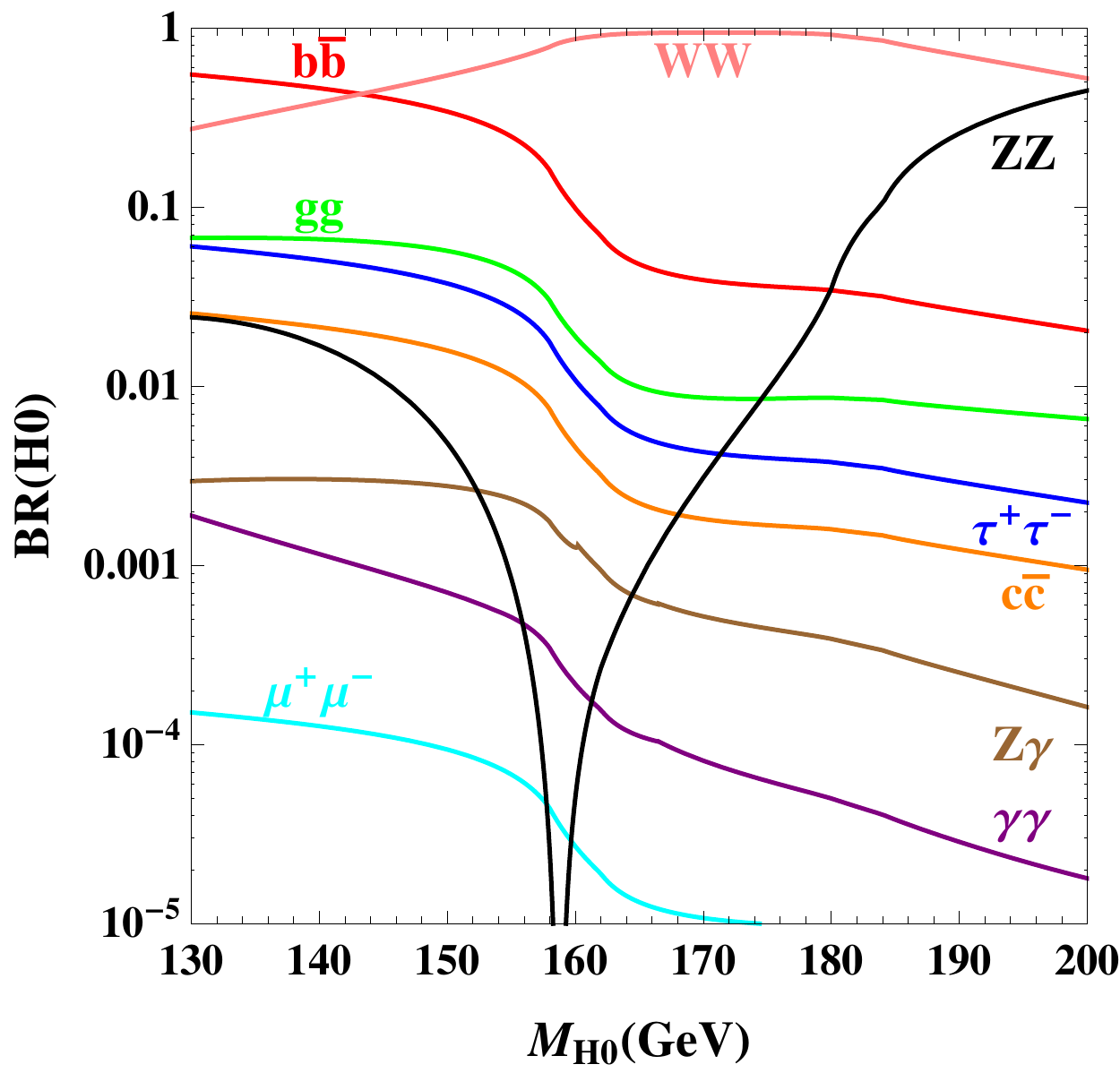}
\includegraphics[width=0.45\linewidth]{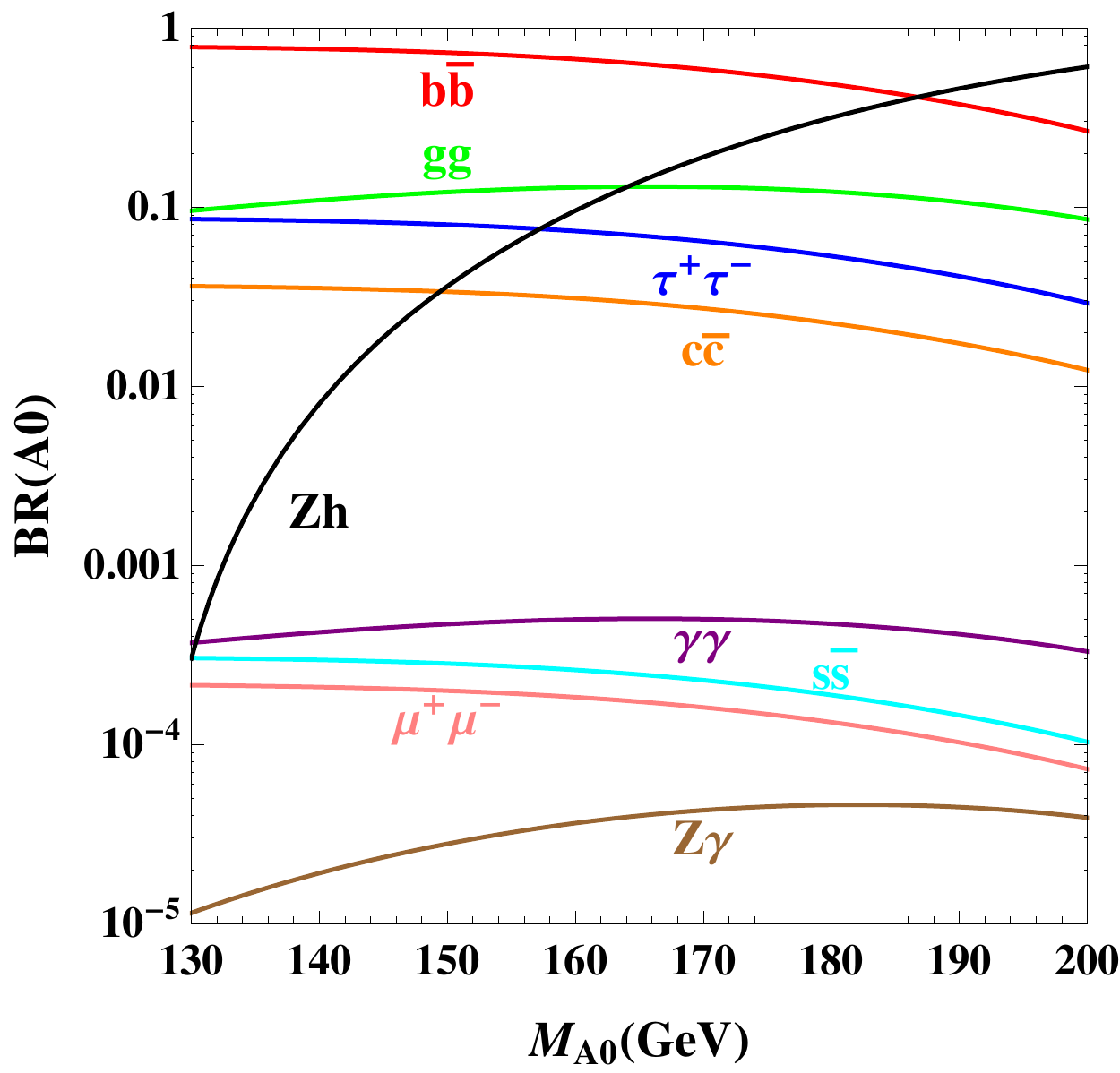}
\end{center}
\caption{Branching ratios of $H^0/A^0$ as a function of $M_{H^0/A^0}$ at the benchmark point in Eq.~(\ref{BP}).
\label{brh0}}
\end{figure}

In the above, we have discussed the decay channels of $H^0$ and $A^0$ separately. We have shown that the scalar self-interactions have a large impact on their branching ratios. In Sec.~\ref{signal}, we will explore their LHC signatures. For this purpose, we choose the following benchmark values:
\begin{equation}\label{BP}
v_{\Delta}=0.001~\GeV, ~\Delta M=5~\GeV, ~\lambda_2=\lambda_3=0.1, ~\lambda_4=0.25.
\end{equation}
The reason that we set relatively small values of $v_{\Delta}$ and $\Delta M$ is to obtain large cascade decays of charged scalars as well as a large enhancement of neutral scalar production. In Fig. \ref{brh0}, we display all relevant branching ratios versus $M_{H^0/A^0}$ for this benchmark model, which is to be simulated in Sec.~\ref{signal} for the LHC in the $b\bar{b}\gamma\gamma$, $b\bar{b}\tau^+\tau^-$, and $b\bar{b}W^+W^-$ signal channels.

%%%%%%%%%%%%%%%%%%%%%%%%%%%%%%%%%%%%%%%%%%%%%%%%%%%%%%%%
\section{Production of Neutral Higgs from Cascade Decays}\label{Eh}
%%%%%%%%%%%%%%%%%%%%%%%%%%%%%%%%%%%%%%%%%%%%%%%%%%%%%%%%%%

\begin{figure}[!htbp]
\begin{center}
\includegraphics[width=0.45\linewidth]{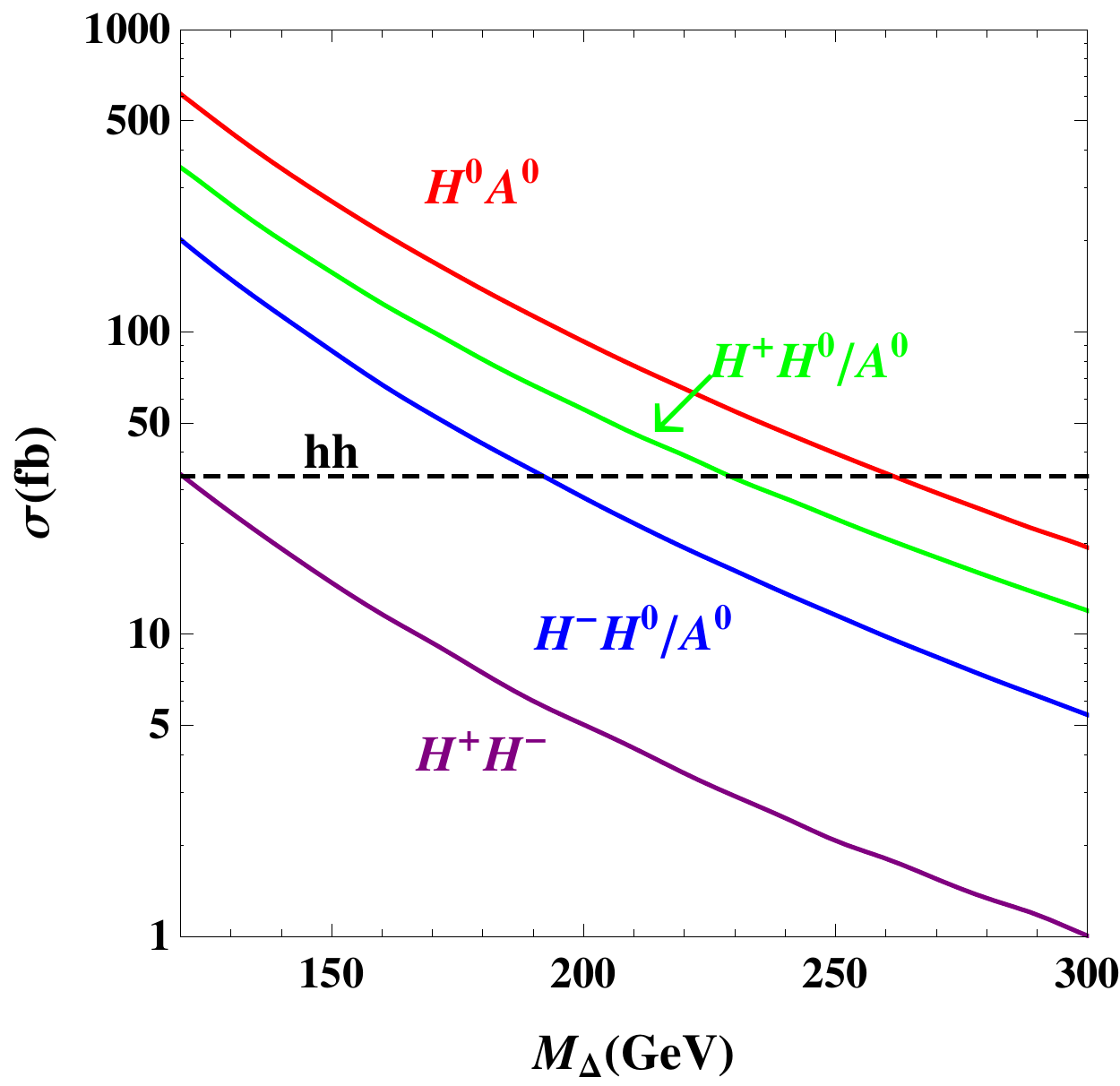}
\includegraphics[width=0.45\linewidth]{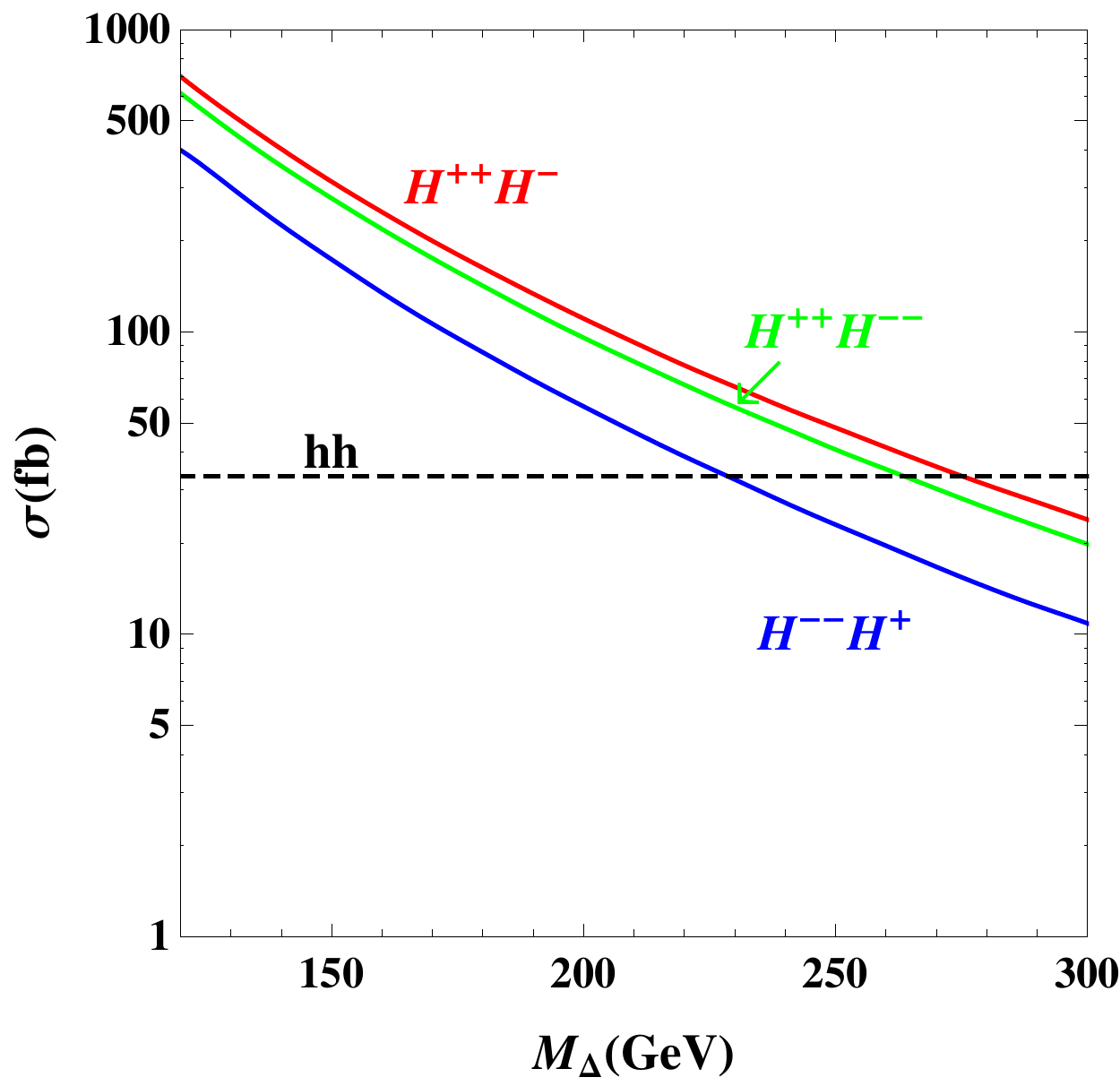}
\end{center}
\caption{Production cross sections for a pair of scalars at LHC14 versus $M_{\Delta}$ for a degenerate spectrum. The black dashed line is for the SM $hh$ production.
\label{cs}}
\end{figure}

We pointed out in Ref.~\cite{Han:2015hba} the importance of the associated $H^0A^0$ production in the nondegenerate case. To estimate the number of signal events, we simulated the signal channel $b\bar{b}\tau^+\tau^-$ at $M_{H^0/A^0}=130~\GeV$. We found that, with a much higher production cross section than the SM Higgs pair ($hh$) production, a $2.9\sigma$ excess in that signal channel is achievable for LHC14@300. In the present work, we are interested in the observability of the associated $H^0A^0$ production in the nondecoupling mass regime $(130$-$200~\GeV)$. In Fig. \ref{cs} we first show the production cross sections for a pair of various scalars at LHC14 versus $M_{\Delta}$ with a degenerate spectrum. As before, we incorporate the next-to-leading-order (NLO) QCD effects by multiplying a $K$-factor of $1.3$ in all $q\bar{q}$ production channels~\cite{Dawson:1998py}. The $hh$ production through gluon-gluon fusion at NLO ($33~\fb$) is also indicated (black dashed line) for comparison. One can see that the cross section for $H^0A^0$ is about $20$-$500~\fb$ in the mass region $130$-$300~\GeV$, which is much larger than the $hh$ production for most of the mass region and thus leads to great discovery potential.

In general, the new scalars are nondegenerate for a nonzero $\lambda_5$. In the positive scenario where $H^{\pm\pm}$ are the lightest, the cascade decays of $H^\pm$ and $H^0/A^0$ can strengthen the observability of $H^{\pm\pm}$~\cite{Akeroyd:2011zza,Chun:2013vma}. For the same reason, in the negative scenario where $H^0/A^0$ are the lightest, the charged scalars contribute instead to the production of $H^0/A^0$ through the cascade decays like $H^{\pm}\to H^0/A^0W^*$. In this work, we study these contributions in the same way as was done for the positive scenario in Refs.~\cite{Akeroyd:2011zza,Chun:2013vma}.

We define the reference cross section $X_0$ for the standard Drell-Yan process
\begin{equation}
X_0=\sigma(pp\to Z^*\to H^0A^0),
\end{equation}
which is independent of the cascade decay parameters $v_{\Delta}$ and $\Delta M$. A detailed study on the $b\bar{b}\tau^+\tau^-$ signal for this process with $M_{\Delta}=130~\GeV$ can be found in Ref.~\cite{Han:2015hba}. Besides the above direct production, neutral scalars can also be produced from cascade decays of charged scalars. These extra production channels include $H^\pm H^0/A^0$, $H^+H^-$, $H^{\pm}H^{\mp\mp}$, and $H^{++}H^{--}$ followed by cascade decays of charged scalars. We consider first the associated $H^\pm H^0/A^0$ production followed by cascade decays of $H^\pm$,
%There are four such processes that involve a pair of neutral scalars in the final state,
\begin{eqnarray}
\nonumber
pp\to W^*\to H^{\pm}H^0 \to H^0H^0 W^* &,&~~
pp\to W^*\to H^{\pm}H^0 \to A^0H^0 W^*,\\
pp\to W^*\to H^{\pm}A^0 \to H^0A^0 W^* &,&~~
pp\to W^*\to H^{\pm}A^0 \to A^0A^0 W^*,
\end{eqnarray}
resulting in three final states classified by a pair of neutral scalars: $A^0H^0$, $H^0H^0$, and $A^0A^0$. Noting that the last two originate only from cascade decays, any detection of such production channels would be a hint of charged scalars being involved. Using the fact that
\begin{eqnarray}
\sigma(pp\to W^*\to H^\pm H^0)&\simeq&\sigma(pp\to W^*\to H^\pm A^0),
\\
\mbox{BR}(H^\pm\to H^0 W^*)&\simeq&\mbox{BR}(H^\pm\to A^0 W^*),
\end{eqnarray}
as well as the narrow width approximation, we calculate the production cross sections for these three final states:
\begin{eqnarray}
H^0A^0:X_1&=&2[\sigma(pp\to W^+\to H^+ H^0)+\sigma(pp\to W^-\to H^- H^0)]\times \mbox{BR}(H^\pm \to A^0 W^*),
\\
H^0H^0:Y_1&=&[\sigma(pp\to W^+\to H^+ H^0)+\sigma(pp\to W^-\to H^- H^0)]\times \mbox{BR}(H^\pm \to H^0 W^*),
\\
A^0A^0:Z_1&=&[\sigma(pp\to W^+\to H^+ A^0)+\sigma(pp\to W^-\to H^- A^0)]\times \mbox{BR}(H^\pm \to A^0 W^*).
\end{eqnarray}
The factor 2 in $X_1$ accounts for the equal contribution from the process with $H^0$ and $A^0$ interchanged. The relations $X_1=2Y_1=2Z_1$ actually hold true for all of the four production channels, since for a given channel the same branching ratios (such as for $H^{\pm}\to H^0/A^0W^*$) are involved,
\begin{equation}\label{XXX}
X_i=2Y_i=2Z_i,~(i=1,2,3,4),
\end{equation}
where $X_i,~Y_i$, and $Z_i$ refer to the cross sections for $H^0A^0$, $H^0H^0$, and $A^0A^0$ production with the subscript $i=1,2,3,4$ denoting the production channels $H^\pm H^0/A^0$, $H^{+}H^{-}$, $H^{\pm}H^{\mp\mp}$, and $H^{++}H^{--}$, respectively. The relations imply that we may concentrate on the cross section of $H^0A^0$ production.

Naively, one would expect the next important channel to be $H^+H^-$ since it only involves two cascade decays:
\begin{equation}
X_2=2\sigma(pp\to \gamma^*/Z^*\to H^+H^-)\times \mbox{BR}(H^{\pm}\to H^0 W^*)\mbox{BR}(H^{\pm}\to A^0 W^*).
\end{equation}
But as already mentioned in Ref. \cite{Akeroyd:2011zza}, a smaller coupling and destructive interference between the $\gamma^*$ and $Z^*$ exchange make the cross section of $H^+H^-$ production an order of magnitude smaller than that of $H^0A^0$ even for a degenerate spectrum. Considering further suppression due to cascade decays, $X_2$ is not important for the enhancement of $H^0A^0$ production and can be safely neglected in the numerical analysis.

\begin{figure}[!htbp]
\begin{center}
\includegraphics[width=0.45\linewidth]{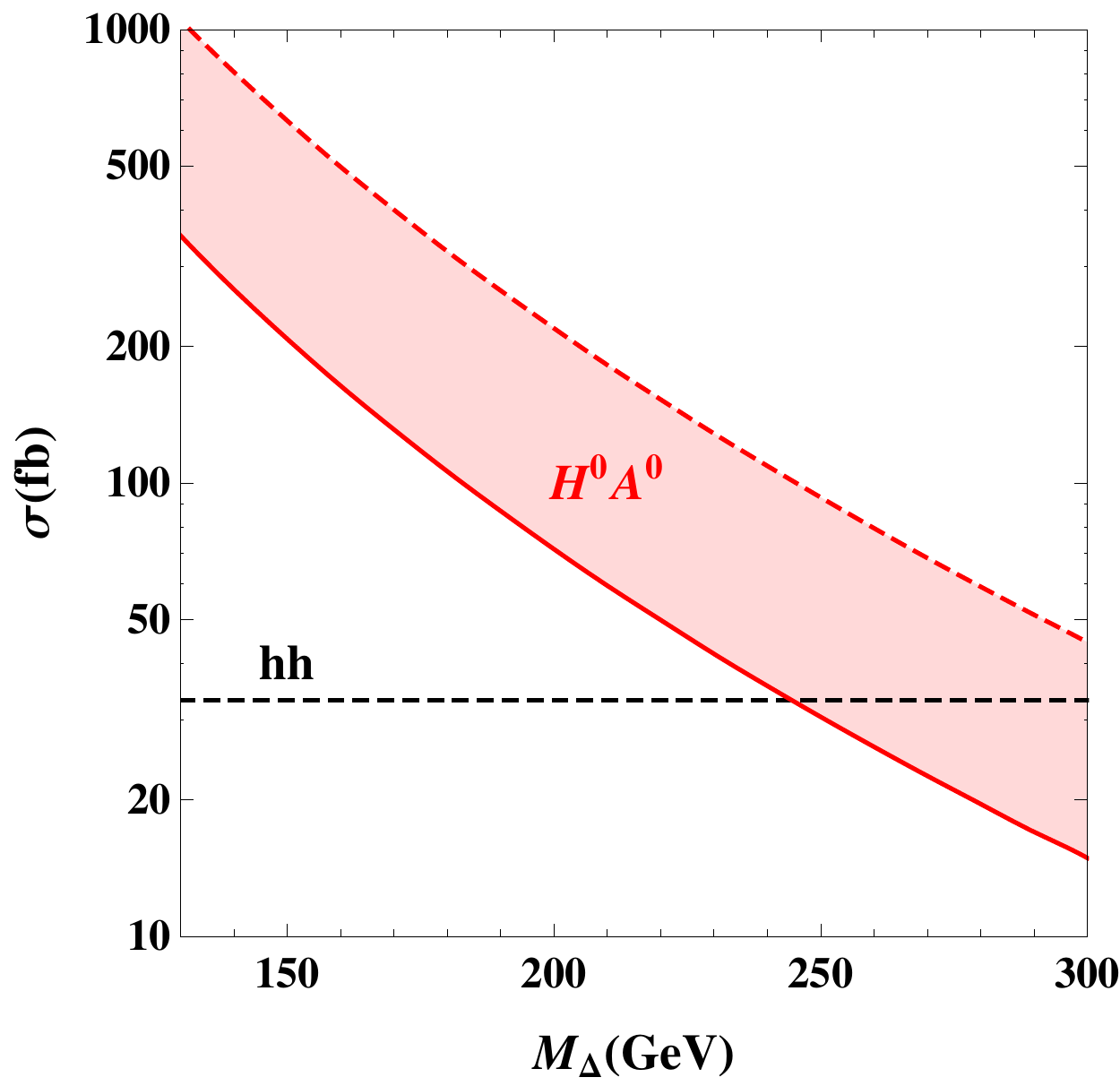}
\includegraphics[width=0.45\linewidth]{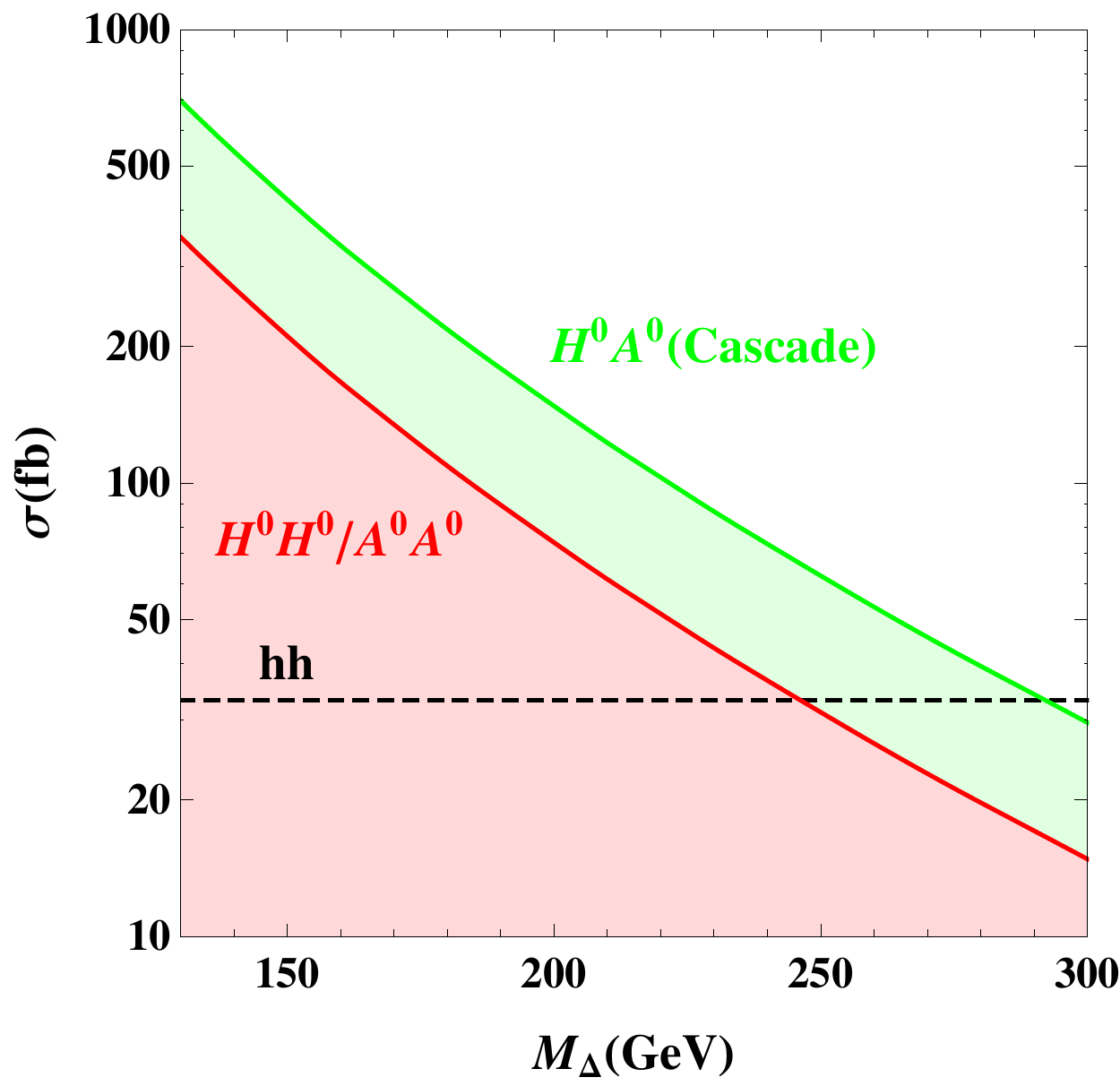}
\end{center}
\caption{Production cross sections for a pair of neutral scalars versus $M_{\Delta}$ at LHC14 and with $\Delta M=5~\GeV$, $v_{\Delta}=0.001~\GeV$. Left: The red solid (dashed) line corresponds to $X_0$ ($X$). Right: The red line corresponds to $H^0H^0/A^0A^0$ from cascade decays $Y/Z$, and the green line to $H^0A^0$ from cascade decays ($X_C$). The shaded regions are filled by scanning over $\Delta M$ and $v_\Delta$.
\label{csx}}
\end{figure}

The contribution from $H^{\pm}H^{\mp\mp}$ is more important despite the fact that it involves three cascade decays:
\begin{eqnarray}
X_3&=& 2[\sigma(pp\to W^{-*}\to H^+H^{--})+\sigma(pp\to W^{+*}\to H^-H^{++})]\times\\ \nonumber
&&\mbox{BR}(H^{\pm\pm}\to H^\pm W^*)\mbox{BR}(H^{\pm}\to H^0 W^*)\mbox{BR}(H^{\pm}\to A^0 W^*).
\end{eqnarray}
As shown in Fig.~\ref{cs}, $\sigma(pp\to W^*\to H^\pm H^{\mp\mp})$ is the largest for a degenerate mass spectrum. When cascade decays are dominant, the phase-space suppression of heavy charged scalars will be important. So we expect that the $H^0A^0$ production receives considerable enhancement from $H^{\pm}H^{\mp\mp}$ when the mass splitting is small and cascade decays are dominant.

Finally, the last mechanism is $H^{++}H^{--}$, which involves four cascade decays:
\begin{eqnarray}\nonumber
X_4 &=&2\sigma(pp\to \gamma^*/Z^* \to H^{++}H^{--})\times\mbox{BR}(H^{\pm\pm}\to H^\pm W^*)^2\\
&&\times\mbox{BR}(H^{\pm}\to H^0 W^*)\mbox{BR}(H^{\pm}\to A^0 W^*)~.
\end{eqnarray}
This mechanism is also promising since the cross section of $H^{++}H^{--}$ production is slightly larger than $H^{0}A^0$ production for a degenerate mass spectrum. The phase-space suppression of $X_4$ is more severe than that of $X_3$, because a pair of the heaviest $H^{\pm\pm}$ are produced.

Summing over all four of the above channels yields the contribution to the $H^0A^0$ production from cascade decays,
\begin{equation}
X_C=X_1+X_2+X_3+X_4,
\end{equation}
and the total production cross section of $H^0A^0$ is then $X=X_0+X_C$. Using Eq.~(\ref{XXX}), the total cross sections for the pair production $H^0H^0/A^0A^0$, $Y=\sum_iY_i$, $Z=\sum_iZ_i$, are given by
\begin{equation}
Y=Z=\frac{1}{2}X_C.
\end{equation}
Since the enhancement from cascade decays depends on a not severely suppressed phase space and a larger branching ratio of cascade decays, we choose to work with a relatively smaller mass splitting and triplet vev as shown in Eq. (\ref{BP}). Figure \ref{csx} displays the cross sections of the $H^0A^0$, $H^0H^0$, and $A^0A^0$ production as a function of $M_{\Delta}$. As can be seen from the figure, the production of $H^0A^0$ can be enhanced by a factor of 3, while the $H^0H^0/A^0A^0$ production at the maximal enhancement can reach the level of $X_0$. This could make the detection of neutral scalar pair productions very promising in the negative scenario.

%%%%%%%%%%%%%%%%%%%%%%%%%%%%%%%%
\section{LHC Signatures of Neutral Scalar Production}\label{signal}
%%%%%%%%%%%%%%%%%%%%%%%%%%%%%%%%

In this section we investigate the signatures of neutral scalar production at the LHC. From previous studies on the SM $hh$ production, we already know that the most promising signal is $b\bar{b}\gamma\gamma$, and $b\bar{b}\tau^+\tau^-$ is next to it, while both semileptonic and dileptonic decays of $W$'s in the $b\bar{b}W^+W^-$ channel are challenging. In this work we analyze all three of the signals---$b\bar{b}\gamma\gamma$, $b\bar{b}\tau^+\tau^-$, and $b\bar{b}W^+W^-\to b\bar{b}\ell^+\ell^-2\nu$ ($\ell=e,~\mu$ for collider identification)---as well as their backgrounds based on the benchmark model presented in Eq. (\ref{BP}).

In Sec.~\ref{Eh} we discussed the Drell-Yan production of $H^0A^0$ and the enhanced pair and associated production of neutral scalars $H^0/A^0$ due to cascade decays of charged scalars $H^\pm,~H^{\pm\pm}$. We are now ready to incorporate the branching ratios of $H^0/A^0$ decays for a specific signal channel. For instance, the cross sections for the $b\bar{b}\gamma\gamma$ signal channel can be written as
\begin{eqnarray}\label{S0}
S_0(b\bar{b}\gamma\gamma)& = & X_0\times\left[\mbox{BR}(H^0\to b\bar{b})\mbox{BR}(A^0\to\gamma\gamma)
+\mbox{BR}(H^0\to\gamma\gamma)\mbox{BR}(A^0\to b\bar{b})\right],\\
S(b\bar{b}\gamma\gamma) & = & X\times\left[\mbox{BR}(H^0\to b\bar{b})\mbox{BR}(A^0\to\gamma\gamma)
+\mbox{BR}(H^0\to\gamma\gamma)\mbox{BR}(A^0\to b\bar{b})\right]\\\nonumber
&&+2Y\hspace{-0.25em}\times\mbox{BR}(H^0\to b\bar{b})\mbox{BR}(H^0\to\gamma\gamma)
+2Z\hspace{-0.35em}\times\mbox{BR}(A^0\to b\bar{b})\mbox{BR}(A^0\to\gamma\gamma).
\end{eqnarray}
Here $S_0$ denotes the signal from the direct production $pp\to Z^*\to H^0A^0$ alone, and $S$ includes contributions from cascade decays. $S_{(0)}(b\bar{b}\tau^+\tau^-)$ has a similar expression as $S_{(0)}(b\bar{b}\gamma\gamma)$, while $S_{(0)}(b\bar{b}\ell^+\ell^-2\nu)$ is simpler since the decay mode $A^0\to W^+W^-$ is absent.

\begin{figure}[!htbp]
\begin{center}
\includegraphics[width=0.45\linewidth]{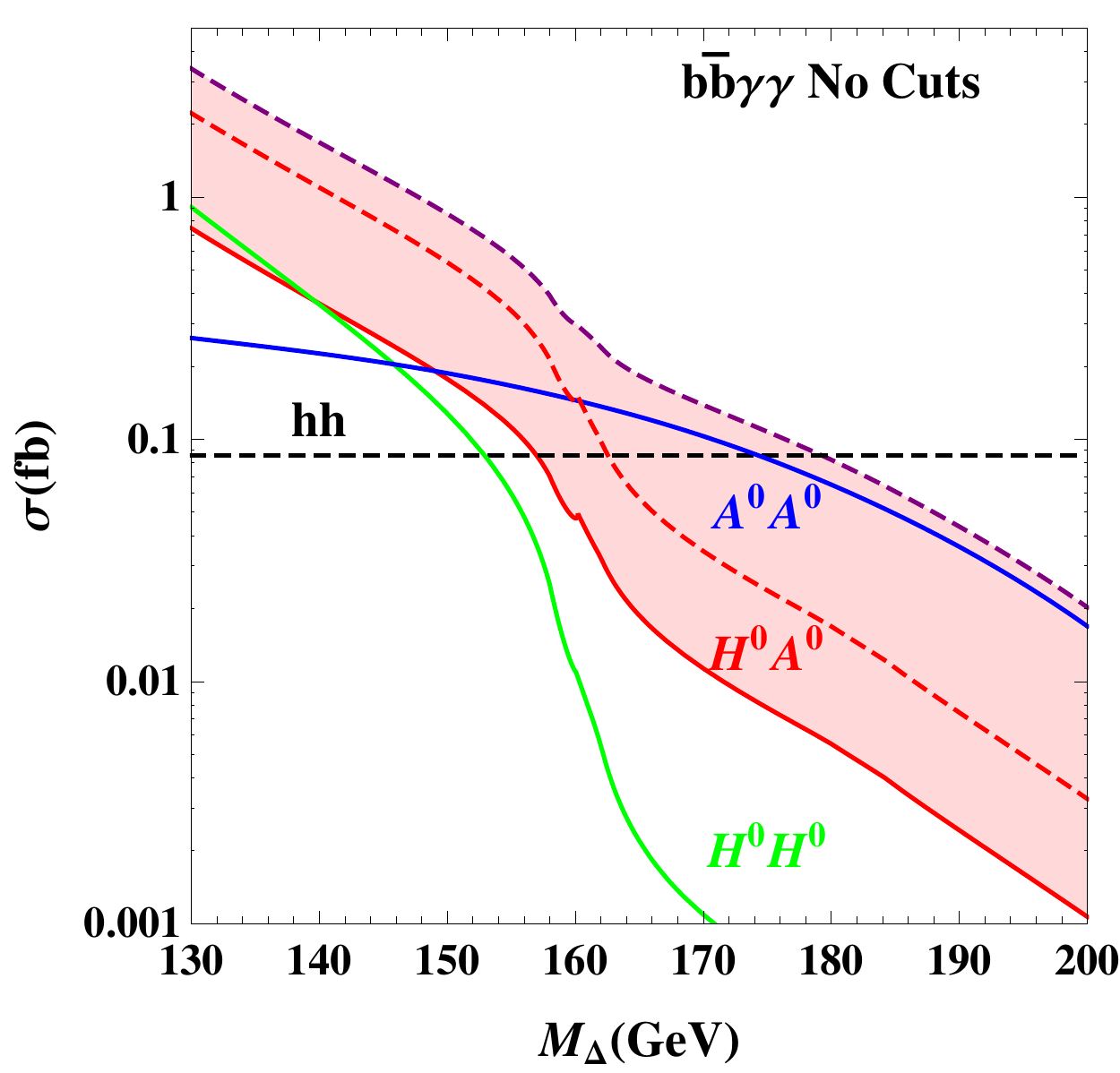}
\includegraphics[width=0.44\linewidth]{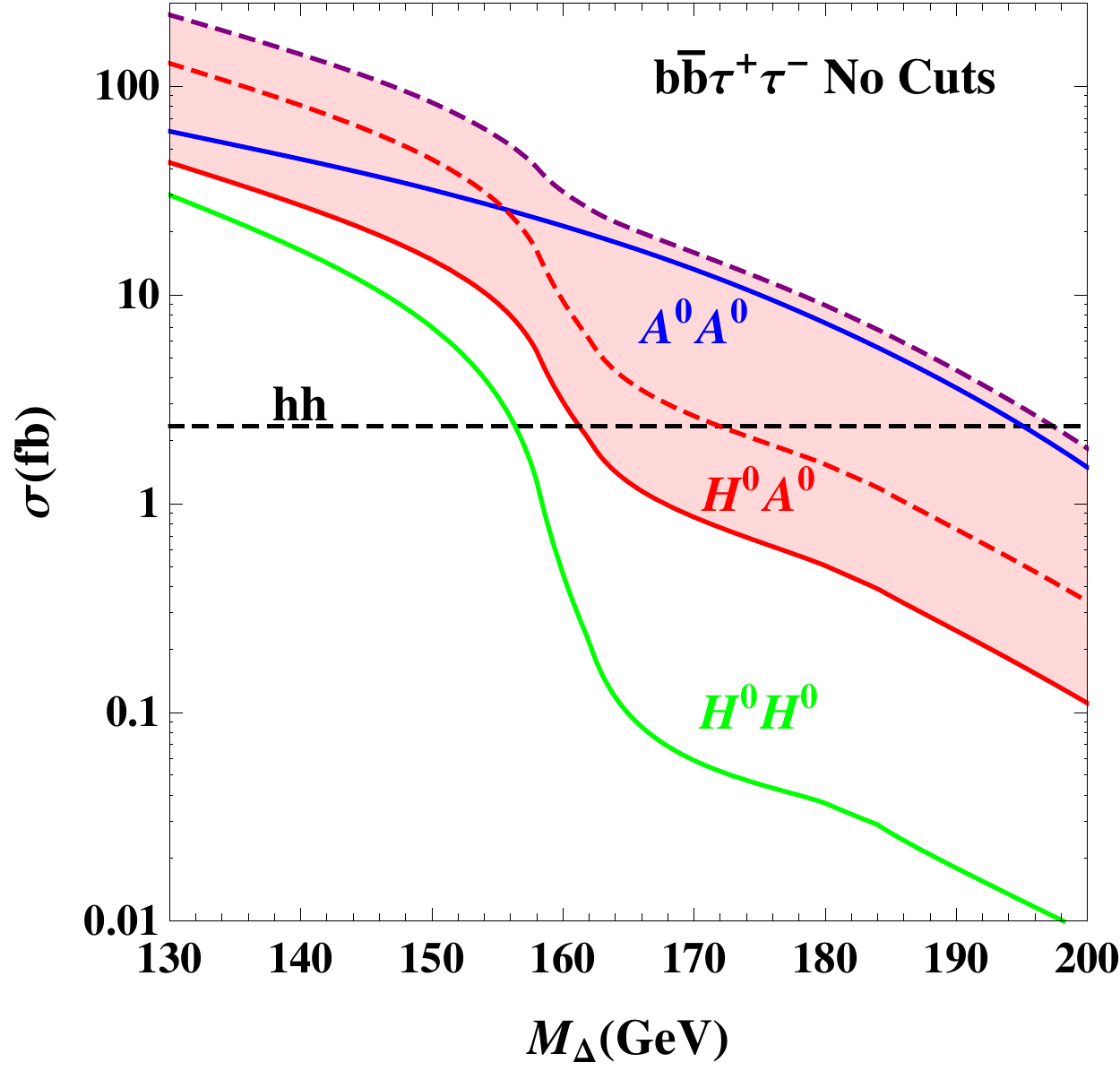}
\includegraphics[width=0.44\linewidth]{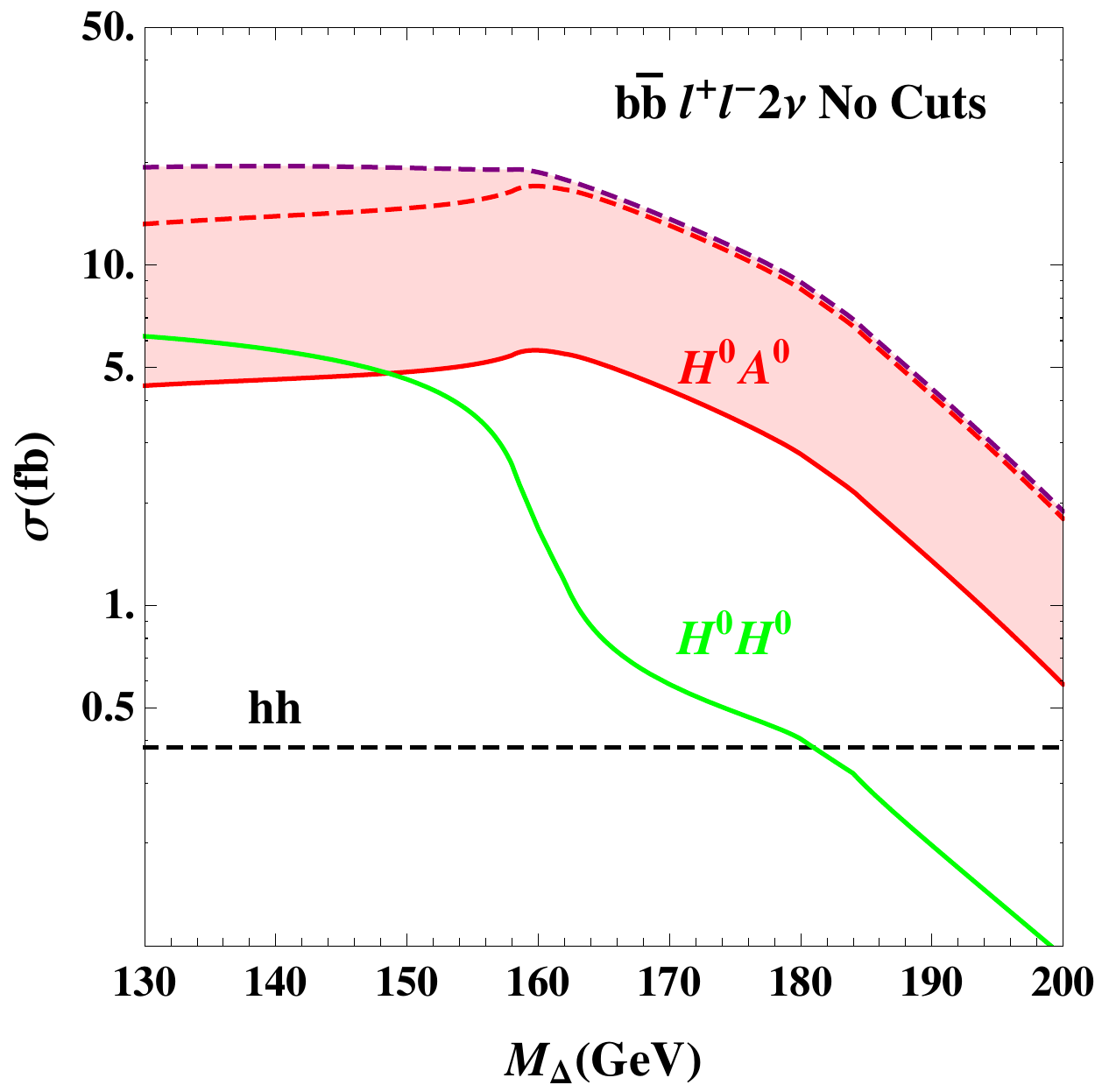}
\includegraphics[width=0.44\linewidth]{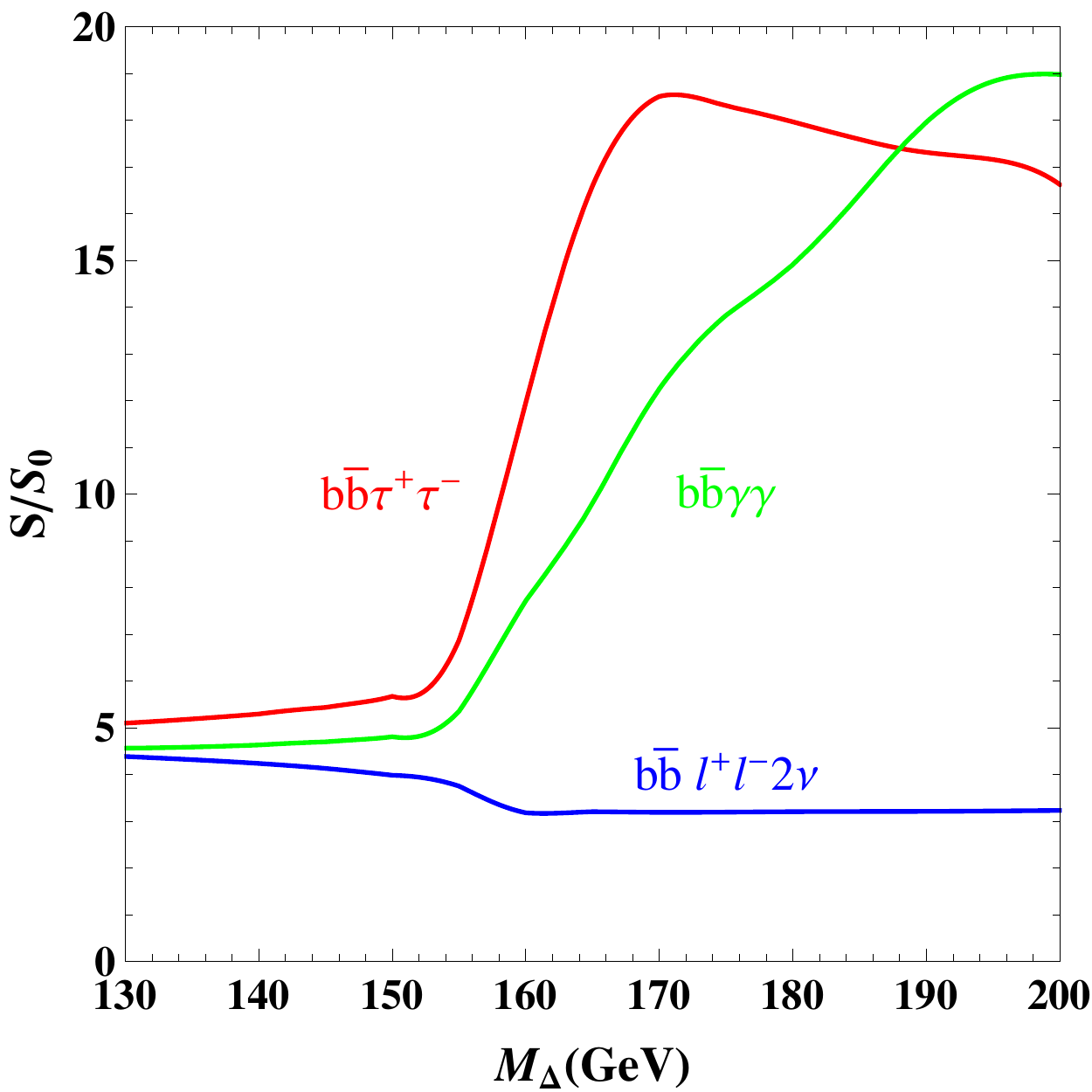}
\end{center}
\caption{Theoretical cross sections of $b\bar{b}\gamma\gamma$, $b\bar{b}\tau^+\tau^-$, and $b\bar{b}\ell^+\ell^-2\nu$ signal channels at LHC14. The red solid (dashed) line corresponds to the signal from $X_0$ ($X$), the green (blue) solid line corresponds to the signal from $Y~(Z)$, and the purple dashed line shows the total cross section $S$ for the signal. The SM $hh$ cross section is shown for comparison. The lower right panel shows the enhancement factor $S/S_0$ in the three signal channels.
\label{sgn}}
\end{figure}

The theoretical cross sections for the $b\bar{b}\gamma\gamma$, $b\bar{b}\tau^+\tau^-$, and $b\bar{b}\ell^+\ell^-2\nu$ signal channels are plotted in Fig.~\ref{sgn}. The cross section $S_0(b\bar{b}\gamma\gamma)/S_0(b\bar{b}\tau^+\tau^-)$ is larger than that of the SM $hh$ production until $M_{\Delta}=159/161~\GeV$; taking into account cascade enhancement pushes the corresponding $M_{\Delta}$ further to $179/197~\GeV$. $S_{0}(b\bar{b}\ell^+\ell^-2\nu)$ is always larger than that of $hh$ in the mass region $130$-$200~\GeV$, and interestingly, it keeps about the same value when $M_{\Delta}<160~\GeV$. The signal from $H^0H^0$ is comparable with $S_0$ in these three channels only for $M_{\Delta}<160~\GeV$, while in contrast the signal from $A^0A^0$ becomes dominant for the $b\bar{b}\gamma\gamma$ and $b\bar{b}\tau^+\tau^-$ channels when $M_{\Delta}>160~\GeV$. Therefore, we have a chance to probe the $A^0A^0$ pair production in these two channels. Also shown in Fig.~\ref{sgn} is the enhancement factor $S/S_0$ for the three signal channels at the benchmark point (\ref{BP}) as a function of $M_\Delta$, which will help us understand the simulation results.

\subsection{$b\bar{b}\gamma\gamma$ signal channel}
\label{sec:bbgg}

In our simulation, the parton-level signal and background events are generated with {\bf\scriptsize MADGRAPH5}~\cite{MG5}. We perform parton shower and fast detector simulations with {\bf\scriptsize PYTHIA}~\cite{Sjostrand:2006za} and {\bf\scriptsize DELPHES3}~\cite{Delphes}. Finally, {\bf\scriptsize MADANALYSIS5}~\cite{Conte:2012fm} is responsible for data analysis and plotting. We take a flat $b$-tagging efficiency of 70\%, and mistagging rates of 10\% for $c$ jets and 1\% for light-flavor jets, respectively. Jet reconstruction is done using the anti-$k_T$ algorithm with a radius parameter of $R=0.5$. We further assume a photon identification efficiency of $85\%$ and a jet-faking-photon rate of $1.2\times 10^{-4}$~\cite{Aad:2009wy}.

The main SM backgrounds to the signal are as follows:
\begin{eqnarray}
b\bar{b}\gamma\gamma: p p &\to& b\bar{b}\gamma\gamma,\\
t\bar{t}h: p p &\to& t\bar{t}h \to b\ell^+\nu~\bar{b}\ell^-\nu~\gamma\gamma~(\ell^\pm~\mbox{missed}),\\
Zh: pp &\to& Zh \to b\bar{b} \gamma\gamma.
\end{eqnarray}
Among them, $b\bar{b}\gamma\gamma$ and $Zh$ are irreducible, while $t\bar{t}h$ is reducible and can be reduced by vetoing the additional $\ell$'s with $p_T^{\ell}>20~\GeV$ and $|\eta_\ell|<2.4$. In addition, there exist many reducible sources of fake $b\bar{b}\gamma\gamma$:
\begin{eqnarray}
\nonumber
pp\to b\bar{b}jj\nrightarrow b\bar{b}\gamma\gamma,pp\to b\bar{b}j\gamma \nrightarrow b\bar{b}\gamma\gamma, \ldots \\
pp\to c\bar{c}\gamma\gamma\nrightarrow b\bar{b}\gamma\gamma,pp\to j\bar{j}\gamma\gamma\nrightarrow b\bar{b}\gamma\gamma, \ldots,
\end{eqnarray}
where $x\nrightarrow y$ stands for a final-state $x$ misidentified as $y$. The remaining fake
sources are subdominant and are thus not included in our simulation. The QCD corrections to the
backgrounds are included by a multiplicative $K$-factor of 1.10 and 1.33 for the leading cross sections of $t\bar{t}h$ and $Zh$ at LHC14~\cite{Dittmaier:2011ti}, respectively. The cross section of the $b\bar{b}\gamma\gamma$ background has been normalized to include fake sources and does not take NLO corrections into account.

\begin{figure}[!htbp]
\begin{center}
\includegraphics[width=0.45\linewidth]{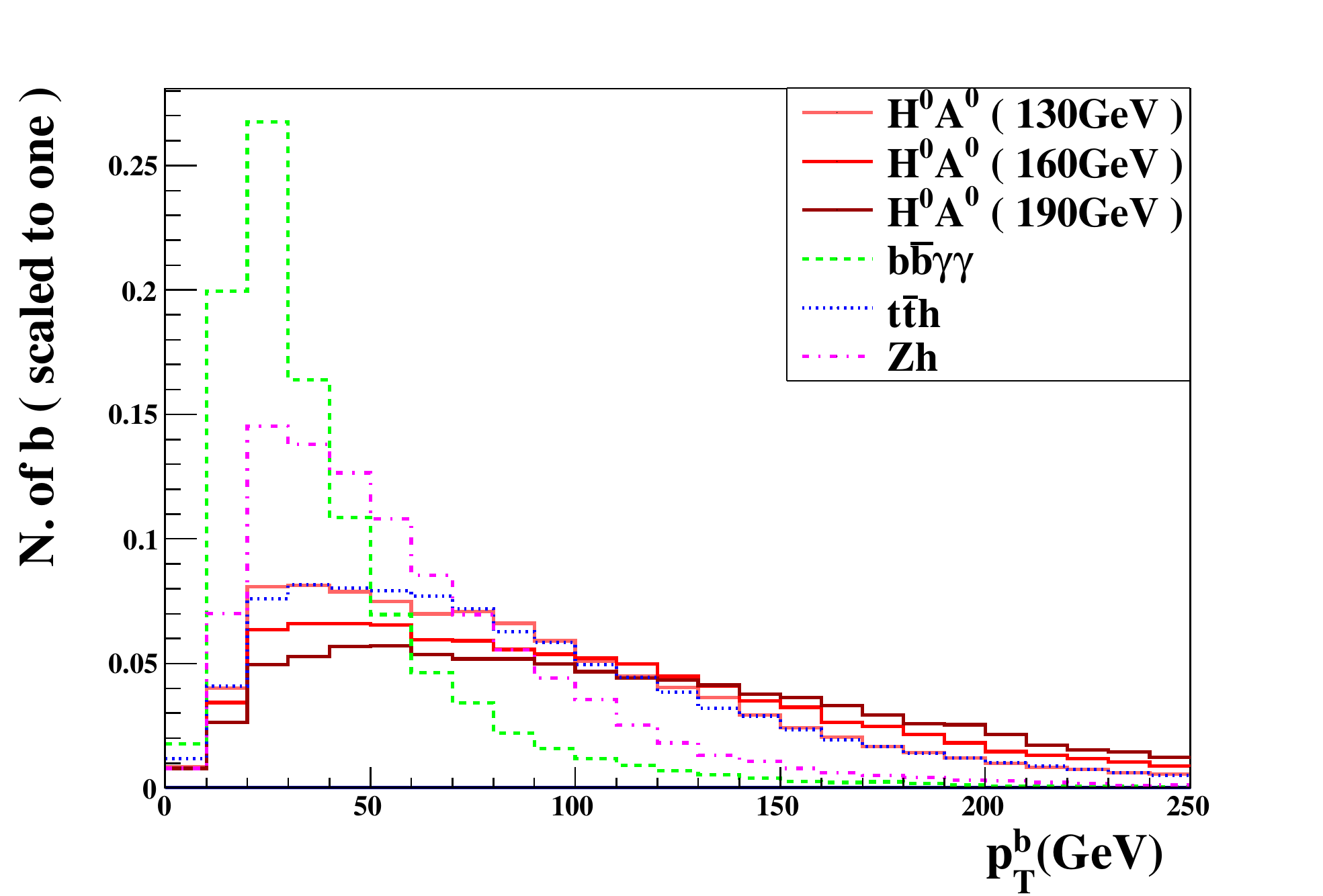}
\includegraphics[width=0.45\linewidth]{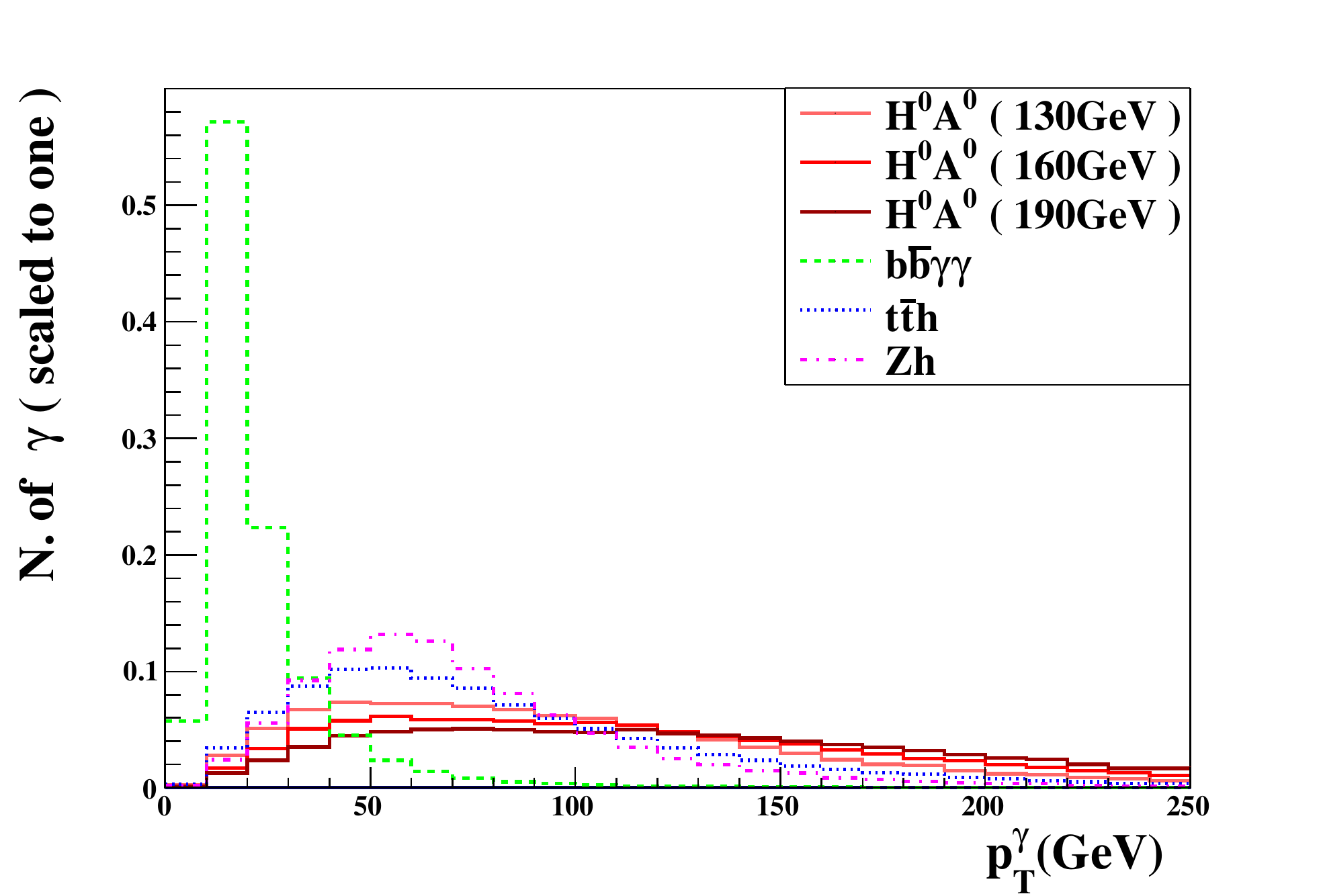}
\includegraphics[width=0.45\linewidth]{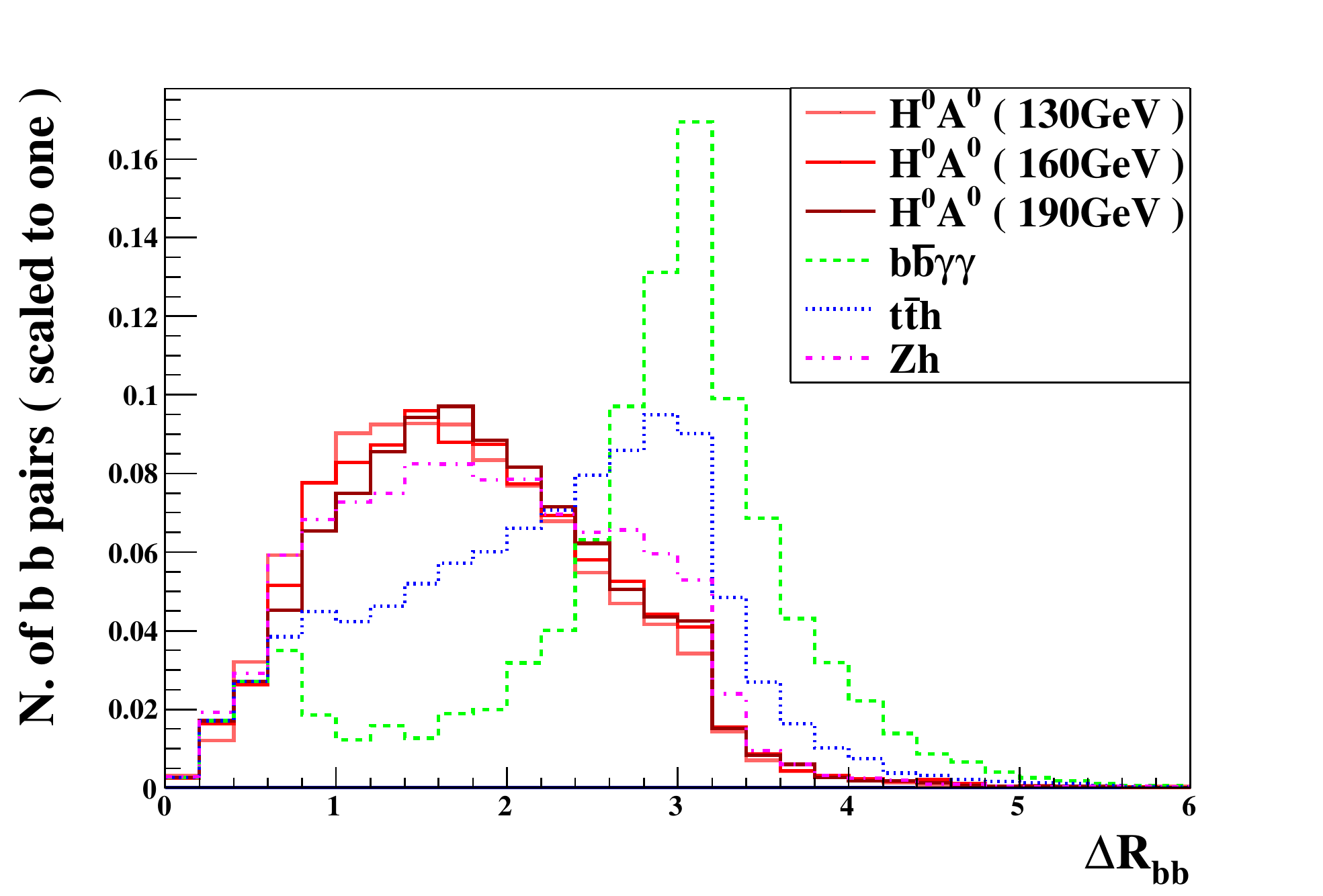}
\includegraphics[width=0.45\linewidth]{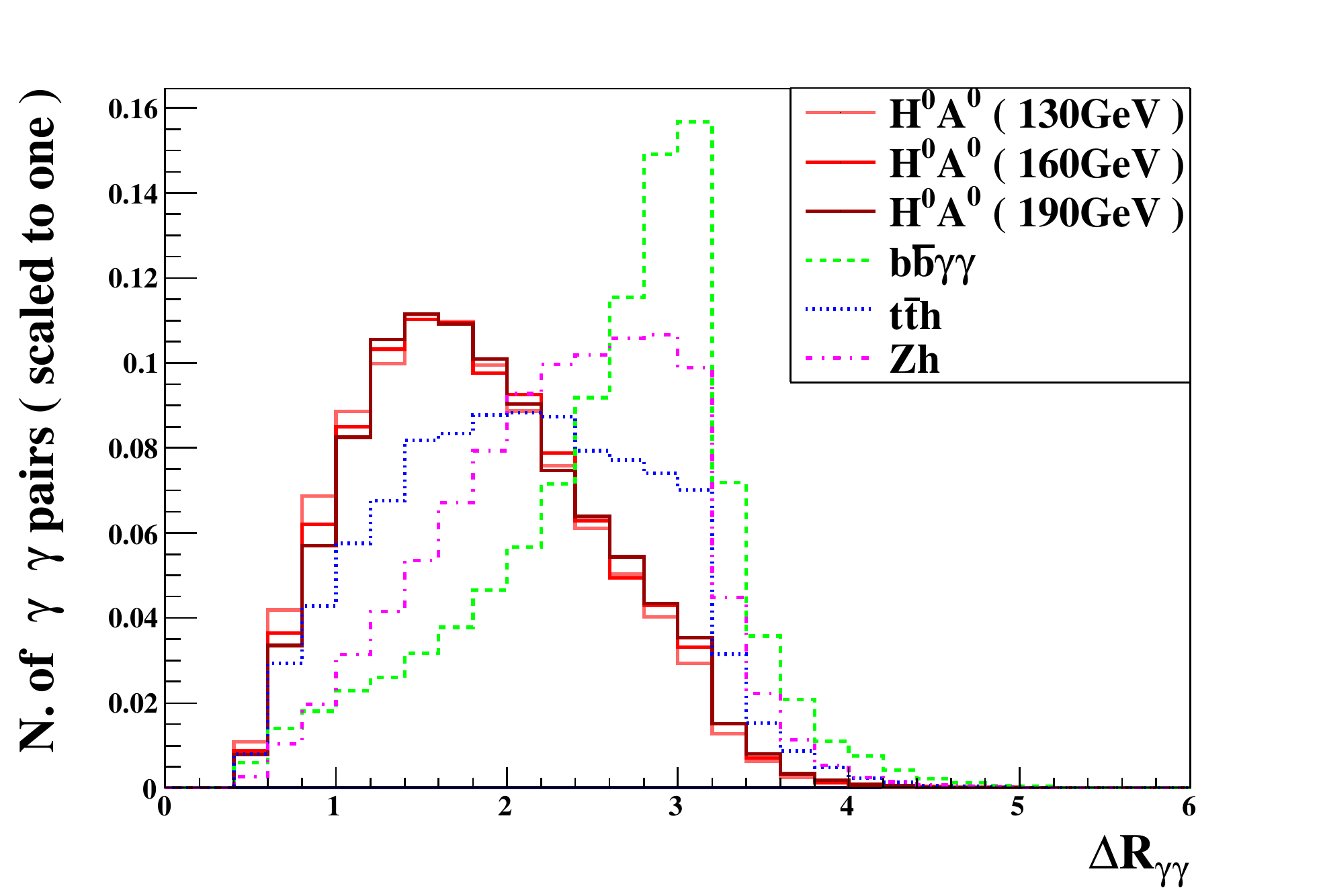}
\includegraphics[width=0.45\linewidth]{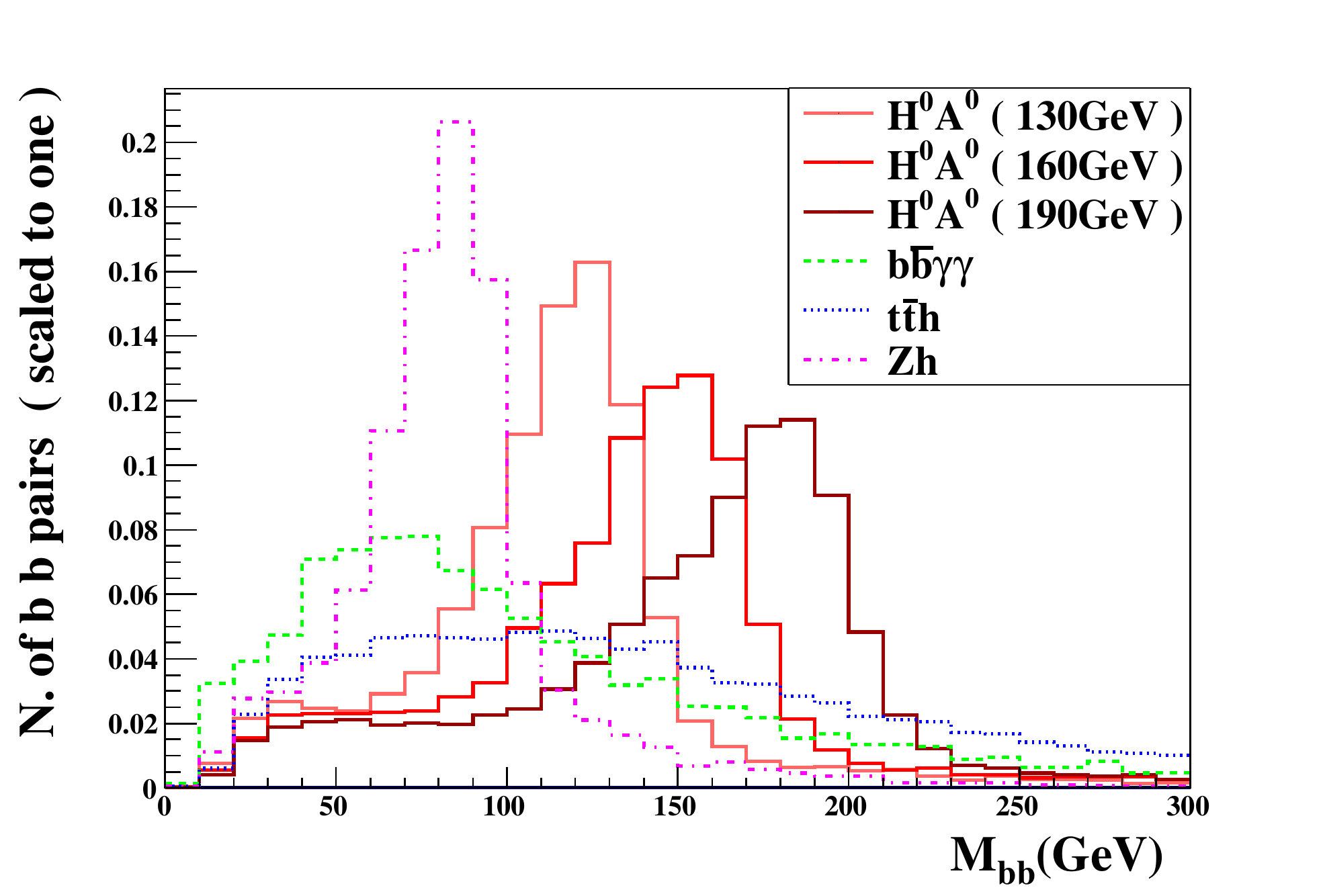}
\includegraphics[width=0.45\linewidth]{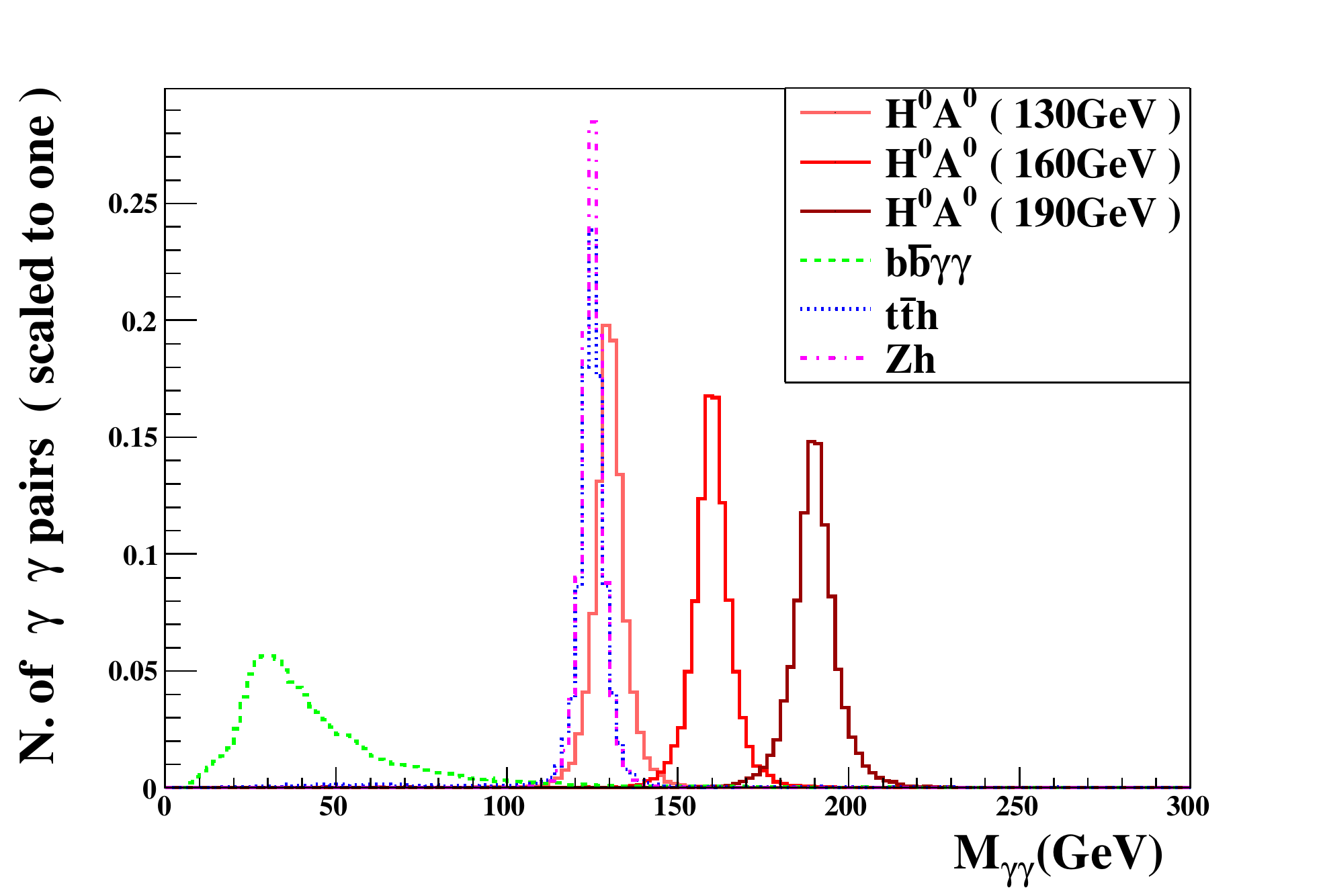}
\includegraphics[width=0.45\linewidth]{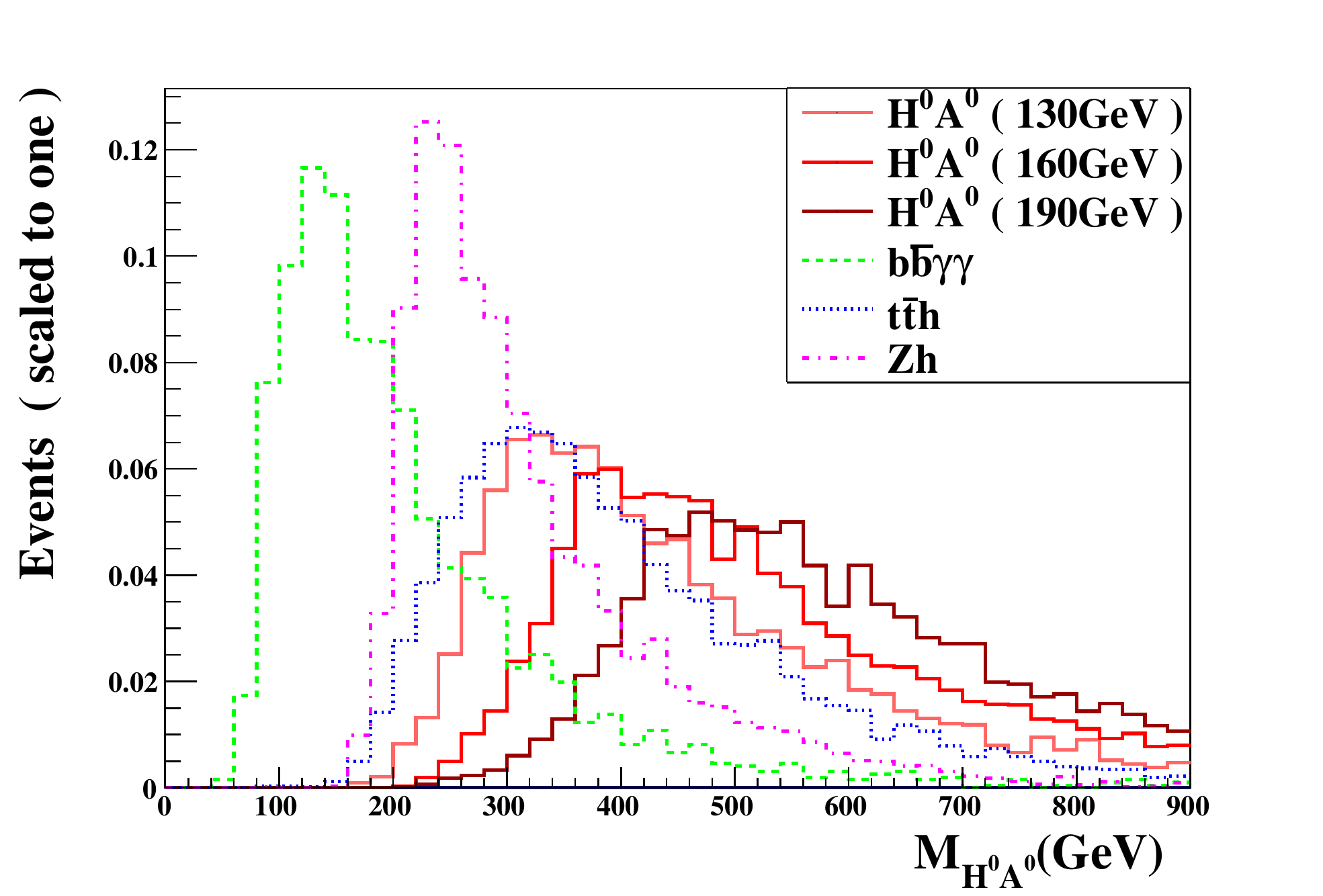}
\includegraphics[width=0.45\linewidth]{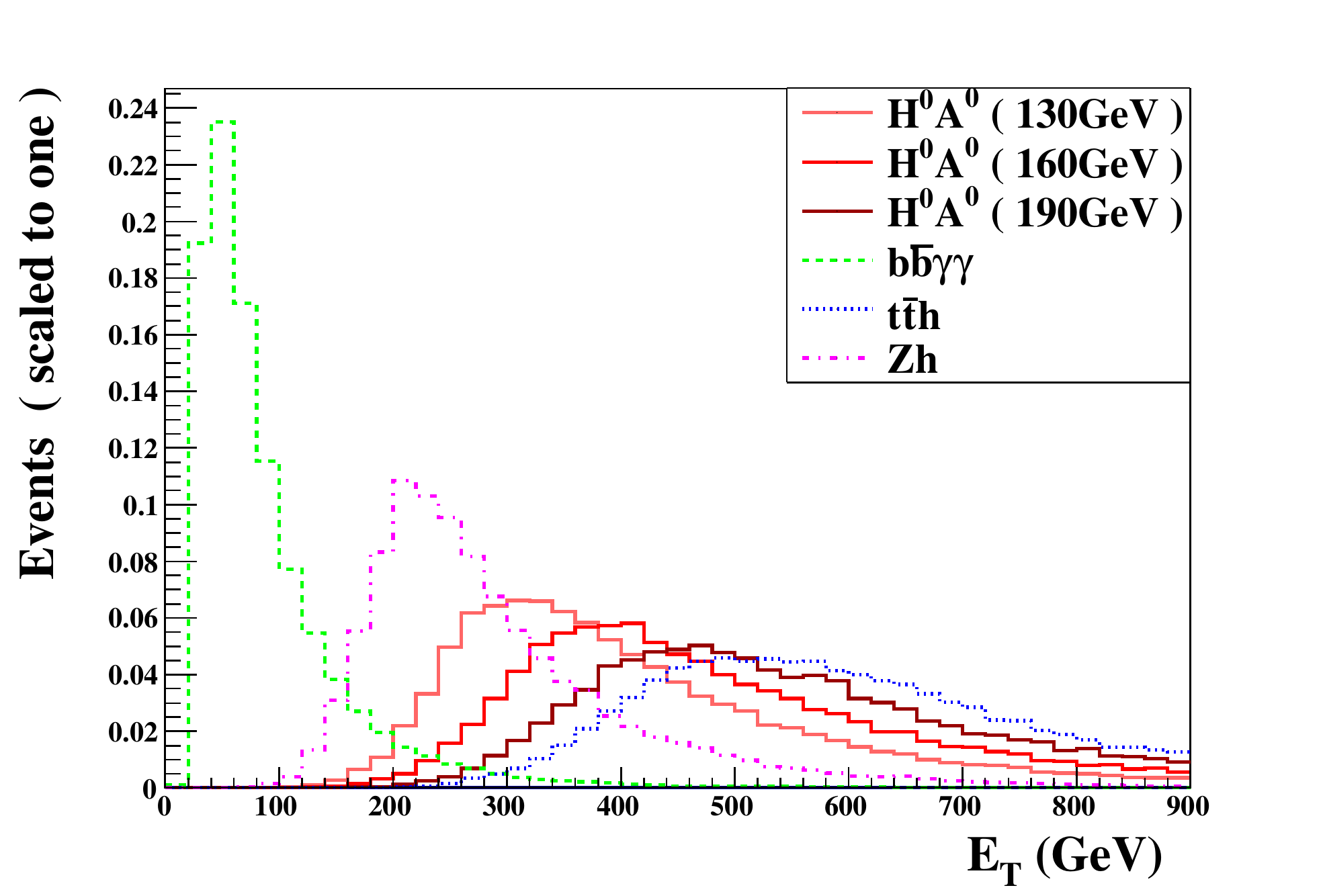}
\end{center}
\caption{Distributions of $p_T^{b,\gamma},~\Delta R_{bb,\gamma\gamma},~M_{bb,\gamma\gamma,H^0A^0}$, and $E_T$ for the signal $b\bar{b}\gamma\gamma$ and its backgrounds before applying any cuts at LHC14.
\label{fig:bbaa}}
\end{figure}

The distributions of some kinematical variables before applying any cuts are shown in Fig. \ref{fig:bbaa}, where we assume $M_{\Delta}=130,~160,~190~\GeV$. In our analysis, we require that the final states include exactly one $b$-jet pair and one $\gamma$ pair and satisfy the following basic cuts:
\begin{eqnarray}
p_T^{b,\gamma}>30~\GeV,~|\eta_{b,\gamma}|<2.4,~\Delta R_{bb,\gamma\gamma,b\gamma}>0.4,
\end{eqnarray}
where $\Delta R=\sqrt{(\Delta \phi)^2+(\Delta \eta)^2}$ is the particle separation, with $\Delta \phi$ and $\Delta \eta$ being the separation in the azimuthal angle and rapidity, respectively. Here we employ a tighter $p_T$ cut than is usually applied to suppress the QCD-electroweak $b\bar{b}\gamma\gamma$ background. The $b$-jet pair and $\gamma$ pair are then required to fall in the following windows on the invariant masses and fulfill the $\Delta R$ cut criteria:
\begin{eqnarray}
\Delta R_{bb}<2.5,&&~|M_{bb}-M_{\Delta}|<15~\GeV,\\
\nonumber
\Delta R_{\gamma\gamma}<2.5,&&~|M_{\gamma\gamma}-M_{\Delta}|<10~\GeV.
\end{eqnarray}

\begin{table} [!htbp]
\begin{center}
\begin{tabular}{|c|c|c|c|c|c|c|}
\hline
 $M_{\Delta}=130~\GeV$  &  $H^0A^0(S_0)$  &  $b\bar{b}\gamma\gamma$ & $t\bar{t}h$  & $Zh$ & $S/B$  & $\mathcal{S}(S,B)$
\\
\hline
Cross section at NLO& $8.01\times10^{-1}$ & $5.92\times10^3$~~ &1.18& $2.99\times10^{-1}$ & $1.39\times10^{-4}$ & $5.75\times10^{-1}$
\\
Basic cuts & $1.22\times10^{-1}$ & $4.16\times10^{1}$~~ & $1.03\times10^{-1}$ & $3.41\times10^{-2}$ & $2.92\times10^{-3}$ & 1.03
\\
Reconstruct scalars from $b$s & $6.99\times10^{-2}$ & 7.07 & $1.50\times10^{-2}$ & $9.61\times10^{-4}$ & $9.87\times10^{-3}$ & 1.44
\\
Reconstruct scalars from $\gamma$s & $5.28\times10^{-2}$ & $1.03\times10^{-1}$ & $1.08\times10^{-2}$ & $7.32\times10^{-4}$ & $4.63\times10^{-1}$ & 8.01
\\
Cut on $M_{H^0A^0}$ & $4.21\times10^{-2}$ & $2.04\times10^{-2}$ & $4.69\times10^{-3}$ & $3.23\times10^{-4}$ & 1.65 & $12.0$
\\
Cut on $E_T$ & $3.31\times10^{-2}$ & $6.58\times10^{-3}$ & $4.68\times10^{-3}$ & $2.27\times10^{-4}$ & 2.88 & $12.8$
\\
\hline
Cascade enhanced & $1.51\times10^{-1}$ & $-$ & $-$ & $-$ & $13.1$ & $41.0$
\\
\hline
\hline
$M_{\Delta}=160~\GeV$  &  $H^0A^0(S_0)$  &  $b\bar{b}\gamma\gamma$ & $t\bar{t}h$  & $Zh$ & $S/B$  & $\mathcal{S}(S,B)$
\\
\hline
Cross section at NLO & $5.10\times10^{-2}$ & $5.92\times10^3$~~ &1.18& $2.99\times10^{-1}$  & $8.61\times10^{-6}$ & $3.63\times10^{-2}$
\\
Basic cuts & $8.78\times10^{-3}$ & $4.16\times10^{1}$~~ & $1.03\times10^{-1}$ & $3.41\times10^{-2}$  & $2.10\times10^{-4}$ & $7.44\times10^{-2}$
\\
Reconstruct scalars from $b$s & $4.11\times10^{-3}$ & $5.06$ & $1.34\times10^{-2}$ & $2.36\times10^{-4}$  & $8.11\times10^{-4}$ & $9.99\times10^{-2}$
\\
Reconstruct scalars from $\gamma$s & $3.27\times10^{-3}$ & $3.42\times10^{-2}$ & $1.57\times10^{-5}$ & 0.00  & $9.56\times10^{-2}$ & $9.53\times10^{-1}$
\\
Cut on $M_{H^0A^0}$ & $2.57\times10^{-3}$ & $1.12\times10^{-2}$ & $1.18\times10^{-5}$  & 0.00 & $2.30\times10^{-1}$ & 1.28
\\
Cut on $E_T$ & $1.73\times10^{-3}$ & $3.95\times10^{-3}$ & $1.03\times10^{-5}$ & 0.00 & $4.37\times10^{-1}$ & 1.41
\\
\hline
Cascade enhanced & $1.10\times10^{-2}$ & $-$ & $-$ & $-$   & 2.77 & 7.29
\\
\hline
\hline
$M_{\Delta}=190~\GeV$  &  $H^0A^0(S_0)$  &  $b\bar{b}\gamma\gamma$ & $t\bar{t}h$  & $Zh$ & $S/B$  & $\mathcal{S}(S,B)$
\\
\hline
Cross section at NLO& $2.68\times10^{-3}$ & $5.92\times10^3$~~ &1.18& $2.99\times10^{-1}$  & $4.53\times10^{-7}$  &  $1.91\times10^{-3}$
\\
Basic cuts & $5.33\times10^{-4}$ & $4.16\times10^{1}$~~ & $1.03\times10^{-1}$ & $3.41\times10^{-2}$  & $1.28\times10^{-5}$ & $4.52\times10^{-3}$
\\
Reconstruct scalars from $b$s & $2.27\times10^{-4}$  & 3.61  &  $1.05\times10^{-2}$ & $1.24\times10^{-4}$  & $6.27\times10^{-5}$  &  $6.53\times10^{-3}$
\\
Reconstruct scalars from $\gamma$s & $1.81\times10^{-4}$  & $2.47\times10^{-2}$  & $3.93\times10^{-6}$  & 0.00  & $7.34\times10^{-3}$  &  $6.30\times10^{-2}$
\\
Cut on $M_{H^0A^0}$ & $1.55\times10^{-4}$  & $9.87\times10^{-3}$  & $3.93\times10^{-6}$   & 0.00  & $1.57\times10^{-2}$  &  $8.52\times10^{-2}$
\\
Cut on $E_T$ & $8.35\times10^{-5}$  & $1.48\times10^{-3}$  & $3.93\times10^{-6}$  & 0.00  & $5.63\times10^{-2}$  &  $1.18\times10^{-1}$
\\
\hline
Cascade enhanced & $1.50\times10^{-3}$  & $-$  & $-$  &  $-$   & 1.01  &  1.87
\\
\hline
\end{tabular}
\end{center}
\caption{Evolution of signal and background cross sections (in $\fb$) at LHC14 for the $b\bar{b}\gamma\gamma$ signal channel upon imposing the cuts one by one. For the cascade-enhanced signal only the cross section passing all the cuts is shown. The last two columns assume an integrated luminosity of $3000~\fb^{-1}$.
\label{tab:bbaacut}}
\end{table}

As shown in Fig. \ref{fig:bbaa}, the $\Delta R_{bb,\gamma\gamma}$ distributions of the signal are clearly more compact as they are more likely coming from the same particles. Thus the $\Delta R$ cuts can effectively suppress the background. More specific cuts are necessary for further analysis. A useful variable is the invariant mass of the neutral scalar pair $M_{H^0A^0}$, and the total transverse energy $E_T$ is also distinctive. The peak of $M_{H^0A^0}$ increases with $M_{\Delta}$, and similarly for $E_T$. For simplicity, we adopt for the cuts a linear shift between $M_{H^0A^0},~E_T$ and $M_{\Delta}$:
\begin{equation}
M_{H^0A^0}>2M_{\Delta}+90~\GeV,~E_T>2M_{\Delta}-60~\GeV.
\end{equation}
For instance, we apply $M_{H^0A^0}>350~\GeV$, $E_T>200~\GeV$ at the benchmark point $M_{\Delta}=130~\GeV$.

To estimate the observability quantitatively, we adopt the following significance measurement:
\begin{equation}
\mathcal{S}(S,B)=\sqrt{2\left((S\cdot\mathcal{L}+B\cdot\mathcal{L})
\log\left(1+\frac{S}{B}\right)-S\cdot\mathcal{L}\right)},
\end{equation}
which is more suitable than the usual definition of $S/\sqrt{B}$ or $S/\sqrt{S+B}$ for Monte Carlo analysis~\cite{Cowan:2010js}. Here $S$ and $B$ are the signal and background cross sections, and $\mathcal{L}$ is the integrated luminosity. The survival cross sections of the signal from the Drell-Yan process and of the backgrounds upon imposing cuts step by step are summarized in Table~\ref{tab:bbaacut} at the benchmark point (\ref{BP}) for $M_\Delta=130,~160,~190~\GeV$ respectively. For the cascade-enhanced signal, only the cross section passing all the cuts is shown. The last two columns in the table show the signal-to-background ratio $S/B$ and the statistical significance $\mathcal{S}(S,B)$.

For $M_{\Delta}=130~\GeV$, the $b\bar{b}\gamma\gamma$ channel is very promising. Without (with) cascade enhancement, the final significance can reach 12.8 (41) for LHC14@3000, corresponding to 99 (453) events. For $M_{\Delta}=160~\GeV$, the channel becomes challenging since the cross section has decreased by a factor of 15.7 compared with the case of $M_{\Delta}=130~\GeV$. But the cuts we applied are efficient to suppress the SM background, and with cascade enhancement the significance could still reach 7.29 for $3000~\fb^{-1}$, corresponding to 33 events in the most optimistic case. For $M_{\Delta}=190~\GeV$, it looks hopeless even with maximal cascade enhancement in our benchmark model: to achieve 10 signal events, an integrated luminosity of at least $6670~\fb^{-1}$ is required, which is beyond the reach of the future LHC.

\subsection{$b\bar{b}\tau^+\tau^-$ signal channel}

For this signal channel, an important part of the analysis depends on the ability to reconstruct the $b$ pair and the $\tau$ pair. Here we consider the hadronic decays of the $\tau$ lepton and assume a $\tau$-tagging efficiency of $70\%$ with a negligible fake rate.

\begin{figure}[!htbp]
\begin{center}
\includegraphics[width=0.45\linewidth]{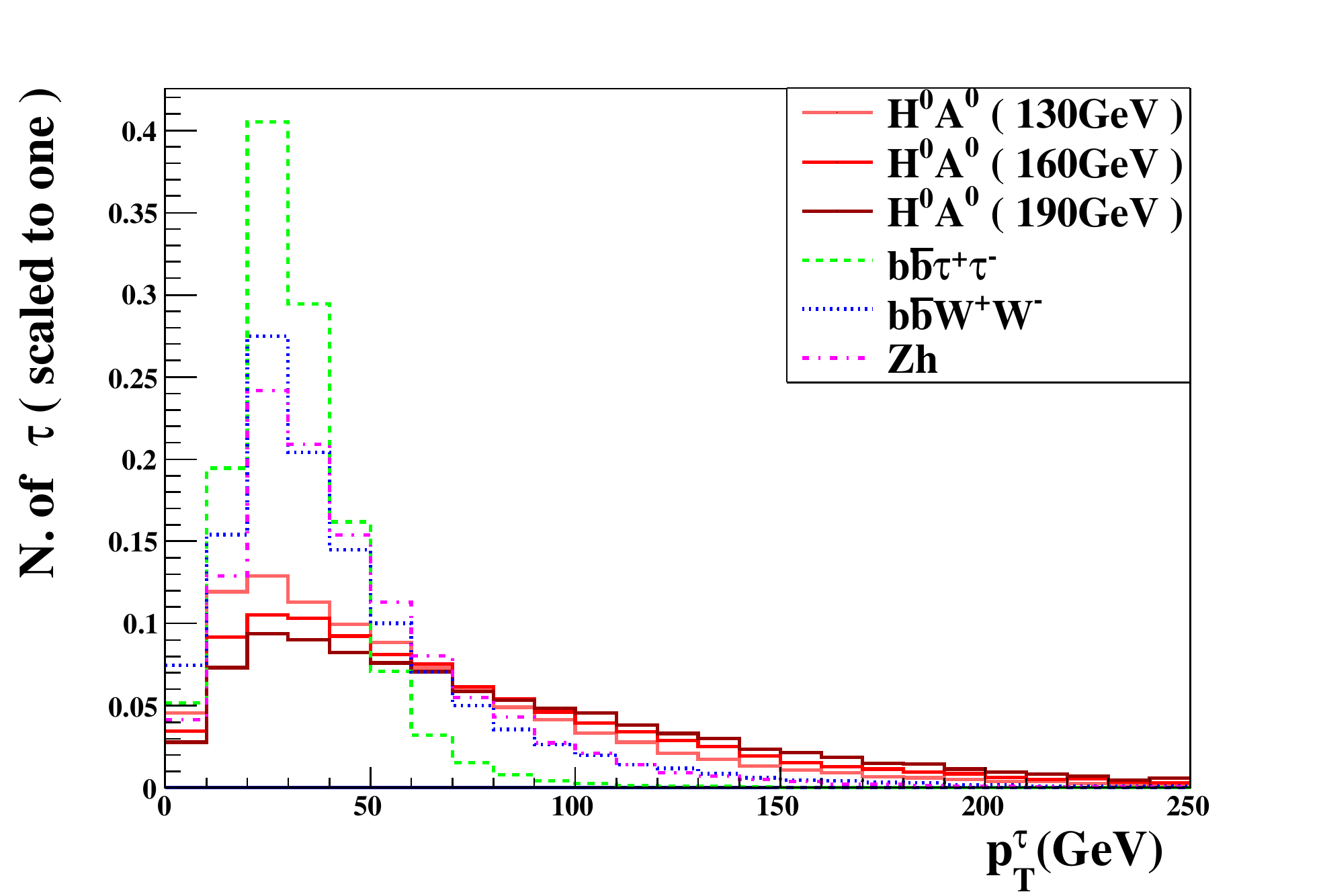}
\includegraphics[width=0.45\linewidth]{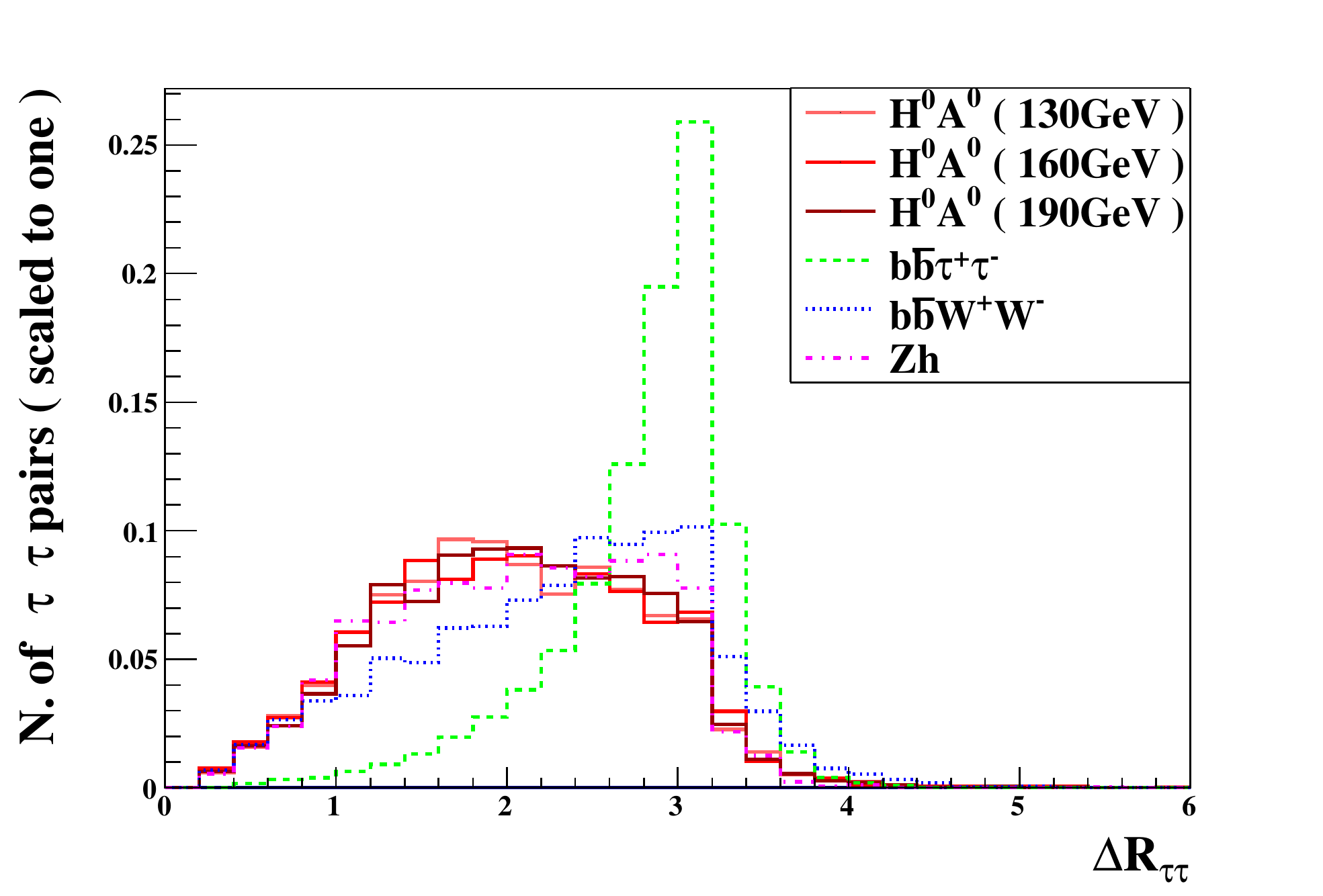}
\includegraphics[width=0.45\linewidth]{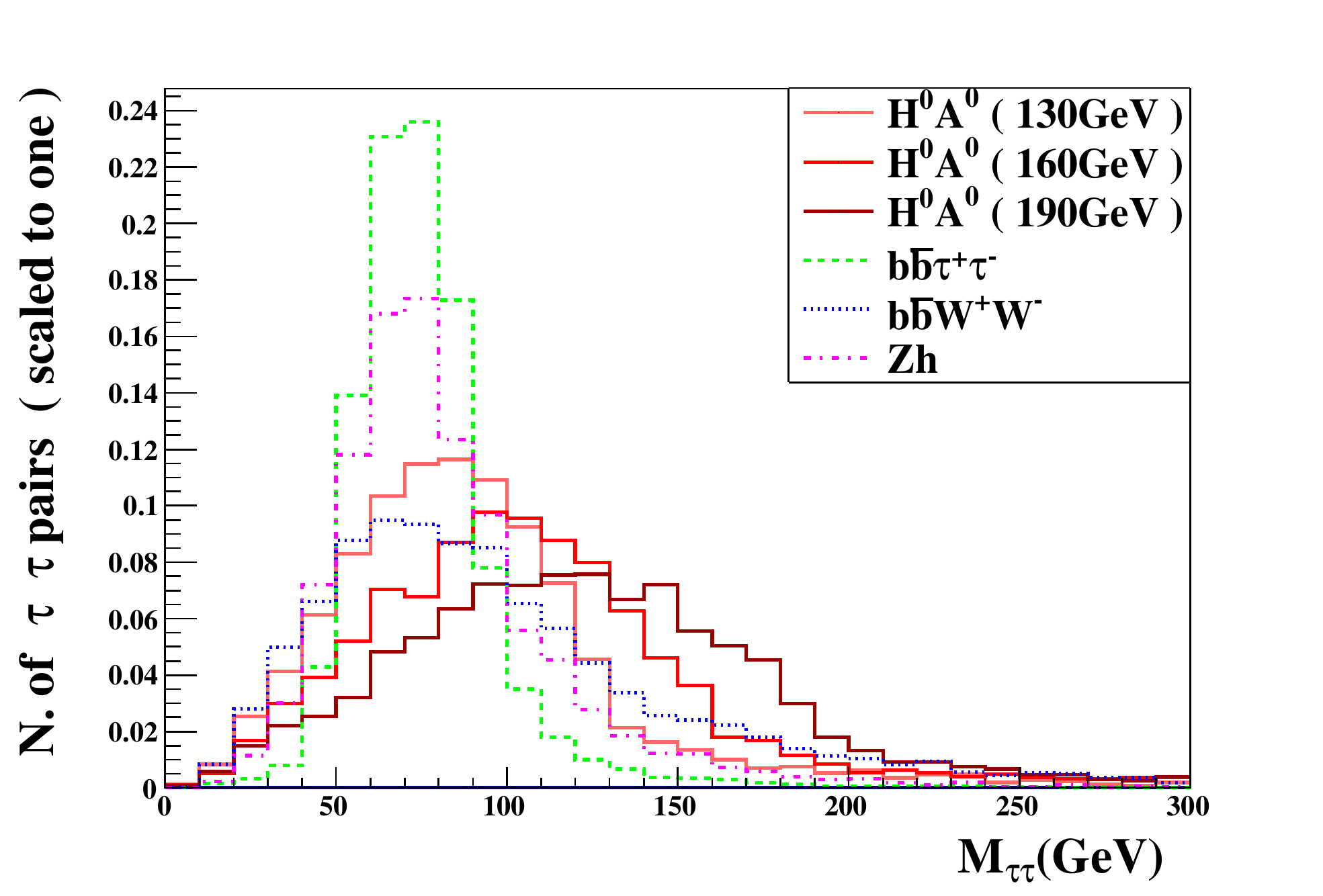}
\includegraphics[width=0.45\linewidth]{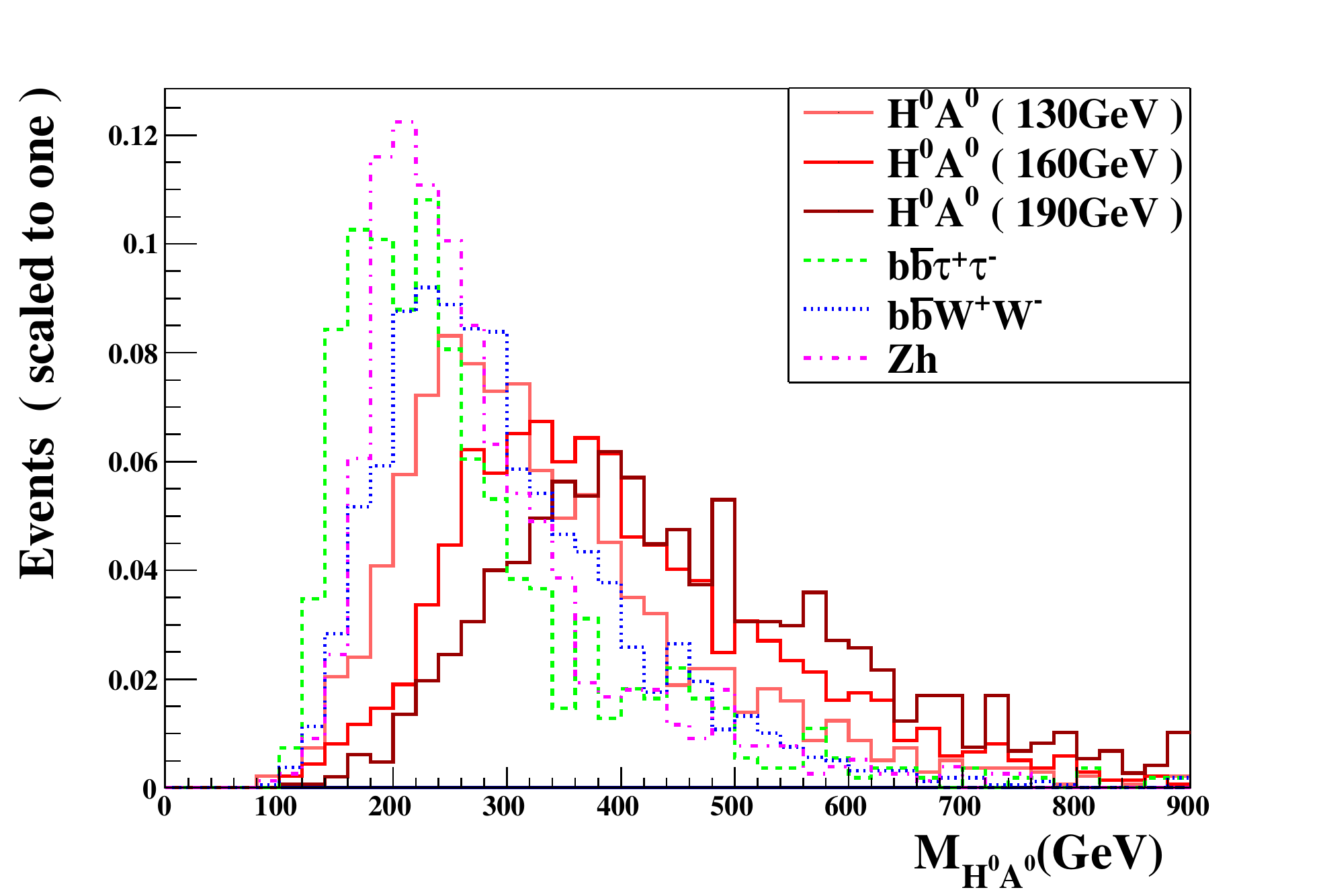}
\includegraphics[width=0.45\linewidth]{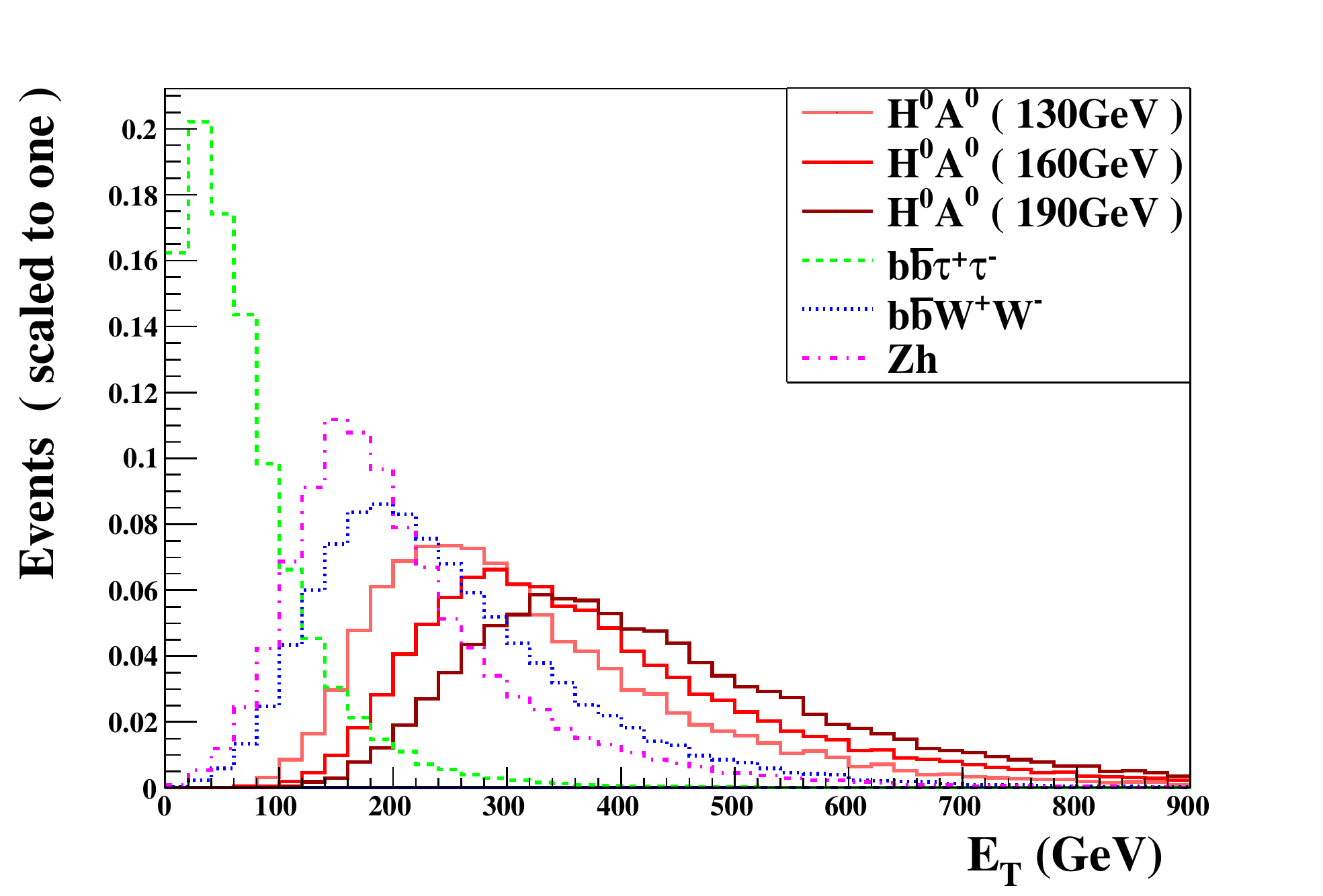}
\includegraphics[width=0.45\linewidth]{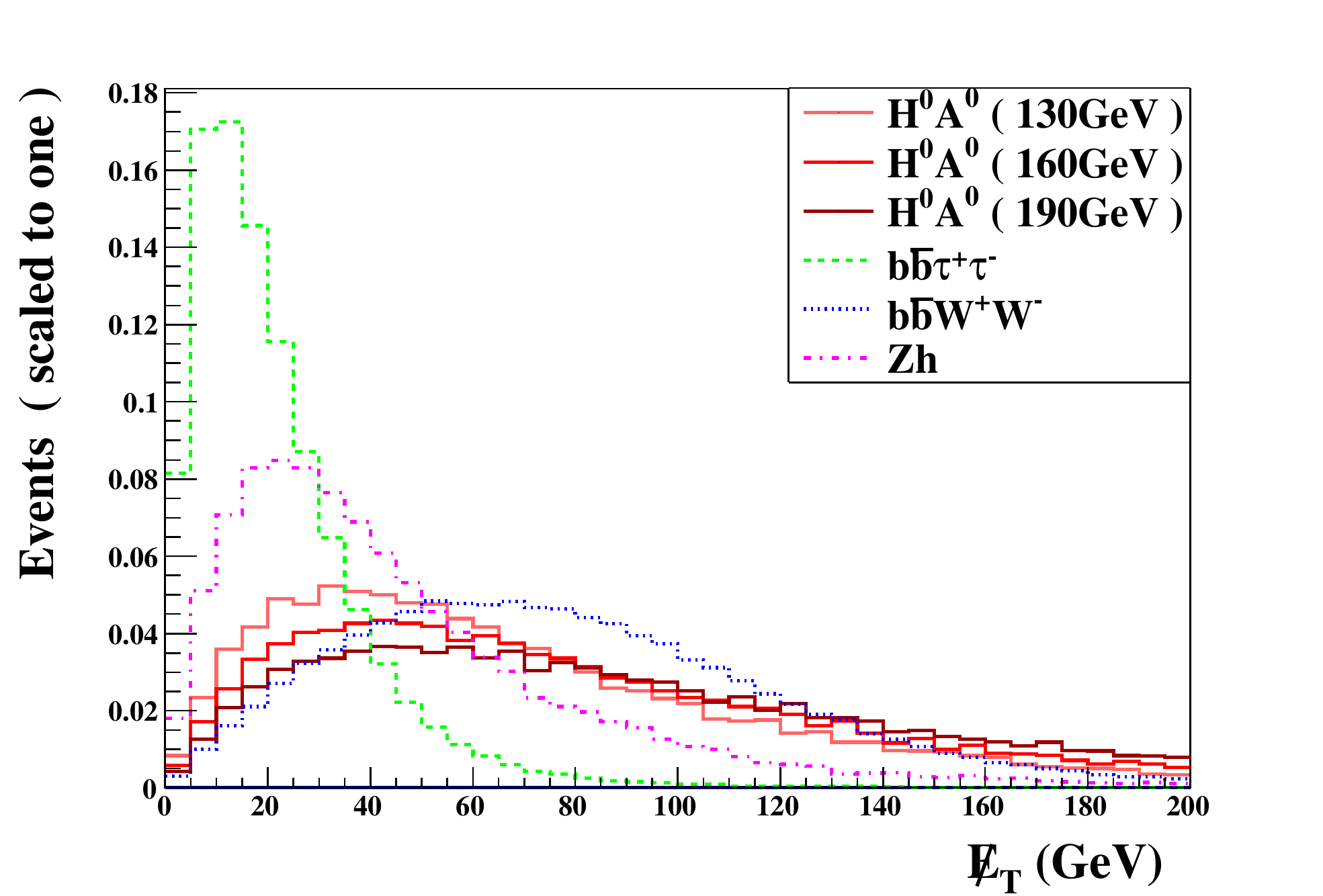}
\end{center}
\caption{Distributions of $p_T^\tau,~\Delta R_{\tau\tau},~M_{\tau\tau,H^0A^0},~E_T$, and $\cancel{E}_T$ for the signal $b\bar{b}\tau^+\tau^-$ and its backgrounds before applying any cuts at LHC14.
\label{fig:bbtata}}
\end{figure}

The main SM backgrounds are as follows:
\begin{eqnarray}
b\bar{b}\tau^+\tau^- : p p &\to& b\bar{b}Z/\gamma^*/h \to b\bar{b}\tau^+\tau^-,
\\
b\bar{b}W^+W^- : p p &\to &b\bar{b}W^+W^- \to b\bar{b}\tau^+ \nu_{\tau} \tau^- \bar{\nu}_{\tau},
\\
Zh : p p &\to& Zh \to b\bar{b}\tau^+\tau^-.
\end{eqnarray}
The irreducible QCD-electroweak background comes from $b\bar{b}\tau^+\tau^-$, where the $\tau$ pair originates from the decays of $Z/\gamma^*/h$. Since the hadronic decays of $\tau$ always contain neutrinos, we also include the SM background $b\bar{b}W^+W^-$, which contributes to the $b\bar{b}\tau^+ \nu_{\tau} \tau^- \bar{\nu}_{\tau}$ final state. The $b\bar{b}W^+W^-$ background mainly originates from $t\bar{t}$ production with subsequent decays $t\to b W$ and $W\to \tau \nu_{\tau}$. Moreover, the associated $Zh$ production gets involved through the subsequent decays $h\to b\bar{b}$ and $Z\to\tau^+\tau^-$ or vice versa. The QCD corrections to the backgrounds are included by a multiplicative $K$-factor of 1.21, 1.35, and 1.33 to the leading cross section of $b\bar{b}\tau^+\tau^-$~\cite{Campbell:2000bg}, $t\bar{t}$ \cite{tt}, and $Zh$ \cite{Dittmaier:2011ti} at LHC14.

\begin{table} [!htbp]
\begin{center}
\begin{tabular}{|c|c|c|c|c|c|c|}
\hline
 $M_{\Delta}=130~\GeV$  &  $H^0A^0(S_0)$  &  $b\bar{b}\tau^+\tau^-$ & $b\bar{b}W^+W^-$  & $Zh$ & $S/B$  & $\mathcal{S}(S,B)$
\\
\hline
Cross section at NLO & $4.31\times10^{1}$~~ & $3.10\times10^{4}$~~ & $7.92\times10^{3}$~~ & $2.21\times10^{1}$~~~& $1.11\times10^{-3}$ & $12.0$
\\
Basic cuts & $7.75\times10^{-1}$ & $4.49\times10^{1}$~~ & $8.97\times10^{1}$~~ & $2.91\times10^{-1}$ & $5.74\times10^{-3}$ & 3.65
\\
Reconstruct scalars from $\tau$s & $5.14\times10^{-1}$ & $1.19\times10^{1}$~~ & $3.57\times10^{1}$~~ & $1.06\times10^{-1}$ & $1.08\times10^{-2}$ & 4.06
\\
Reconstruct scalars from $b$s & $2.14\times10^{-1}$ & 4.34  & $9.44\times10^{-1}$ & $2.28\times10^{-2}$  & $4.04\times10^{-2}$  & 5.06
\\
Cut on $M_{H^0A^0}$ & $1.29\times10^{-1}$ & $1.96\times10^{-1}$  & $1.51\times10^{-1}$ &$8.10\times10^{-3}$ & $3.64\times10^{-1}$ & $11.3$
\\
Cut on $E_T$ & $1.03\times10^{-1}$ & $9.87\times10^{-2}$ & $7.35\times10^{-2}$ & $5.89\times10^{-3}$  & $5.81\times10^{-1}$ & $12.4$
\\
\hline
Cascade enhanced & $5.27\times10^{-1}$ & $-$ & $-$ & $-$  & $2.96$ & $51.6$
\\
\hline
\hline
$M_{\Delta}=160~\GeV$  &  $H^0A^0(S_0)$  &  $b\bar{b}\tau^+\tau^-$ & $b\bar{b}W^+W^-$  & $Zh$ & $S/B$  & $\mathcal{S}(S,B)$
\\
\hline
Cross section at NLO & $3.08$~~ & $3.10\times10^{4}$~~ & $7.92\times10^{3}$~~ & $2.21\times10^{1}$~~~& $7.92\times10^{-5}$ & $8.55\times10^{-1}$
\\
Basic cuts & $6.81\times10^{-2}$ & $4.49\times10^{1}$~~ & $8.97\times10^{1}$~~ & $2.91\times10^{-1}$ & $5.05\times10^{-4}$ & $3.21\times10^{-1}$
\\
Reconstruct scalars from $\tau$s & $3.14\times10^{-2}$ & $1.52\times10^{1}$~~ & $2.46\times10^{2}$~~ & $3.17\times10^{-2}$ & $1.20\times10^{-3}$ & $3.36\times10^{-1}$
\\
Reconstruct scalars from $b$s & $1.2\times10^{-2}$ & 2.47  & $1.06\times10^{-1}$ & 0.00 & $4.80\times10^{-3}$  & $4.10\times10^{-1}$
\\
Cut on $M_{H^0A^0}$ & $6.99\times10^{-3}$ & $1.22\times10^{-1}$  & $2.06\times10^{-2}$ & 0.00 & $4.89\times10^{-2}$ & 1.00
\\
Cut on $E_T$ & $5.04\times10^{-3}$ & $4.72\times10^{-2}$ & $5.88\times10^{-3}$ & 0.00  & $9.48\times10^{-2}$ & 1.18
\\
\hline
Cascade enhanced & $5.11\times10^{-2}$ & $-$ & $-$ & $-$  & $9.63\times10^{-1}$ & $10.7$
\\
\hline
\hline
$M_{\Delta}=190~\GeV$  &  $H^0A^0(S_0)$  &  $b\bar{b}\tau^+\tau^-$ & $b\bar{b}W^+W^-$  & $Zh$ & $S/B$  & $\mathcal{S}(S,B)$
\\
\hline
Cross section at NLO & $2.47\times10^{-1}$  & $3.10\times10^{4}$~~  & $7.92\times10^{3}$~~  & $2.21\times10^{1}$~~~ & $6.34\times10^{-6}$  &  $6.86\times10^{-2}$
\\
Basic cuts & $6.54\times10^{-3}$ & $4.49\times10^{1}$~~ & $8.97\times10^{1}$~~ & $2.91\times10^{-1}$ & $4.86\times10^{-5}$ & $3.09\times10^{-2}$
\\
Reconstruct scalars from $\tau$s & $2.32\times10^{-3}$  & $4.60\times10^{-1}$ & $1.47\times10^{1}$~~  & $5.89\times10^{-3}$  & $1.53\times10^{-4}$  &  $3.26\times10^{-2}$
\\
Reconstruct scalars from $b$s & $7.66\times10^{-4}$ &  $2.06\times10^{-2}$  & 1.21  & 0.00  & $6.25\times10^{-4}$ &  $3.78\times10^{-2}$
\\
Cut on $M_{H^0A^0}$ & $6.34\times10^{-4}$  & $1.47\times10^{-3}$  & $9.03\times10^{-2}$  & 0.00 & $6.91\times10^{-3}$  &  $1.14\times10^{-1}$
\\
Cut on $E_T$ & $3.87\times10^{-4}$  & 0.00  & $2.64\times10^{-2}$  & 0.00  & $1.47\times10^{-2}$  &  $1.30\times10^{-1}$
\\
\hline
Cascade enhanced & $6.85\times10^{-3}$  & $-$  & $-$  & $-$  & $2.60\times10^{-1}$  &  2.22
\\
\hline
\end{tabular}
\end{center}
\caption{Similar to Table~\ref{tab:bbaacut}, but for the $b\bar{b}\tau^+\tau^-$ signal channel.}
\label{tab:bbtatacut}
\end{table}

The kinematical distributions similar to the $b\bar b\gamma\gamma$ channel are shown in Fig.~\ref{fig:bbtata}. As one can see from the figure, the $\tau$ jets are less energetic than the $b$ jets (similar to those in the $b\bar{b}\gamma\gamma$ signal channel) due to missing neutrinos in the final state. We first employ the following selection cuts to pick up signals with exactly one $b$ pair and one $\tau$ pair:
\begin{eqnarray}
p_T^{b,\tau}>30~\GeV,~|\eta_{b,\tau}|<2.4,~\Delta R_{bb,b\tau,\tau\tau}>0.4,
\end{eqnarray}
and no cut on $\cancel{E}_T$ is adopted. After the selection, the $\tau$ and $b$ pairs are required to fulfill the cuts on the invariant masses and separations:
\begin{eqnarray}
\Delta R_{\tau\tau}<2.5,M_{\Delta}-40~\GeV&<&M_{\tau\tau}<M_{\Delta},\\
\nonumber
\Delta R_{bb}<2.5,~|M_{bb}-M_{\Delta}|&<&15~\GeV.
\end{eqnarray}
The different mass shift between $M_{\tau\tau}$ and $M_{bb}$ is owing to the missing neutrinos in $\tau$ decays resulting in a wider distribution of $M_{\tau\tau}$. For the reconstructed neutral scalars, we further adopt similar cuts on $M_{H^0A^0}$ and $E_T$ as in the $b\bar{b}\gamma\gamma$ channel:
\begin{equation}
M_{H^0A^0}>2M_{\Delta}+70~\GeV,~E_T>2M_{\Delta}-80~\GeV.
\end{equation}
Both $M_{H^0A^0}$ and $E_T$ cuts are reduced by $20~\GeV$ compared with the $b\bar{b}\gamma\gamma$ channel, which again results from neutrinos in the final state. The corresponding results are summarized in Table \ref{tab:bbtatacut}.

The $b\bar{b}\tau^+\tau^-$ is also promising for $M_{\Delta}=130~\GeV$ even without enhancement from cascade decays. The final significance is 12.4 and the corresponding number of signal events is 309 for LHC14@3000. Including the cascade enhancement, the significance is improved to 51.6, which is even better than the $b\bar{b}\gamma\gamma$ signal. For $M_{\Delta}=160~\GeV$, the biggest challenge is also the small production cross section of the signal. But in the most optimistic case, the cascade decays can increase the signal by a factor of 10.1, making this channel feasible. Finally, neutral scalars as heavy as $190~\GeV$ are difficult to detect at LHC14 in this channel.

\subsection{$b\bar{b}W^+W^-$ signal channel}
\label{sec:bbWW}

It is difficult to search for the SM Higgs pair production in this channel due to missing energy brought about by neutrinos in leptonic decays of the $W$ boson, which makes one of the two Higgs bosons not fully reconstructible~\cite{Gouzevitch:2013qca,Baglio:2012np}. The situation is ameliorated in our scenario because, the production rate of $H^0A^0$ can be an order of magnitude larger than that of $hh$ and the di-$W$ decay branching ratio of $H^0$ can also be larger than $h$ in the vast parameter space. This considerably improves the signal events and partially compensates the deficiency of the detection capability.

\begin{figure}[!htbp]
\begin{center}
\includegraphics[width=0.45\linewidth]{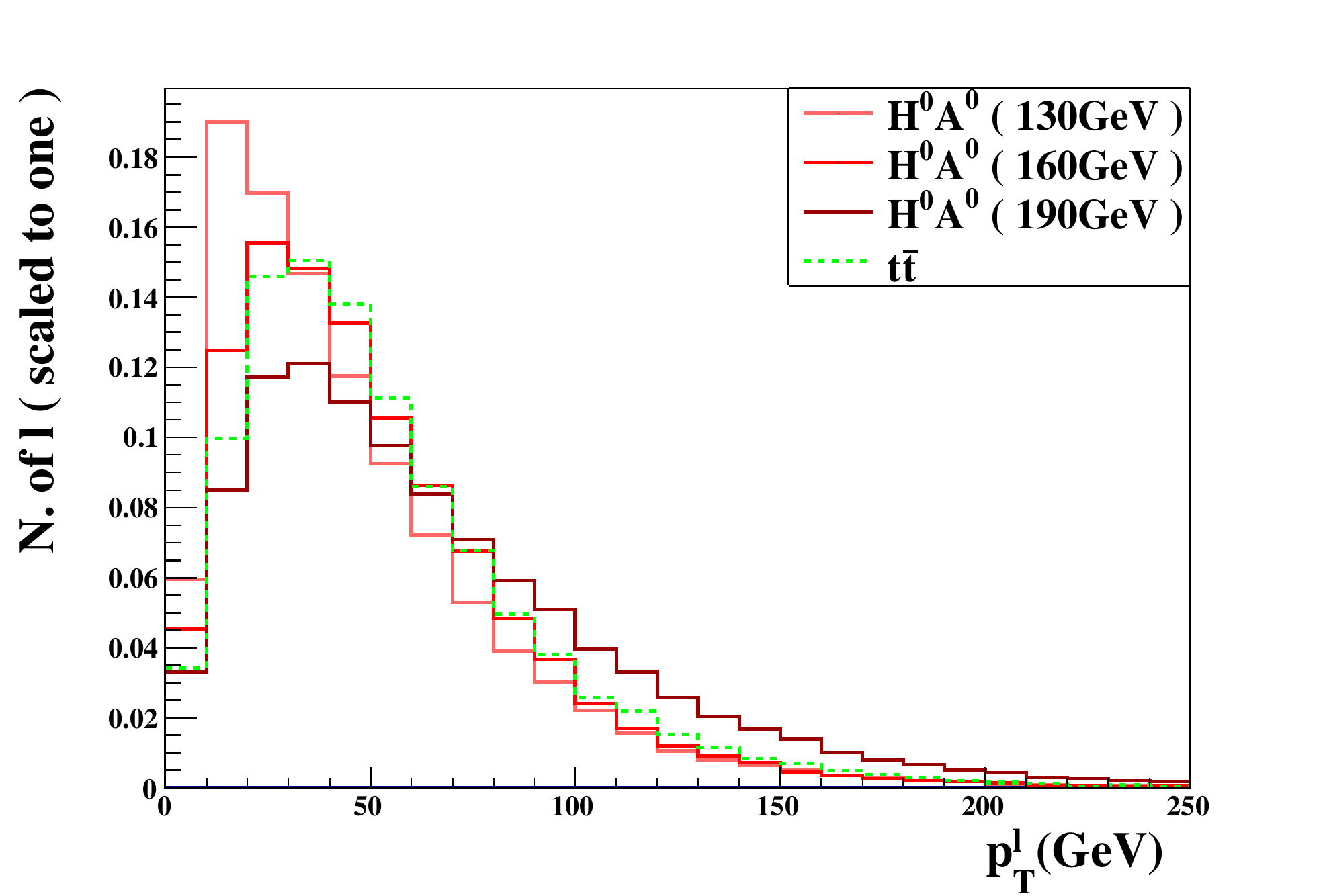}
\includegraphics[width=0.45\linewidth]{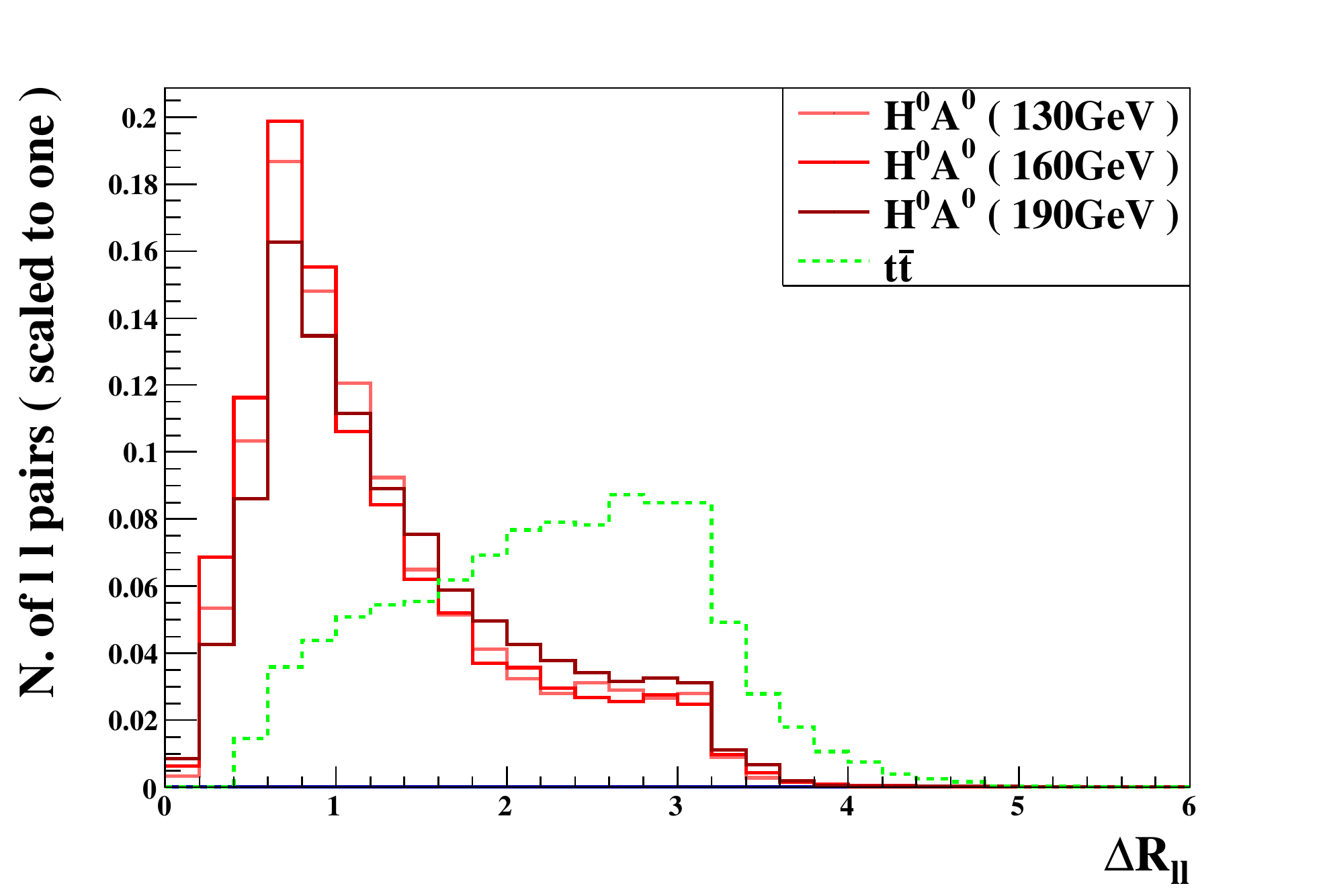}
\includegraphics[width=0.45\linewidth]{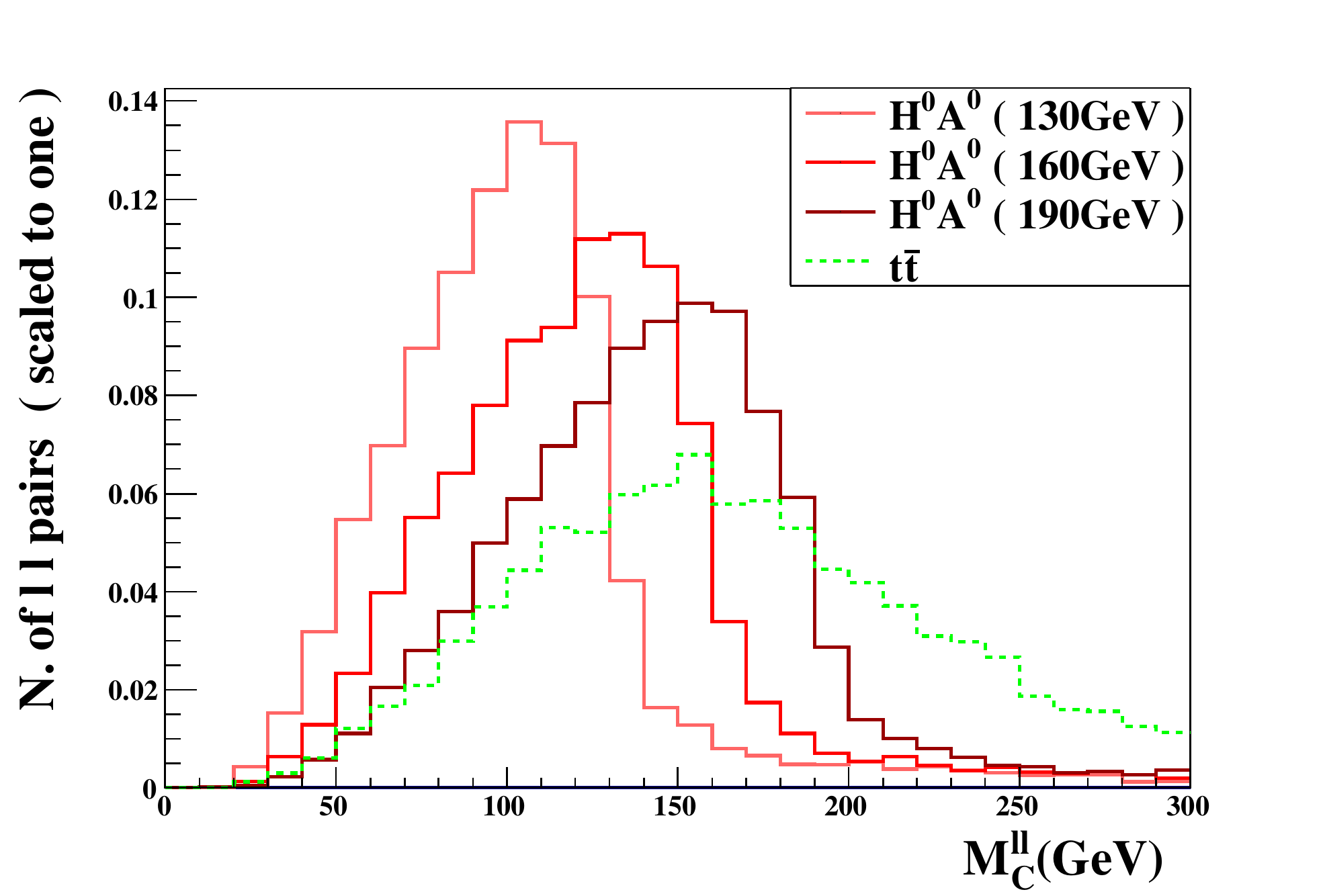}
\includegraphics[width=0.45\linewidth]{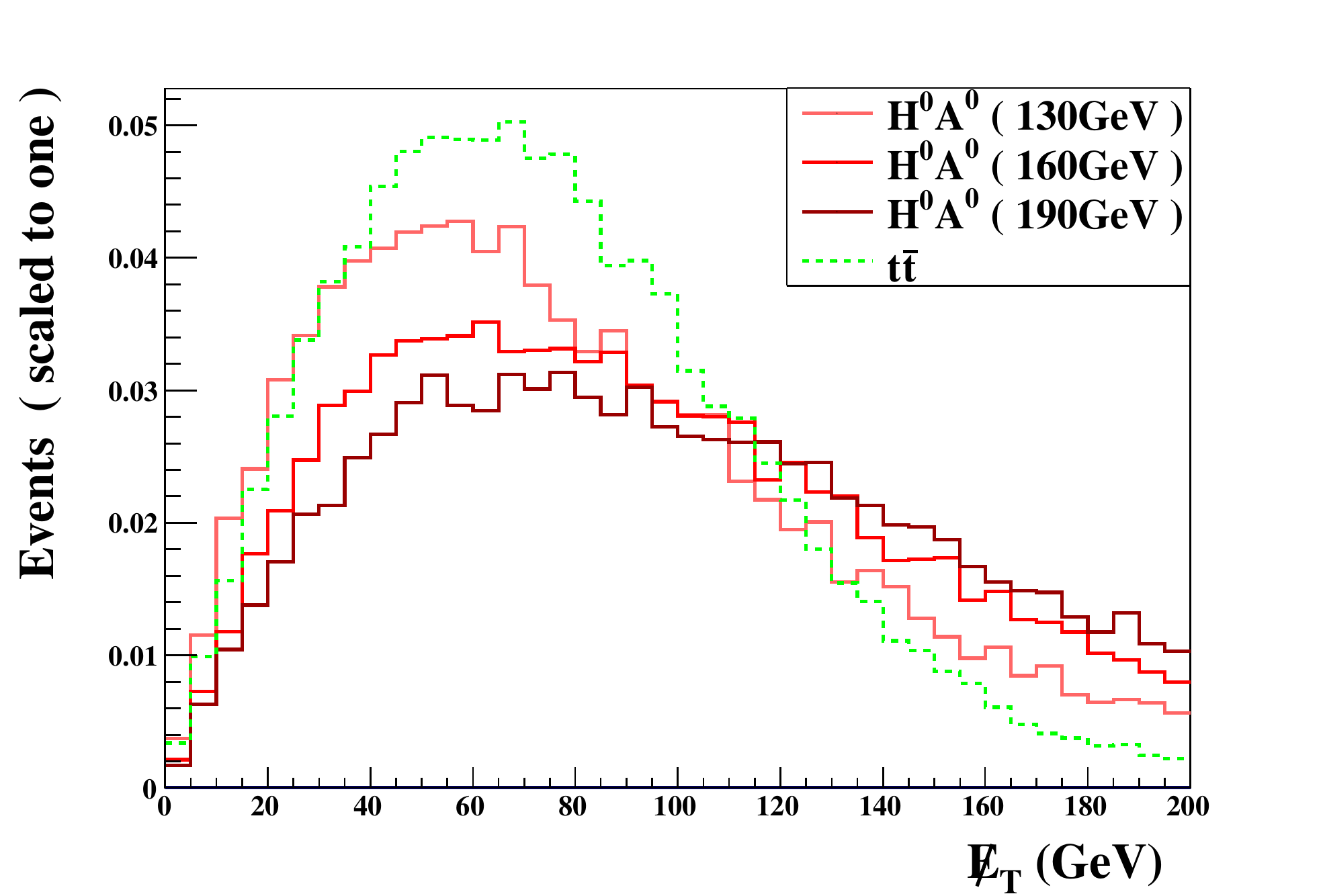}
\includegraphics[width=0.45\linewidth]{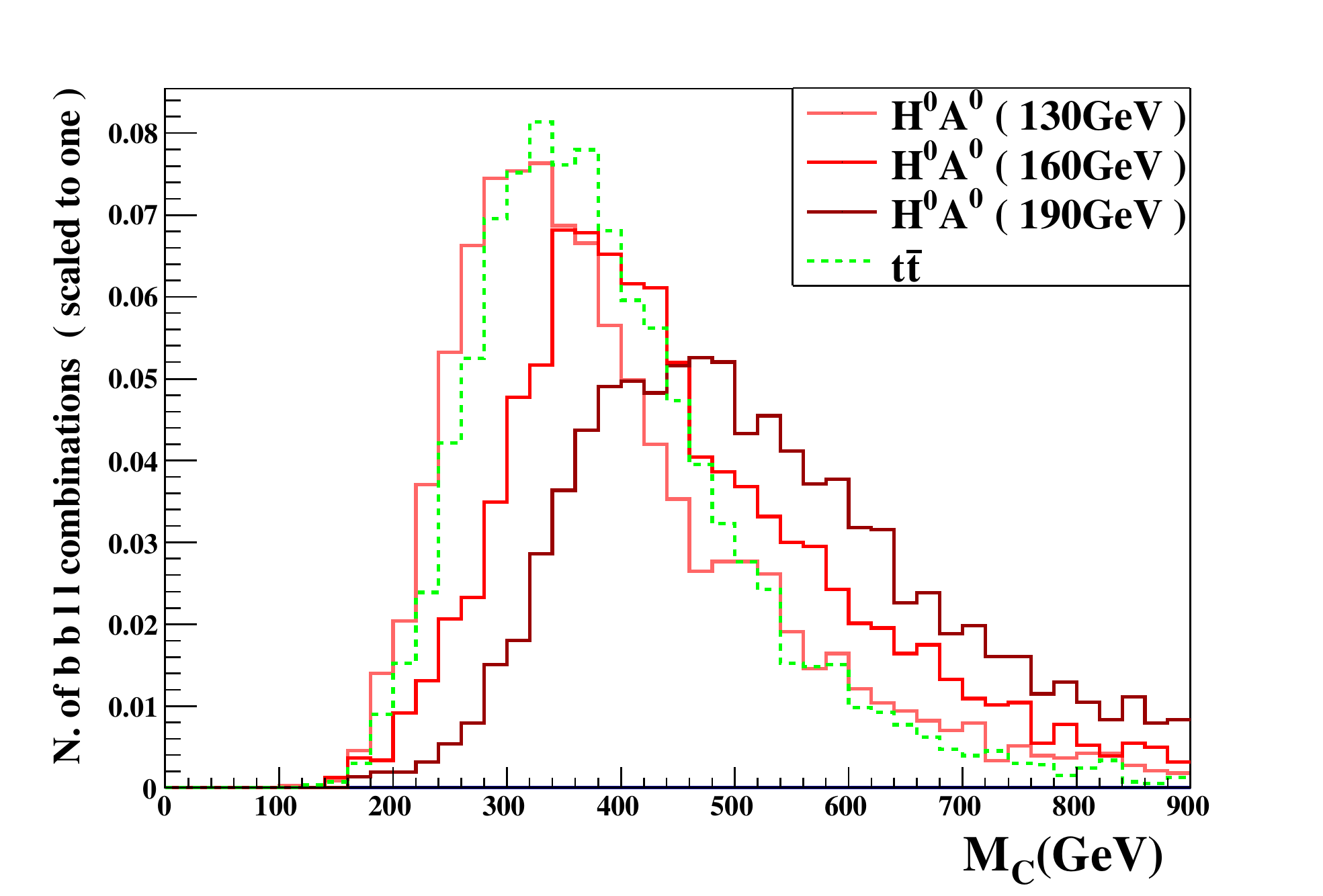}
\includegraphics[width=0.45\linewidth]{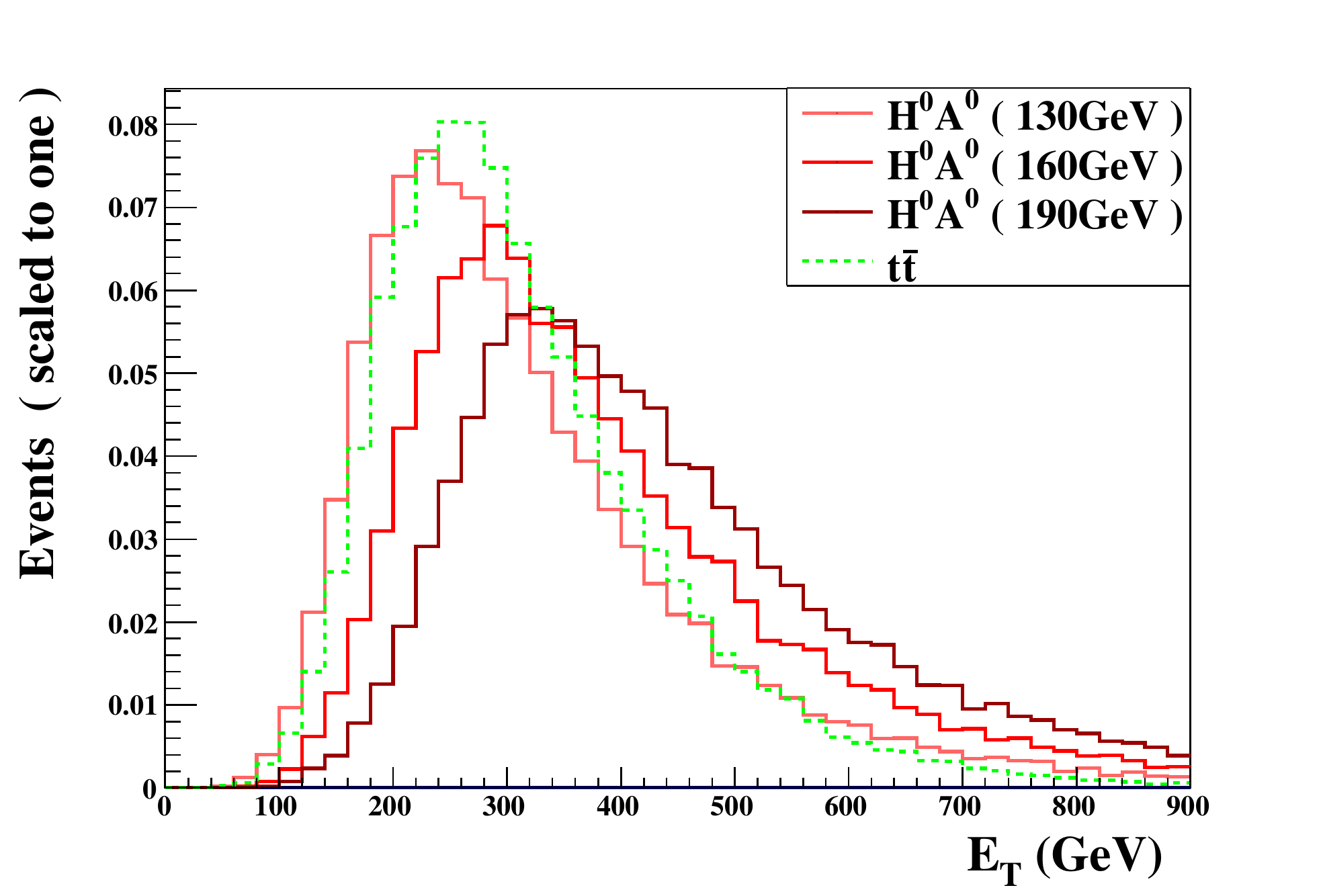}
\end{center}
\caption{Distributions of $p_T^\ell,~\Delta R_{\ell\ell},~M^{\ell\ell}_C,~\cancel{E}_T,~M_C$, and $E_T$ for the signal $b\bar{b}\ell^+\ell^-\cancel{E}_T$ and its backgrounds before applying any cuts at LHC14.
\label{fig:bbll2v}}
\end{figure}

\begin{table} [!htbp]
\begin{center}
\begin{tabular}{|c|c|c|c|c|}
\hline
 $M_{\Delta}=130~\GeV$  &  $H^0A^0(S_0)$  &  $t\bar{t}$ &  $S/B$  & $\mathcal{S}(S,B)$
\\
\hline
Cross section at NLO & 3.91 & $2.38\times10^4$ ~~& $1.69\times10^{-4}$ & 1.41
\\
Basic cuts & $1.51$ & $4.04\times10^3$~~  & $3.74\times10^{-4}$ & $1.30$
\\
Reconstruct scalars from $b$s & $3.29\times10^{-1}$ & $3.35\times10^2$~~  & $9.82\times10^{-4}$ & $0.984$
\\
Cut on $M_C^{\ell\ell}$ & $3.21\times10^{-1}$ & $2.14\times10^2$~~  & $1.50\times10^{-3}$ & 1.20
\\
Cut on $\Delta R_{\ell\ell}$ & $2.64\times10^{-1}$ & $9.26\times10^1$~~ & $2.85\times10^{-3}$ & 1.50
\\
Cut on $\cancel{E}_T$ & $8.45\times10^{-2}$ & $1.48\times10^1$~~  & $5.71\times10^{-3}$ & 1.20
\\
Cut on $M_C$ & $3.30\times10^{-2}$  & $1.69\times10^{-1}$ & $1.95\times10^{-1}$ & 4.26
\\
Cut on $E_T$ & $3.19\times10^{-2}$  & $1.47\times10^{-1}$ & $2.17\times10^{-1}$ & 4.41
\\
\hline
Cascade enhanced& $1.40\times10^{-1}$ & $-$ & $9.53\times10^{-1}$ & $17.7$
\\
\hline
\hline
 $M_{\Delta}=160~\GeV$  &  $H^0A^0(S_0)$  &  $t\bar{t}$ &  $S/B$  & $\mathcal{S}(S,B)$
\\
\hline
Cross section at NLO & 4.95 & $2.38\times10^4$~~ & $2.13\times10^{-4}$ & 1.78
\\
Basic cuts & $2.13$ & $4.04\times10^3$~~  & $5.27\times10^{-4}$ & $1.84$
\\
Reconstruct scalars from $b$s & $4.25\times10^{-1}$ & $2.68\times10^2$~~   & $1.59\times10^{-3}$ & 1.42
\\
Cut on $M_C^{\ell\ell}$ & $3.97\times10^{-1}$ & $1.89\times10^2$~~  & $2.10\times10^{-3}$ & 1.58
\\
Cut on $\Delta R_{\ell\ell}$ & $3.21\times10^{-1}$ & $7.04\times10^1$~~ & $4.56\times10^{-3}$ & 2.09
\\
Cut on $\cancel{E}_T$ & $9.47\times10^{-2}$ & 4.29 & $2.21\times10^{-2}$ & 2.50
\\
Cut on $M_C$ & $3.28\times10^{-2}$ & $4.74\times10^{-2}$ & $6.92\times10^{-1}$ & 7.50
\\
Cut on $E_T$ & $3.02\times10^{-2}$ & $3.62\times10^{-2}$ & $8.34\times10^{-1}$ & 7.78
\\
\hline
Cascade enhanced & $1.01\times10^{-1}$ & $-$  & 3.24 & $23.2$
\\
\hline
\hline
 $M_{\Delta}=190~\GeV$  &  $H^0A^0(S_0)$  &  $t\bar{t}$ &  $S/B$  & $\mathcal{S}(S,B)$
\\
\hline
Cross section at NLO & 1.19  & $2.38\times10^4$~~ & $5.00\times10^{-5}$  &  $0.424$
\\
Basic cuts & $6.44\times10^{-1}$ & $4.04\times10^3$~~  & $1.59\times10^{-4}$ & $0.554$
\\
Reconstruct scalars from $b$s & $1.36\times10^{-1}$  &  $2.26\times10^{2}$~~  & $6.02\times10^{-4}$ &  $0.495$
\\
Cut on $M_C^{\ell\ell}$ & $1.27\times10^{-1}$  & $1.79\times10^{2}$~~ & $7.09\times10^{-4}$  &  $0.520$
\\
Cut on $\Delta R_{\ell\ell}$ & $9.70\times10^{-2}$ & $6.05\times10^{1}$~~ & $1.60\times10^{-3}$  &  $0.683$
\\
Cut on $\cancel{E}_T$ & $2.57\times10^{-2}$ & 1.62 & $1.59\times10^{-2}$  &  1.10
\\
Cut on $M_C$ & $8.85\times10^{-3}$  & $1.89\times10^{-2}$  & $4.68\times10^{-1}$  & 3.29
\\
Cut on $E_T$ & $8.37\times10^{-3}$  & $1.40\times10^{-2}$  & $5.98\times10^{-1}$  & 3.56
\\
\hline
Cascade enhanced & $2.69\times10^{-2}$  & $-$  & 1.92  &  $10.1$~~
\\
\hline
\end{tabular}
\end{center}
\caption{Similar to Table~\ref{tab:bbaacut}, but for the $b\bar{b}\ell^+\ell^-\cancel{E}_T$ signal channel.}
\label{tab:bbllcut}
\end{table}

With both $W$'s decaying leptonically, the final state appears as $b\bar{b}\ell^+\ell^-\cancel{E}_T$. The dominant SM backgrounds are as follows:
\begin{equation}
t\bar{t}:pp\to t\bar{t}\to bW^+\bar{b}W^-\to b\bar{b}\ell^+\ell^-\cancel{E}_T.
\end{equation}
As before, the QCD correction is included by a multiplicative $K$-factor of 1.35 for the $t\bar{t}$ production~\cite{tt}. We pick up the events that include exactly one $b$-jet pair and one opposite-sign lepton pair and filter them with the basic cuts:
\begin{eqnarray}
p_T^{b}>30~\GeV,~p_T^{\ell}>20~\GeV,~|\eta_{b,\ell}|<2.4,\\
\nonumber
\Delta R_{bb,b\ell,\ell\ell}>0.4,~\cancel{E}_T>20~\GeV.
\end{eqnarray}
The separation and invariant mass of the $b$-jet pair are required to fulfill
\begin{equation}
\Delta R_{bb}<2.5,~|M_{bb}-M_{\Delta}|<15~\GeV.
\end{equation}
For the lepton pair, we reconstruct the transverse cluster mass $M_C^{\ell\ell}$:
\begin{equation}
M_C^{\ell\ell}=\sqrt{\left(\sqrt{p_{T,\ell\ell}^2+M^2_{\ell\ell}}+\cancel{E}_T\right)^2
+\left(\vec{p}_{T,\ell\ell}+\vec{\cancel{E}}_T\right)^2}.
\end{equation}
The distributions of $M_C^{\ell\ell}$, $\Delta R_{\ell\ell}$, and $\cancel{E}_T$ are shown in Fig.~\ref{fig:bbll2v}. The peak of $M_C^{\ell\ell}$ is always lower than $M_{\Delta}$ by about $30$-$40~\GeV$, and the lepton separation $\Delta R_{\ell\ell}$ in the signal is much smaller than in the $t\bar{t}$ background. Accordingly, we set a wide window on $M_{C}^{\ell\ell}$ while tightening up the cuts on $\Delta R_{\ell\ell}$ and $\cancel{E}_T$:
\begin{equation}
M_{\Delta}-80~\GeV<M_C^{\ell\ell}<M_{\Delta},~\Delta R_{\ell\ell}<1.2,~\cancel{E}_T>0.9M_{\Delta}.
\end{equation}
We find that $M_{C}^{\ell\ell}$ is least efficient around $M_{\Delta}\sim190~\GeV$, where the peak of $M_{C}^{\ell\ell}$ for the $t\bar{t}$ background is around $150~\GeV$. The very tight cuts on $\Delta R_{\ell\ell}$ and $\cancel{E}_T$ are sufficient to suppress the background by 1 or 2 orders of magnitude, while keeping the number of signal events as large as possible. We further combine the $b$-jet pair and the lepton pair into a cluster and construct the transverse cluster mass:
\begin{equation}
M_C=\sqrt{\left(\sqrt{p_{T,bb\ell\ell}^2+M_{bb\ell\ell}^2}+\cancel{E}_T\right)^2-\left( \vec{p}_{T,bb\ell\ell}+\vec{\cancel{E}}_T\right)^2},
\end{equation}
\begin{figure}[!htbp]
\begin{center}
\includegraphics[width=0.44\linewidth]{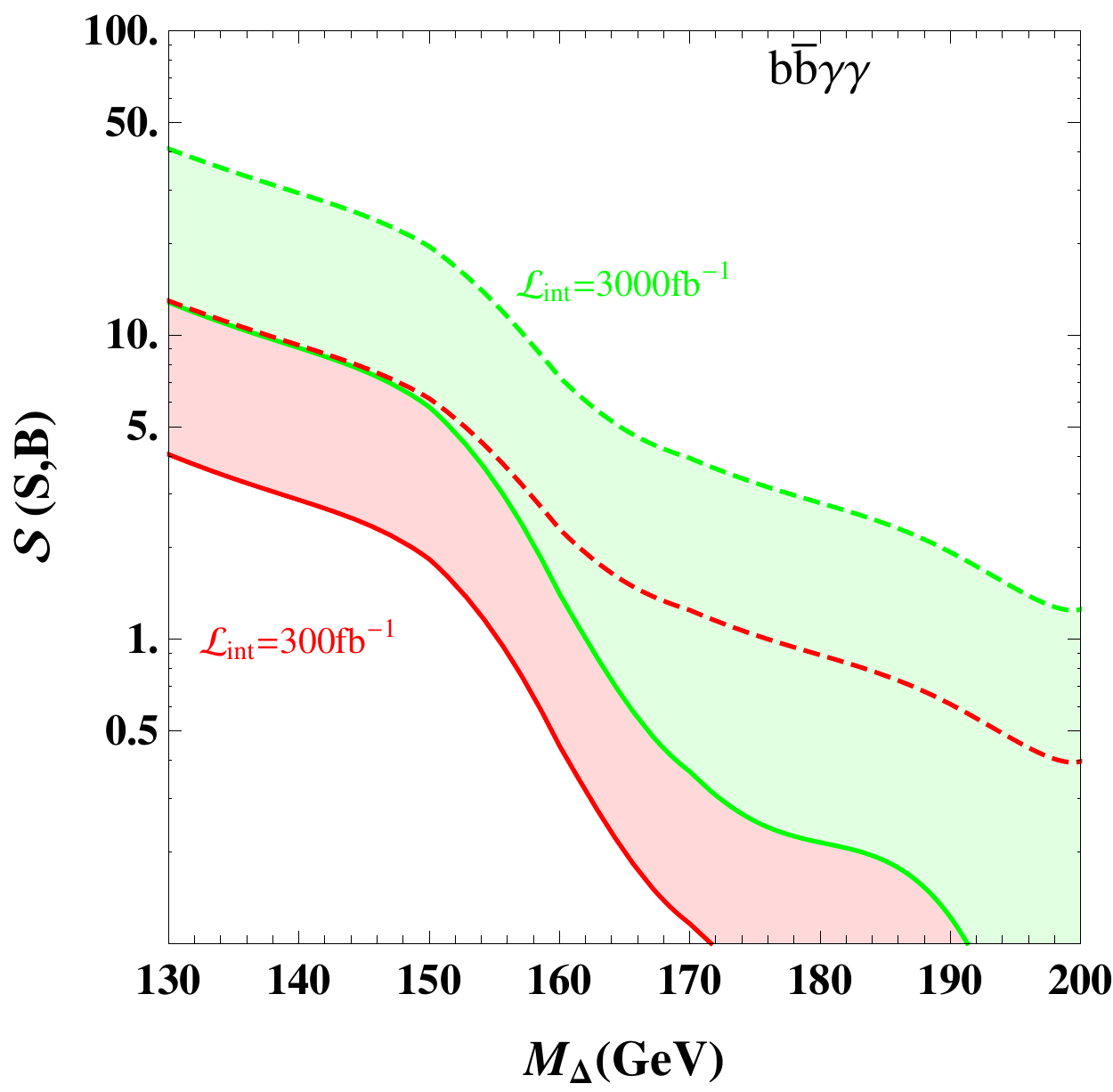}
\includegraphics[width=0.45\linewidth]{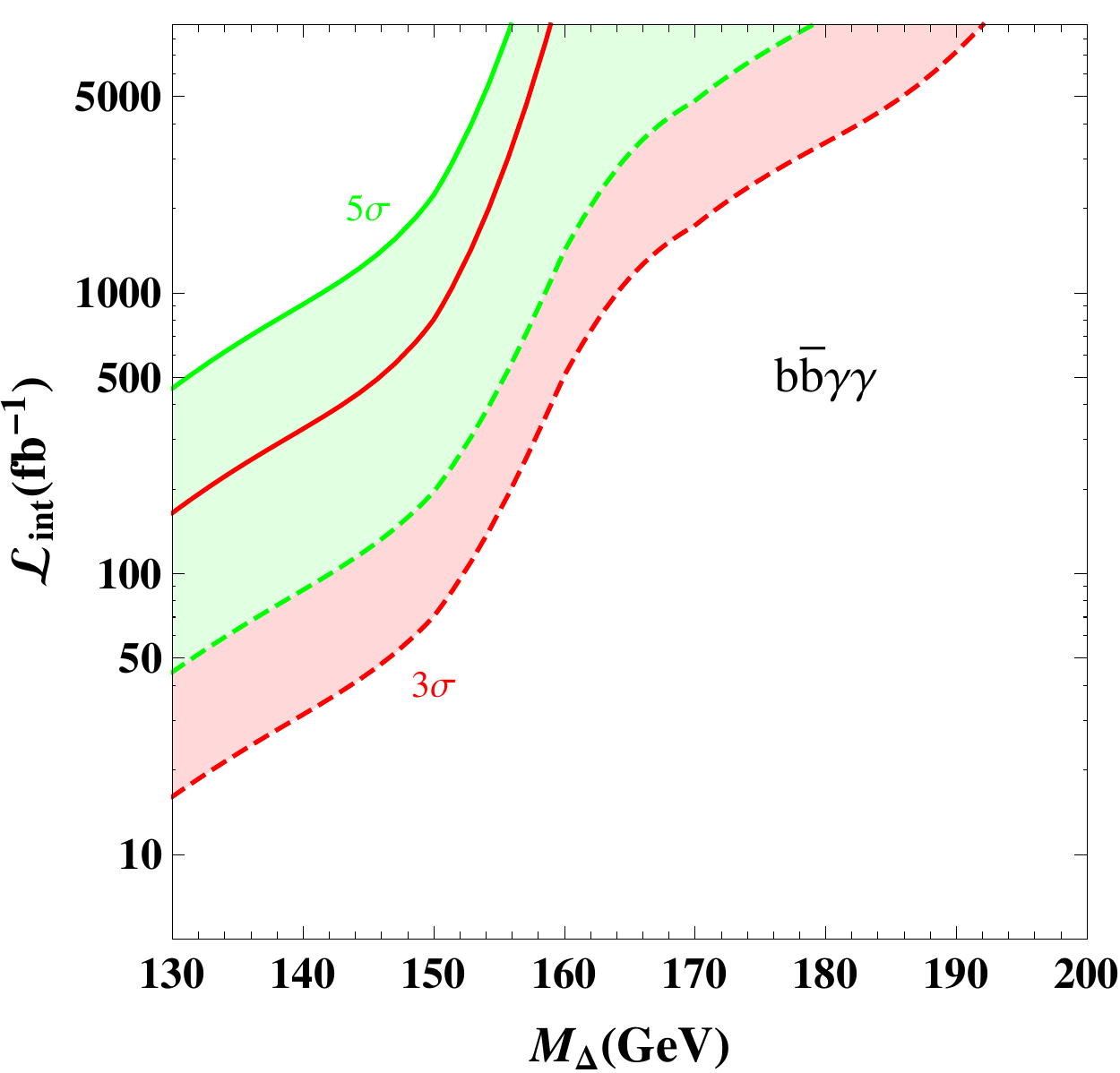}
\end{center}
\caption{Left: Significance $\mathcal{S}(S,B)$ of the $b\bar{b}\gamma\gamma$ channel versus $M_{\Delta}$ reachable at LHC14@300 (red region) and LHC14@3000 (green). Right: Required luminosity to reach a $3\sigma$ (red region) and $5\sigma$ (green) significance in the $b\bar{b}\gamma\gamma$ channel versus $M_{\Delta}$ at LHC14. The solid line corresponds to the signal from $X_0$ alone, and the dashed line corresponds to the total signal including cascade enhancement. \label{bbaa_sen}}
\end{figure}
which is an analog of $M_{H^0A^0}$ in the previous subsection. The distribution of $M_C$ is displayed in Fig.~\ref{fig:bbll2v}, which is very similar to that of $M_{H^0A^0}$ in the $b\bar{b}\gamma\gamma$ channel. Although it looks from the $M_C$ distributions (before any cuts are made) that the $t\bar{t}$ background has a large overlap with the signal, the cuts on $M_C^{\ell\ell}$, $\Delta R_{\ell\ell}$, and $\cancel{E}_T$ actually modify them remarkably, so that a further cut on $M_C$ could improve the significance efficiently. We apply a cut on $M_C$ as we did with $M_{H^0A^0}$, as well as one on $E_T$:
\begin{equation}
M_C>2M_{\Delta}+90~\GeV,~E_T>2M_{\Delta}-60~\GeV.
\end{equation}
The results following the cutflow are summarized in Table \ref{tab:bbllcut}. For $M_{\Delta}=130~\GeV$, the final significance is 4.41 (17.7) without (with) cascade enhancement. With cascade enhancement this should be enough to discover the neutral scalars. The signal channel is more promising for $M_{\Delta}=160~\GeV$ due to a slightly larger cross section and higher cut efficiencies. The final significance is 7.78 (23.2), which is also better than the $b\bar{b}\gamma\gamma$ and $b\bar{b}\tau^+\tau^-$ channels with the same mass. Finally, for $M_{\Delta}=190~\GeV$, the significance becomes 3.56 (10.1). Therefore, for our benchmark model, the only promising signal for such heavy neutral scalars ($\sim190~\GeV$) comes from the $b\bar{b}W^+W^-$ channel.

\subsection{Observability}

\begin{figure}[!htbp]
\begin{center}
\includegraphics[width=0.44\linewidth]{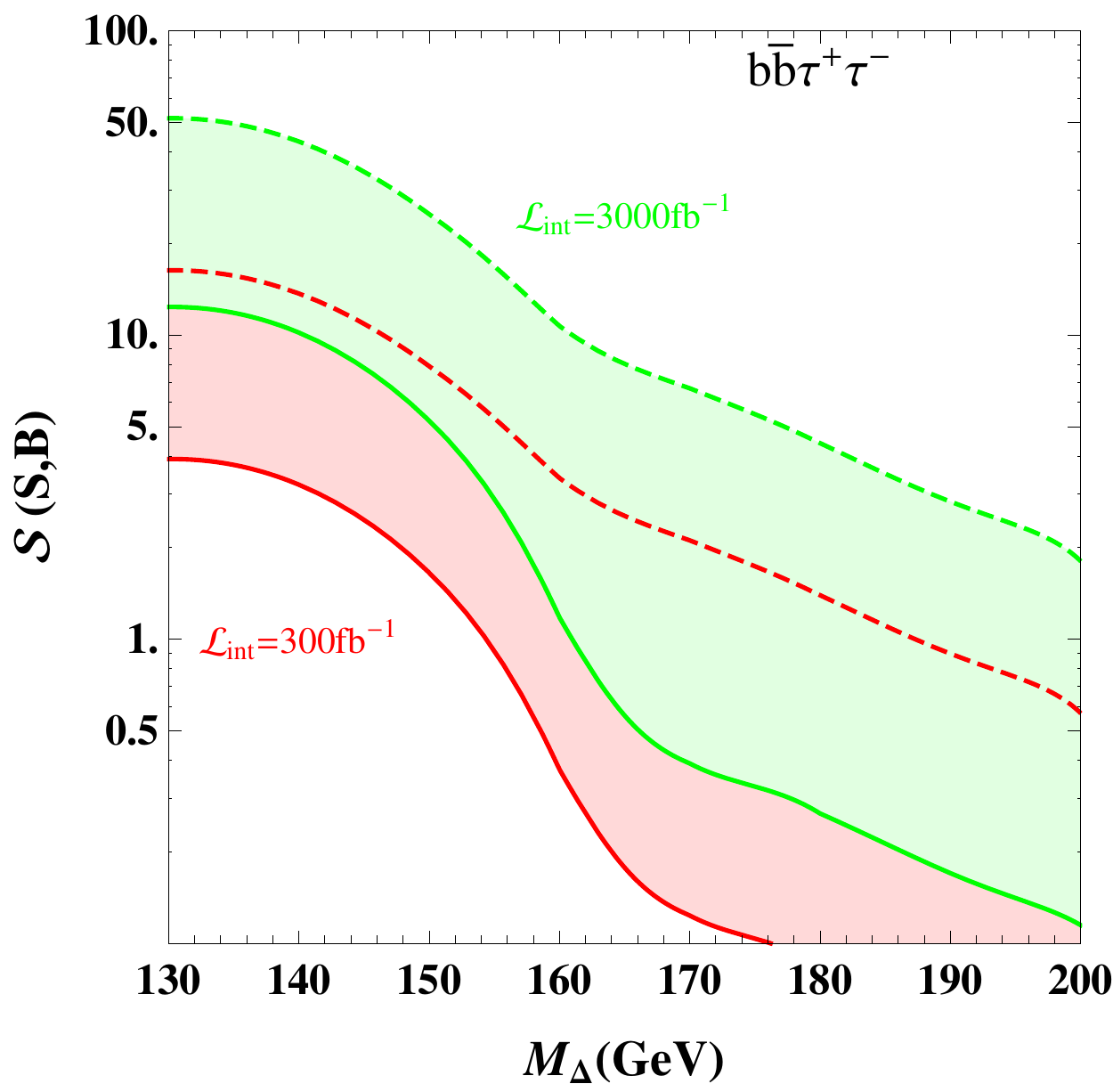}
\includegraphics[width=0.45\linewidth]{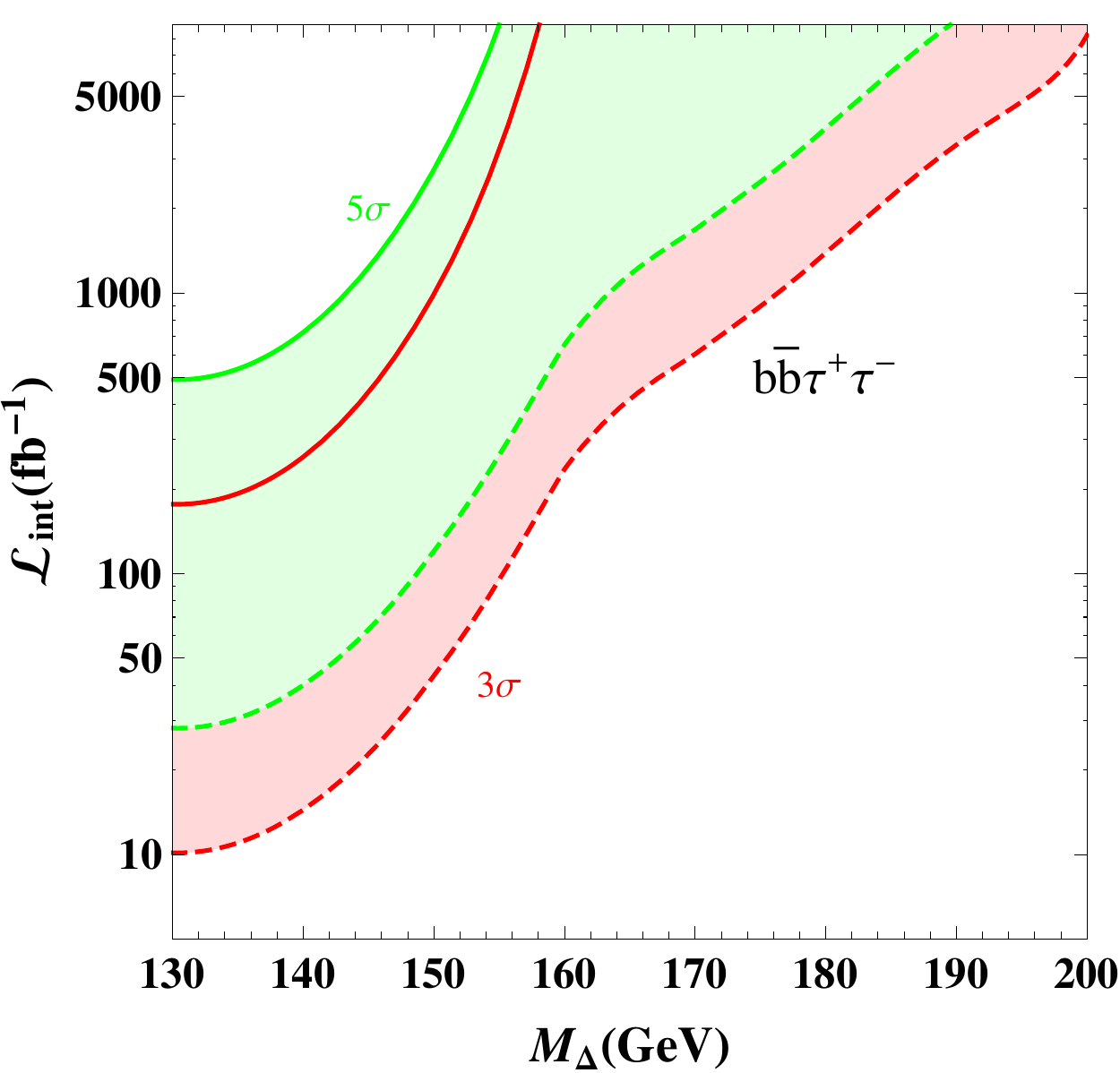}
\end{center}
\caption{Same as Fig.~\ref{bbaa_sen}, but for the $b\bar{b}\tau^+\tau^-$ channel.
\label{bbtata_sen}}
\end{figure}

\begin{figure}[!htbp]
\begin{center}
\includegraphics[width=0.44\linewidth]{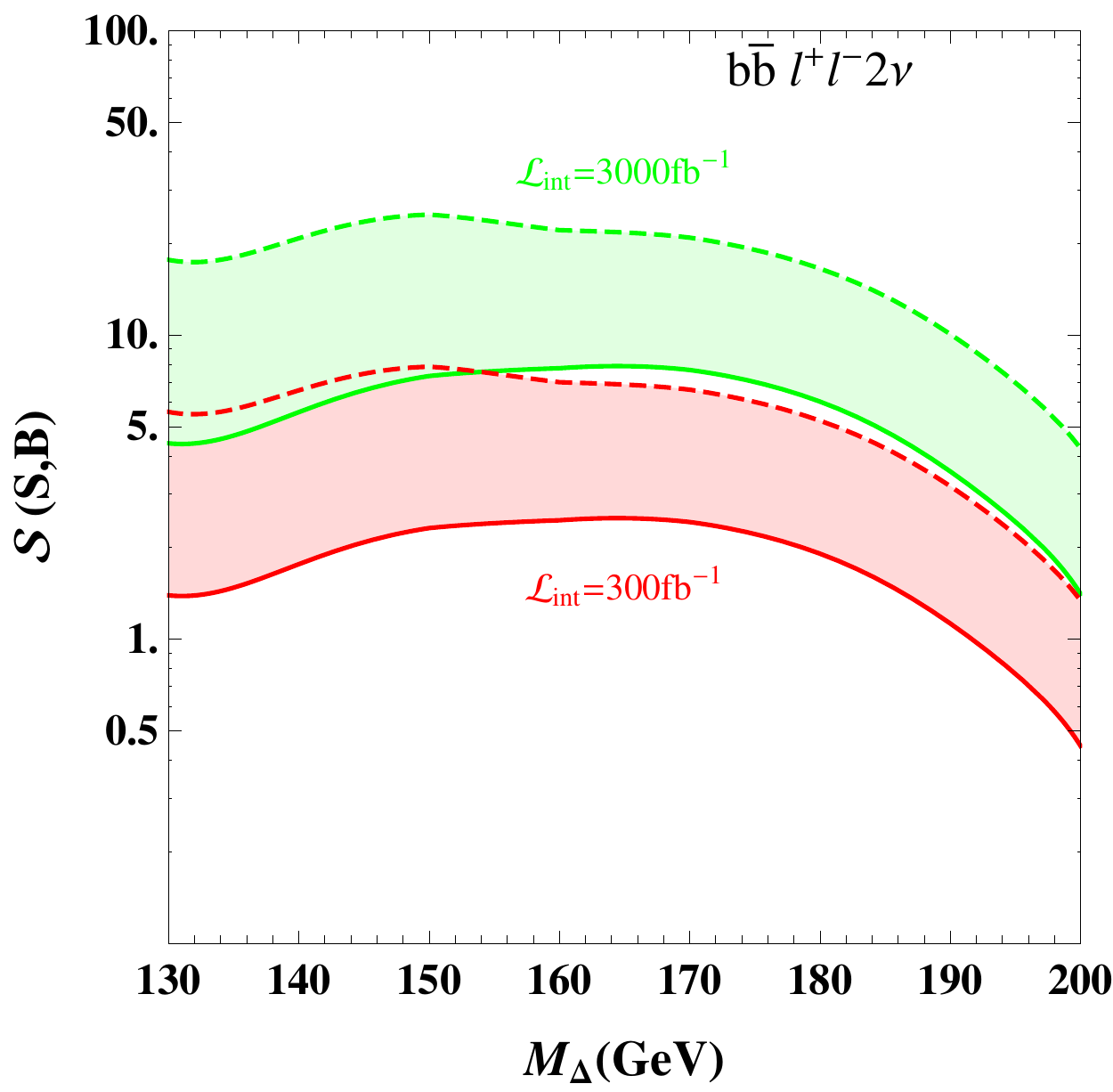}
\includegraphics[width=0.45\linewidth]{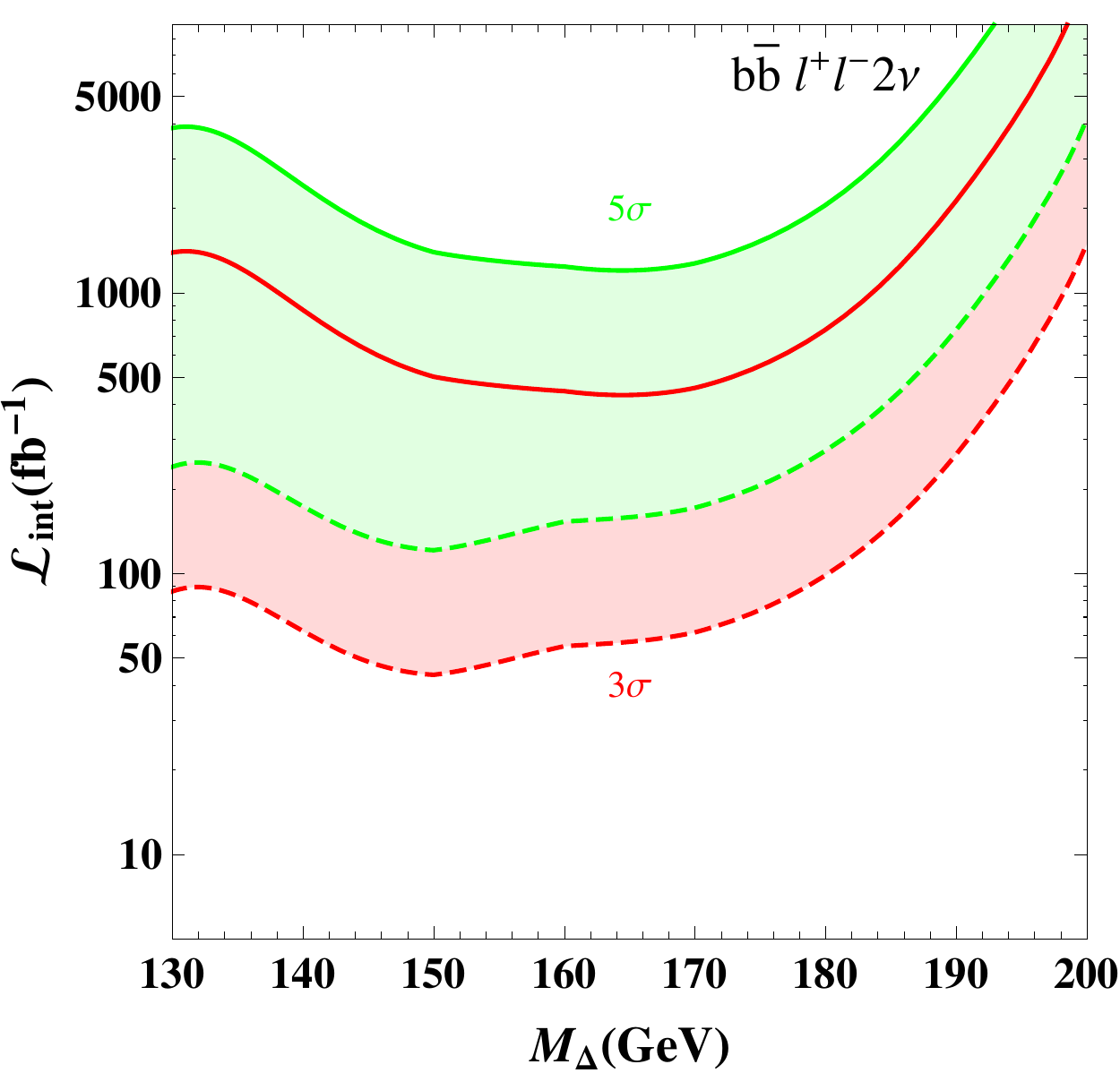}
\end{center}
\caption{Same as Fig.~\ref{bbaa_sen}, but for the $b\bar{b}\ell^+\ell^-\cancel{E}_T$ channel.
\label{bbll2v_sen}}
\end{figure}

Based on our elaborate analysis of signal channels in Secs.~\ref{sec:bbgg}--\ref{sec:bbWW}, we examine the observability of the neutral scalars $H_0,~A_0$ in the mass region $130\sim200~\GeV$ by adopting essentially the same cuts as before. In the left panel of Figs.~\ref{bbaa_sen}, \ref{bbtata_sen}, and \ref{bbll2v_sen} we present the significance $\mathcal{S}(S,B)$ as a function of $M_{\Delta}$ in the three signal channels $b\bar{b}\gamma\gamma$, $b\bar{b}\tau^+\tau^-$, and $b\bar{b}\ell^+\ell^-\cancel{E}_T$ that is reachable for LHC14@300 and LHC14@3000, respectively. The required luminosity to achieve a $3\sigma$ and $5\sigma$ significance is displayed in the right panel of the figures. As was done in our previous analysis, the effect of cascade enhancement is included by a factor $S/S_0$ in the final results.

As shown in Figs.~\ref{bbaa_sen} and \ref{bbtata_sen}, both the $b\bar{b}\gamma\gamma$ and $b\bar{b}\tau^+\tau^-$ channel are typically sensitive to the low-mass region ($M_{\Delta}\lesssim160~\GeV$). In the absence of cascade enhancement, the $3\sigma$ significance would never be reached for $M_{\Delta}\gtrsim138~(142)~\GeV$ in the $b\bar{b}\gamma\gamma$ ($b\bar{b}\tau^+\tau^-$) channel for LHC14@300. However, a cascade enhancement of $S/S_0\sim 4-6$ (as can be seen from Fig.~\ref{sgn}) in this mass region can greatly improve the observability, pushing the $3\sigma$ mass limit up to $157~(162)~\GeV$ in the $b\bar{b}\gamma\gamma$ ($b\bar{b}\tau^+\tau^-$) channel. Moreover, with cascade enhancement, one has a good chance to reach a $5\sigma$ significance if $M_{\Delta}\lesssim 153~(155)~\GeV$. In other words, the cascade enhancement significantly reduces the required luminosity. For instance, to achieve a $3\sigma$ and $5\sigma$ significance in the $b\bar{b}\gamma\gamma$ ($b\bar{b}\tau^+\tau^-$) channel with $M_{\Delta}=130~\GeV$, the required luminosity is as low as $16~(10)~\fb^{-1}$ and $42~(27)~\fb^{-1}$ at LHC14, respectively. The $b\bar{b}\tau^+\tau^-$ channel is more promising, thanks to a relatively larger production rate.

At the future LHC14 with $3000~\fb^{-1}$ data, the heavier mass region can also be probed. With a maximal cascade enhancement, the $3\sigma$ and $5\sigma$ mass reach is pushed to $177$ and $164~\GeV$, respectively, in the $b\bar{b}\gamma\gamma$ channel, which should be compared to $156$ and $151~\GeV$ in the absence of enhancement. For the $b\bar{b}\tau^+\tau^-$ channel, the enhancement factor $S/S_0$ can reach about $18$ above the $W$-pair threshold, upshifting the $3\sigma$ and  $5\sigma$ mass reach to $189$ and $177~\GeV$, respectively, from $154$ and $150~\GeV$ without the enhancement.

The $b\bar{b}\ell^+\ell^-\cancel{E}_T$ channel shown in Fig. \ref{bbll2v_sen} is more special, compared with $b\bar{b}\gamma\gamma$ and $b\bar{b}\tau^+\tau^-$. It is relatively more sensitive to a higher mass between $150$-$180~\GeV$, where the decay mode $H^0\to W^+W^-$ dominates, while its observability deteriorates for $M_{\Delta}<150~\GeV$ due to phase-space suppression in the decay. The cascade enhancement $S/S_0$ at our benchmark point (\ref{BP}) is typically $3$-$4$ in the mass region $130$-$200~\GeV$, and decreases as $M_{\Delta}$ increases. For LHC14@300, the $3\sigma$ and $5\sigma$ mass reach is, respectively, $190$ and $181~\GeV$ with maximal cascade enhancement. These limits would just increase by $2$-$3~\GeV$ for LHC14@3000 if there were no cascade enhancement, while with cascade enhancement the $5\sigma$ limit, for instance, is pushed up to $200~\GeV$. Finally, a $3\sigma$ or $5\sigma$ reach in the mass region $150$-$180~\GeV$ requires an integrated luminosity of $50~\fb^{-1}$ ($450~\fb^{-1}$) or $150~\fb^{-1}$ ($1300~\fb^{-1}$) with (without) cascade enhancement.

\section{Discussions and Conclusions}
\label{Dis}

In this paper, we have systematically investigated the LHC phenomenology of neutral scalar pair production in the negative scenario of the type II seesaw model. To achieve this goal, we first
examined the decay properties of the neutral scalars $H_0/A_0$ and found that the scalar self-couplings $\lambda_i$ have a great impact on the branching ratios of $H^0/A^0$. The coupling $\lambda_4$ is important for tree-level decays of $H^0$ and $A^0$, while one-loop-induced decays of $H^0$ further depend on $\lambda_2$ and $\lambda_3$. We found that the decay $H^0\to W^+W^-$ could dominate for $2M_W<M_{H^0}<2M_h$ with $\lambda_4<0$, while it can be neglected once $M_{H^0}$ is above the light scalar pair threshold $2M_h$. Moreover, the branching ratios of the decays $H^0\to \gamma\gamma,~Z\gamma$ can cross 3 orders of magnitude when varying the couplings $\lambda_i$, and there exist zero points for the $H^0ZZ$, $H^0hh$, and $A^0Zh$ couplings.

The cross section of the Drell-Yan process $pp\to Z^*\to H^0A^0$ for $M_{\Delta}<200~\GeV$ is much larger than that of the SM Higgs pair production driven by gluon fusion. In this paper, we studied the contributions to $H^0/A^0$ production from cascade decays of the charged scalars $H^{\pm}$ and $H^{\pm\pm}$. There are actually three different states for the neutral scalar pair: $H^0A^0$, $H^0H^0$, and $A^0A^0$. Here, $H^0H^0$ and $A^0A^0$ can only arise from cascade decays of charged scalars, and their production rates always stay the same to a good approximation. Further, for a fixed value of $M_{\Delta}$, cascade enhancement is determined by the variables $v_{\Delta}$ and $\Delta M$. By tuning these two variables, the associated production rate of $H^0A^0$ can be maximally enhanced by about a factor of 3, while those of the $H^0H^0$ and $A^0A^0$ pair production can reach the value of $H^0A^0$ production through the pure Drell-Yan process.

We implemented detailed collider simulations of the associated $H^0A^0$ production for three typical signal channels ($b\bar{b}\gamma\gamma$, $b\bar{b}\tau^+\tau^-$, and $b\bar{b}W^+W^-$ with both $W$'s decaying leptonically). The enhancement from cascade decays of charged scalars is quantified by a multiplicative factor $S/S_0$. Due mainly to a larger production rate, all three channels are more promising than the SM Higgs pair case. If there were no cascade enhancement, the $5\sigma$ mass reach of the $b\bar{b}\gamma\gamma$, $b\bar{b}\tau^+\tau^-$, and $b\bar{b}\ell^+\ell^-\cancel{E}_T$ channels would be, respectively, $151$, $150$, and $180~\GeV$ for LHC14@3000. The cascade enhancement pushes these limits up to $164$, $177$, and $200~\GeV$. The $b\bar{b}\gamma\gamma$ and $b\bar{b}\tau^+\tau^-$ channels are more promising in the mass region below about $150~\GeV$, and the required luminosities for $5\sigma$ significance are $42~\fb^{-1}$ and $27~\fb^{-1}$, respectively, at our benchmark point. Compared with these two channels, the $b\bar{b}\ell^+\ell^-\cancel{E}_T$ channel is more advantageous in the relatively higher mass region $150$-$200~\GeV$, and the required luminosity for $5\sigma$ significance is about $150~\fb^{-1}$ with maximal cascade enhancement. Needless to say, for the purpose of a full investigation on the impact of heavy neutral scalars on the SM Higgs pair production, more sophisticated simulations are necessary. We hope that this work may shed some light on further studies in both the phenomenological and experimental communities.

\section*{Acknowledgments}
This work was supported in part by the Grants No. NSFC-11025525, No. NSFC-11575089 and by the CAS Center for Excellence in Particle Physics (CCEPP).

%%%%%%%%%%%%%%%%%%%%%%%%%%%%%


\begin{thebibliography}{000}
%%%%%%%%%%%%%%%%%%%%%%%%%%%%%%

%\cite{Han:2015hba}
\bibitem{Han:2015hba}
  Z.~L.~Han, R.~Ding, and Y.~Liao,
  %``LHC Phenomenology of Type II Seesaw: Nondegenerate Case,''
  Phys.\ Rev.\ D {\bf 91}, 093006 (2015)
  [arXiv:1502.05242 [hep-ph]].
  %%CITATION = ARXIV:1502.05242;%%
  %1 citations counted in INSPIRE as of 04 juin 2015

%\cite{Aad:2013wqa}
\bibitem{Aad:2013wqa}
  G.~Aad {\it et al.}  [ATLAS Collaboration],
  %``Measurements of Higgs boson production and couplings in diboson final states with the ATLAS detector at the LHC,''
  Phys.\ Lett.\ B {\bf 726}, 88 (2013);
  Phys.\ Lett.\ B {\bf 734}, 406 (2014)
  [arXiv:1307.1427 [hep-ex]].
  %%CITATION = ARXIV:1307.1427;%%
  %368 citations counted in INSPIRE as of 19 May 2015

%\cite{Chatrchyan:2013lba}
\bibitem{Chatrchyan:2013lba}
  S.~Chatrchyan {\it et al.}  [CMS Collaboration],
  %``Observation of a new boson with mass near 125 GeV in pp collisions at $\sqrt{s}$ = 7 and 8 TeV,''
  JHEP {\bf 1306}, 081 (2013)
  [arXiv:1303.4571 [hep-ex]].
  %%CITATION = ARXIV:1303.4571;%%
  %296 citations counted in INSPIRE as of 19 May 2015

%\cite{Aad:2012tfa}
\bibitem{Aad:2012tfa}
G.~Aad {\it et al.}  [ATLAS Collaboration],
  %``Observation of a new particle in the search for the Standard Model Higgs boson with the ATLAS detector at the LHC,''
  Phys.\ Lett.\ B {\bf 716}, 1 (2012)
  [arXiv:1207.7214 [hep-ex]].
  %%CITATION = ARXIV:1207.7214;%%
  %4354 citations counted in INSPIRE as of 19 May 2015

%\cite{Chatrchyan:2012ufa}
\bibitem{Chatrchyan:2012ufa}
  S.~Chatrchyan {\it et al.}  [CMS Collaboration],
  %``Observation of a new boson at a mass of 125 GeV with the CMS experiment at the LHC,''
  Phys.\ Lett.\ B {\bf 716}, 30 (2012)
  [arXiv:1207.7235 [hep-ex]].
  %%CITATION = ARXIV:1207.7235;%%
  %4274 citations counted in INSPIRE as of 19 May 2015

%\cite{Plehn:1996wb}
\bibitem{Plehn:1996wb}
  T.~Plehn, M.~Spira, and P.~M.~Zerwas,
  %``Pair production of neutral Higgs particles in gluon-gluon collisions,''
  Nucl.\ Phys.\ B {\bf 479}, 46 (1996)
  [Erratum-ibid.\ B {\bf 531}, 655 (1998)]
  [hep-ph/9603205].
  %%CITATION = HEP-PH/9603205;%%
  %167 citations counted in INSPIRE as of 04 Jul 2014

%\cite{Dawson:1998py}
\bibitem{Dawson:1998py}
  S.~Dawson, S.~Dittmaier, and M.~Spira,
  %``Neutral Higgs boson pair production at hadron colliders: QCD corrections,''
  Phys.\ Rev.\ D {\bf 58}, 115012 (1998)
  [hep-ph/9805244].
  %%CITATION = HEP-PH/9805244;%%
  %142 citations counted in INSPIRE as of 04 Jul 2014

%\cite{Djouadi:1999rca}
\bibitem{Djouadi:1999rca}
  A.~Djouadi, W.~Kilian, M.~Muhlleitner, and P.~M.~Zerwas,
  %``Production of neutral Higgs boson pairs at LHC,''
  Eur.\ Phys.\ J.\ C {\bf 10}, 45 (1999)
  [hep-ph/9904287].
  %%CITATION = HEP-PH/9904287;%%
  %159 citations counted in INSPIRE as of 04 Jul 2014

%\cite{Baur:2002qd}
\bibitem{Baur:2002qd}
  U.~Baur, T.~Plehn, and D.~L.~Rainwater,
  %``Determining the Higgs boson selfcoupling at hadron colliders,''
  Phys.\ Rev.\ D {\bf 67}, 033003 (2003)
  [hep-ph/0211224].
  %%CITATION = HEP-PH/0211224;%%
  %128 citations counted in INSPIRE as of 24 Apr 2015

%\cite{Asakawa:2010xj}
\bibitem{Asakawa:2010xj}
  E.~Asakawa, D.~Harada, S.~Kanemura, Y.~Okada, and K.~Tsumura,
  %``Higgs boson pair production in new physics models at hadron, lepton, and photon colliders,''
  Phys.\ Rev.\ D {\bf 82}, 115002 (2010)
  [arXiv:1009.4670 [hep-ph]].
  %%CITATION = ARXIV:1009.4670;%%
  %34 citations counted in INSPIRE as of 07 Jul 2014

%\cite{Dolan:2012rv}
\bibitem{Dolan:2012rv}
  M.~J.~Dolan, C.~Englert, and M.~Spannowsky,
  %``Higgs self-coupling measurements at the LHC,''
  JHEP {\bf 1210}, 112 (2012)
  [arXiv:1206.5001 [hep-ph]].
  %%CITATION = ARXIV:1206.5001;%%
  %125 citations counted in INSPIRE as of 24 Apr 2015

%\cite{Papaefstathiou:2012qe}
\bibitem{Papaefstathiou:2012qe}
  A.~Papaefstathiou, L.~L.~Yang, and J.~Zurita,
  %``Higgs boson pair production at the LHC in the $b \bar{b} W^+ W^-$ channel,''
  Phys.\ Rev.\ D {\bf 87}, no. 1, 011301 (2013)
  [arXiv:1209.1489 [hep-ph]].
    %%CITATION = ARXIV:1209.1489;%%
  %70 citations counted in INSPIRE as of 24 Apr 2015

%\cite{Goertz:2013kp}
\bibitem{Goertz:2013kp}
  F.~Goertz, A.~Papaefstathiou, L.~L.~Yang, and J.~Zurita,
  %``Higgs Boson self-coupling measurements using ratios of cross sections,''
  JHEP {\bf 1306}, 016 (2013)
  [arXiv:1301.3492 [hep-ph]].
  %%CITATION = ARXIV:1301.3492;%%
  %83 citations counted in INSPIRE as of 24 Apr 2015

%\cite{Gupta:2013zza}
\bibitem{Gupta:2013zza}
  R.~S.~Gupta, H.~Rzehak, and J.~D.~Wells,
  %``How well do we need to measure the Higgs boson mass and self-coupling?,''
  Phys.\ Rev.\ D {\bf 88}, 055024 (2013)
  [arXiv:1305.6397 [hep-ph]].
  %%CITATION = ARXIV:1305.6397;%%
  %35 citations counted in INSPIRE as of 24 Apr 2015

%\cite{Barr:2013tda}
\bibitem{Barr:2013tda}
  A.~J.~Barr, M.~J.~Dolan, C.~Englert, and M.~Spannowsky,
  %``Di-Higgs final states augMT2ed -- selecting $hh$ events at the high luminosity LHC,''
  Phys.\ Lett.\ B {\bf 728}, 308 (2014)
  [arXiv:1309.6318 [hep-ph]].
  %%CITATION = ARXIV:1309.6318;%%
  %36 citations counted in INSPIRE as of 24 Apr 2015

%\cite{deFlorian:2013jea}
\bibitem{deFlorian:2013jea}
  D.~de Florian and J.~Mazzitelli,
  %``Higgs Boson Pair Production at Next-to-Next-to-Leading Order in QCD,''
  Phys.\ Rev.\ Lett.\  {\bf 111}, 201801 (2013)
  [arXiv:1309.6594 [hep-ph]].
  %%CITATION = ARXIV:1309.6594;%%
  %24 citations counted in INSPIRE as of 05 Jul 2014

%\cite{Dolan:2013rja}
\bibitem{Dolan:2013rja}
  M.~J.~Dolan, C.~Englert, N.~Greiner, and M.~Spannowsky,
  %``Further on up the road: $hhjj$ production at the LHC,''
  Phys.\ Rev.\ Lett.\  {\bf 112}, 101802 (2014)
  [arXiv:1310.1084 [hep-ph]].
  %%CITATION = ARXIV:1310.1084;%%
  %34 citations counted in INSPIRE as of 24 Apr 2015

  %\cite{Barger:2013jfa}
\bibitem{Barger:2013jfa}
  V.~Barger, L.~L.~Everett, C.~B.~Jackson, and G.~Shaughnessy,
  %``Higgs-Pair Production and Measurement of the Triscalar Coupling at LHC(8,14),''
  Phys.\ Lett.\ B {\bf 728}, 433 (2014)
  [arXiv:1311.2931 [hep-ph]].
  %%CITATION = ARXIV:1311.2931;%%
  %18 citations counted in INSPIRE as of 05 Aug 2014

%\cite{Englert:2014uqa}
\bibitem{Englert:2014uqa}
  C.~Englert, F.~Krauss, M.~Spannowsky, and J.~Thompson,
  %``Di-Higgs phenomenology in $t\bar{t}hh$: The forgotten channel,''
  Phys.\ Lett.\ B {\bf 743}, 93 (2015)
  [arXiv:1409.8074 [hep-ph]].
  %%CITATION = ARXIV:1409.8074;%%
  %6 citations counted in INSPIRE as of 24 Apr 2015

%\cite{Liu:2014rva}
\bibitem{Liu:2014rva}
  T.~Liu and H.~Zhang,
  %``Measuring Di-Higgs Physics via the $t \bar t hh \to t \bar t b \bar bb\bar b$ Channel,''
  arXiv:1410.1855 [hep-ph].
  %%CITATION = ARXIV:1410.1855;%%
  %4 citations counted in INSPIRE as of 24 Apr 2015

%\cite{deLima:2014dta}
\bibitem{deLima:2014dta}
  D.~E.~Ferreira de Lima, A.~Papaefstathiou, and M.~Spannowsky,
  %``Standard model Higgs boson pair production in the ( $ b\overline{b} $ )( $ b\overline{b} $ ) final state,''
  JHEP {\bf 1408}, 030 (2014)
  [arXiv:1404.7139 [hep-ph]].
  %%CITATION = ARXIV:1404.7139;%%
  %15 citations counted in INSPIRE as of 07 Jan 2015

  %\cite{Barr:2014sga}
\bibitem{Barr:2014sga}
  A.~J.~Barr, M.~J.~Dolan, C.~Englert, D.~E.~F.~de Lima, and M.~Spannowsky,
  %``Higgs Self-Coupling Measurements at a 100 TeV Hadron Collider,''
  arXiv:1412.7154 [hep-ph].
  %%CITATION = ARXIV:1412.7154;%%

%\cite{Dolan:2012ac}
\bibitem{Dolan:2012ac}
  M.~J.~Dolan, C.~Englert, and M.~Spannowsky,
  %``New Physics in LHC Higgs boson pair production,''
  Phys.\ Rev.\ D {\bf 87}, no. 5, 055002 (2013)
  [arXiv:1210.8166 [hep-ph]].
  %%CITATION = ARXIV:1210.8166;%%
  %68 citations counted in INSPIRE as of 24 Apr 2015

%2HDM

%\cite{Arhrib:2009hc}
\bibitem{Arhrib:2009hc}
  A.~Arhrib, R.~Benbrik, C.~-H.~Chen, R.~Guedes, and R.~Santos,
  %``Double Neutral Higgs production in the Two-Higgs doublet model at the LHC,''
  JHEP {\bf 0908}, 035 (2009)
  [arXiv:0906.0387 [hep-ph]].
  %%CITATION = ARXIV:0906.0387;%%
  %18 citations counted in INSPIRE as of 07 Jul 2014

%\cite{Craig:2013hca}
\bibitem{Craig:2013hca}
  N.~Craig, J.~Galloway, and S.~Thomas,
  %``Searching for Signs of the Second Higgs Doublet,''
  arXiv:1305.2424 [hep-ph].
  %%CITATION = ARXIV:1305.2424;%%
  %96 citations counted in INSPIRE as of 24 Apr 2015

%\cite{Hespel:2014sla}
\bibitem{Hespel:2014sla}
  B.~Hespel, D.~Lopez-Val, and E.~Vryonidou,
  %``Higgs pair production via gluon fusion in the Two-Higgs-Doublet Model,''
  JHEP {\bf 1409}, 124 (2014)
  [arXiv:1407.0281 [hep-ph]].
  %%CITATION = ARXIV:1407.0281;%%
  %18 citations counted in INSPIRE as of 24 Apr 2015

%MSSM & NMSSM

%\cite{Kribs:2012kz}
\bibitem{Kribs:2012kz}
  G.~D.~Kribs and A.~Martin,
  %``Enhanced di-Higgs Production through Light Colored Scalars,''
  Phys.\ Rev.\ D {\bf 86}, 095023 (2012)
  [arXiv:1207.4496 [hep-ph]].
  %%CITATION = ARXIV:1207.4496;%%
  %49 citations counted in INSPIRE as of 24 Apr 2015

%\cite{Cao:2013si}
\bibitem{Cao:2013si}
  J.~Cao, Z.~Heng, L.~Shang, P.~Wan, and J.~M.~Yang,
  %``Pair Production of a 125 GeV Higgs Boson in MSSM and NMSSM at the LHC,''
  JHEP {\bf 1304}, 134 (2013)
  [arXiv:1301.6437 [hep-ph]].
  %%CITATION = ARXIV:1301.6437;%%
  %42 citations counted in INSPIRE as of 24 Apr 2015

%\cite{Nhung:2013lpa}
\bibitem{Nhung:2013lpa}
  D.~T.~Nhung, M.~Muhlleitner, J.~Streicher, and K.~Walz,
  %``Higher Order Corrections to the Trilinear Higgs Self-Couplings in the Real NMSSM,''
  JHEP {\bf 1311}, 181 (2013)
  [arXiv:1306.3926 [hep-ph]].
  %%CITATION = ARXIV:1306.3926;%%
  %26 citations counted in INSPIRE as of 24 Apr 2015

%\cite{Ellwanger:2013ova}
\bibitem{Ellwanger:2013ova}
  U.~Ellwanger,
  %``Higgs pair production in the NMSSM at the LHC,''
  JHEP {\bf 1308}, 077 (2013)
  [arXiv:1306.5541 [hep-ph]].
  %%CITATION = ARXIV:1306.5541,;%%
  %44 citations counted in INSPIRE as of 24 Apr 2015

%\cite{Bhattacherjee:2014bca}
\bibitem{Bhattacherjee:2014bca}
  B.~Bhattacherjee and A.~Choudhury,
  %``Role of supersymmetric heavy Higgs boson production in the self-coupling measurement of 125 GeV Higgs boson at the LHC,''
  Phys.\ Rev.\ D {\bf 91}, 073015 (2015)
  [arXiv:1407.6866 [hep-ph]].
  %%CITATION = ARXIV:1407.6866;%%
  %11 citations counted in INSPIRE as of 24 Apr 2015

  %\cite{Christensen:2012si}
\bibitem{Christensen:2012si}
  N.~D.~Christensen, T.~Han, and T.~Li,
  %``Pair Production of MSSM Higgs Bosons in the Non-decoupling Region at the LHC,''
  Phys.\ Rev.\ D {\bf 86}, 074003 (2012)
  [arXiv:1206.5816 [hep-ph]].
  %%CITATION = ARXIV:1206.5816;%%
  %16 citations counted in INSPIRE as of 05 Aug 2014

%\cite{Wu:2015nba}
\bibitem{Wu:2015nba}
  L.~Wu, J.~M.~Yang, C.~P.~Yuan, and M.~Zhang,
  %``Higgs self-coupling in the MSSM and NMSSM after the LHC Run 1,''
  Phys.\ Lett.\ B {\bf 747}, 378 (2015)
  [arXiv:1504.06932 [hep-ph]].
  %%CITATION = ARXIV:1504.06932;%%
  %2 citations counted in INSPIRE as of 24 Apr 2015

%\cite{Cao:2014kya}
\bibitem{Cao:2014kya}
  J.~Cao, D.~Li, L.~Shang, P.~Wu, and Y.~Zhang,
  %``Exploring the Higgs Sector of a Most Natural NMSSM and its Prediction on Higgs Pair Production at the LHC,''
  JHEP {\bf 1412}, 026 (2014)
  [arXiv:1409.8431 [hep-ph]].
  %%CITATION = ARXIV:1409.8431;%%
  %6 citations counted in INSPIRE as of 24 Apr 2015

%\cite{Han:2013sga}
\bibitem{Han:2013sga}
  C.~Han, X.~Ji, L.~Wu, P.~Wu, and J.~M.~Yang,
  %``Higgs pair production with SUSY QCD correction: revisited under current experimental constraints,''
  JHEP {\bf 1404}, 003 (2014)
  [arXiv:1307.3790 [hep-ph]].
  %%CITATION = ARXIV:1307.3790;%%
  %23 citations counted in INSPIRE as of 24 Apr 2015

%Composite Higgs

%\cite{Gouzevitch:2013qca}
\bibitem{Gouzevitch:2013qca}
  M.~Gouzevitch, A.~Oliveira, J.~Rojo, R.~Rosenfeld, G.~P.~Salam, and V.~Sanz,
  %``Scale-invariant resonance tagging in multijet events and new physics in Higgs pair production,''
  JHEP {\bf 1307}, 148 (2013)
  [arXiv:1303.6636 [hep-ph]].
  %%CITATION = ARXIV:1303.6636;%%
  %38 citations counted in INSPIRE as of 05 Aug 2014

%\cite{No:2013wsa}
\bibitem{No:2013wsa}
  J.~M.~No and M.~Ramsey-Musolf,
  %``Probing the Higgs Portal at the LHC Through Resonant di-Higgs Production,''
  Phys.\ Rev.\ D {\bf 89}, 095031 (2014)
  [arXiv:1310.6035 [hep-ph]].
  %%CITATION = ARXIV:1310.6035;%%
  %29 citations counted in INSPIRE as of 24 Apr 2015

%\cite{Grober:2010yv}
\bibitem{Grober:2010yv}
  R.~Grober and M.~Muhlleitner,
  %``Composite Higgs Boson Pair Production at the LHC,''
  JHEP {\bf 1106}, 020 (2011)
  [arXiv:1012.1562 [hep-ph]].
  %%CITATION = ARXIV:1012.1562;%%
  %86 citations counted in INSPIRE as of 24 Apr 2015

%\cite{Gillioz:2012se}
\bibitem{Gillioz:2012se}
  M.~Gillioz, R.~Grober, C.~Grojean, M.~Muhlleitner, and E.~Salvioni,
  %``Higgs Low-Energy Theorem (and its corrections) in Composite Models,''
  JHEP {\bf 1210}, 004 (2012)
  [arXiv:1206.7120 [hep-ph]].
  %%CITATION = ARXIV:1206.7120;%%
  %75 citations counted in INSPIRE as of 24 Apr 2015

%\cite{Liu:2013woa}
\bibitem{Liu:2013woa}
  J.~Liu, X.~P.~Wang, and S.~h.~Zhu,
  %``Discovering extra Higgs boson via pair production of the SM-like Higgs bosons,''
  arXiv:1310.3634 [hep-ph].
  %%CITATION = ARXIV:1310.3634;%%
  %21 citations counted in INSPIRE as of 24 Apr 2015

%\cite{Arhrib:2008pw}
\bibitem{Arhrib:2008pw}
  A.~Arhrib, R.~Benbrik, R.~B.~Guedes, and R.~Santos,
  %``Search for a light fermiophobic Higgs boson produced via gluon fusion at Hadron Colliders,''
  Phys.\ Rev.\ D {\bf 78}, 075002 (2008)
  [arXiv:0805.1603 [hep-ph]].
  %%CITATION = ARXIV:0805.1603;%%
  %10 citations counted in INSPIRE as of 07 Jul 2014

%\cite{Heng:2013cya}
\bibitem{Heng:2013cya}
  Z.~Heng, L.~Shang, Y.~Zhang, and J.~Zhu,
  %``Pair production of 125 GeV Higgs boson in the SM extension with color-octet scalars at the LHC,''
  JHEP {\bf 1402}, 083 (2014)
  [arXiv:1312.4260 [hep-ph]].
  %%CITATION = ARXIV:1312.4260;%%
  %2 citations counted in INSPIRE as of 07 Jul 2014

%\cite{Dawson:2012mk}
\bibitem{Dawson:2012mk}
  S.~Dawson, E.~Furlan, and I.~Lewis,
  %``Unravelling an extended quark sector through multiple Higgs production?,''
  Phys.\ Rev.\ D {\bf 87}, 014007 (2013)
  [arXiv:1210.6663 [hep-ph]].
  %%CITATION = ARXIV:1210.6663;%%
  %41 citations counted in INSPIRE as of 07 Jul 2014

%\cite{Chen:2014xwa}
\bibitem{Chen:2014xwa}
  C.~Y.~Chen, S.~Dawson, and I.~M.~Lewis,
  %``Top Partners and Higgs Boson Production,''
  Phys.\ Rev.\ D {\bf 90}, 035016 (2014)
  [arXiv:1406.3349 [hep-ph]].
  %%CITATION = ARXIV:1406.3349;%%
  %5 citations counted in INSPIRE as of 07 Jan 2015

%\cite{Dib:2005re}
\bibitem{Dib:2005re}
  C.~O.~Dib, R.~Rosenfeld, and A.~Zerwekh,
  %``Double Higgs production and quadratic divergence cancellation in little Higgs models with T parity,''
  JHEP {\bf 0605}, 074 (2006)
  [hep-ph/0509179].
  %%CITATION = HEP-PH/0509179;%%
  %39 citations counted in INSPIRE as of 07 Jul 2014

  %\cite{Yang:2014gca}
\bibitem{Yang:2014gca}
  B.~Yang, Z.~Liu, N.~Liu, and J.~Han,
  %``Double Higgs production in the littlest Higgs Model with T-parity at high energy $e^{+}e^{-}$ Colliders,''
  Eur.\ Phys.\ J.\ C {\bf 74}, 3203 (2014)
  [arXiv:1408.4295 [hep-ph]].
  %%CITATION = ARXIV:1408.4295;%%

  %\cite{Chen:2014ask}
\bibitem{Chen:2014ask}
  C.~Y.~Chen, S.~Dawson, and I.~M.~Lewis,
  %``Exploring Resonant di-Higgs production in the Higgs Singlet Model,''
  Phys.\ Rev.\  D {\bf 91}, 035015 (2015)
  [arXiv:1410.5488 [hep-ph]].
  %%CITATION = ARXIV:1410.5488;%%
  %2 citations counted in INSPIRE as of 14 Dec 2014

%Anomalous Couplings

%\cite{Contino:2012xk}
\bibitem{Contino:2012xk}
  R.~Contino, M.~Ghezzi, M.~Moretti, G.~Panico, F.~Piccinini, and A.~Wulzer,
  %``Anomalous Couplings in Double Higgs Production,''
  JHEP {\bf 1208}, 154 (2012)
  [arXiv:1205.5444 [hep-ph]].
  %%CITATION = ARXIV:1205.5444;%%
  %57 citations counted in INSPIRE as of 24 Apr 2015

%\cite{Nishiwaki:2013cma}
\bibitem{Nishiwaki:2013cma}
  K.~Nishiwaki, S.~Niyogi, and A.~Shivaji,
  %``$ttH$ Anomalous Coupling in Double Higgs Production,''
  JHEP {\bf 1404}, 011 (2014)
  [arXiv:1309.6907 [hep-ph]].
  %%CITATION = ARXIV:1309.6907;%%
  %20 citations counted in INSPIRE as of 24 Apr 2015

  %\cite{Liu:2014rba}
\bibitem{Liu:2014rba}
  N.~Liu, S.~Hu, B.~Yang, and J.~Han,
  %``Impact of top-Higgs couplings on di-Higgs production at future colliders,''
  JHEP {\bf 1501}, 008 (2015) 
  [arXiv:1408.4191 [hep-ph]].
  %%CITATION = ARXIV:1408.4191;%%

%\cite{Dawson:2015oha}
\bibitem{Dawson:2015oha}
  S.~Dawson, A.~Ismail, and I.~Low,
  %``What's in the Loop? The Anatomy of Double Higgs Production,''
  Phys.\ Rev.\  D {\bf 91}, 115008 (2015)
  [arXiv:1504.05596 [hep-ph]].
  %%CITATION = ARXIV:1504.05596;%%
  %1 citations counted in INSPIRE as of 24 Apr 2015

%Effective operator

%\cite{Azatov:2015oxa}
\bibitem{Azatov:2015oxa}
  A.~Azatov, R.~Contino, G.~Panico, and M.~Son,
  %``Effective field theory analysis of double Higgs production via gluon fusion,''
  Phys.\ Rev.\  D {\bf 92}, 035001 (2015)
  [arXiv:1502.00539 [hep-ph]].
  %%CITATION = ARXIV:1502.00539;%%
  %9 citations counted in INSPIRE as of 24 Apr 2015

%\cite{Goertz:2014qta}
\bibitem{Goertz:2014qta}
  F.~Goertz, A.~Papaefstathiou, L.~L.~Yang, and J.~Zurita,
  %``Higgs boson pair production in the D=6 extension of the SM,''
  JHEP {\bf 1504}, 167 (2015)
  [arXiv:1410.3471 [hep-ph]].
  %%CITATION = ARXIV:1410.3471;%%
  %12 citations counted in INSPIRE as of 24 Apr 2015
%\cite{Djouadi:1999rca}

  %\cite{Pierce:2006dh}
\bibitem{Pierce:2006dh}
  A.~Pierce, J.~Thaler, and L.~-T.~Wang,
  %``Disentangling Dimension Six Operators through Di-Higgs Boson Production,''
  JHEP {\bf 0705}, 070 (2007)
  [hep-ph/0609049].
  %%CITATION = HEP-PH/0609049;%%
  %28 citations counted in INSPIRE as of 07 Jul 2014

%\cite{Kang:2015nga}
\bibitem{Kang:2015nga}
  Z.~Kang, P.~Ko, and J.~Li,
  %``New Physics Opportunities in the Boosted Di-Higgs plus \ET Signature,''
  arXiv:1504.04128 [hep-ph].
  %%CITATION = ARXIV:1504.04128;%%

%\cite{He:2015spf}
\bibitem{He:2015spf}
  H.~J.~He, J.~Ren, and W.~Yao,
  %``Probing New Physics of Cubic Higgs Interaction via Higgs Pair Production at Hadron Colliders,''
  arXiv:1506.03302 [hep-ph].
  %%CITATION = ARXIV:1506.03302;%%

%\cite{Baglio:2012np}
\bibitem{Baglio:2012np}
  J.~Baglio, A.~Djouadi, R.~Gr\"{o}ber, M.~M.~M\"{u}hlleitner, J.~Quevillon, and M.~Spira,
  %``The measurement of the Higgs self-coupling at the LHC: theoretical status,''
  JHEP {\bf 1304}, 151 (2013)
  [arXiv:1212.5581 [hep-ph]].
  %%CITATION = ARXIV:1212.5581;%%
  %64 citations counted in INSPIRE as of 04 Jun 2014

%\cite{Dawson:2013bba}
\bibitem{Dawson:2013bba}
  S.~Dawson {\it et al.},
  %``Working Group Report: Higgs Boson,''
  arXiv:1310.8361 [hep-ex].
  %%CITATION = ARXIV:1310.8361;%%
  %144 citations counted in INSPIRE as of 24 Apr 2015

\bibitem{typeII}
%\cite{Cheng:1980qt}
%\bibitem{Cheng:1980qt}
 T.~P.~Cheng and L.~F.~Li,
 %``Neutrino Masses, Mixings And Oscillations In SU(2) X U(1) Models Of
 %Electroweak Interactions,''
 Phys.\ Rev.\  D {\bf 22}, 2860 (1980);
 J.~Schechter and J.~W.~F.~Valle,
 %``Neutrino Masses In SU(2) X U(1) Theories,''
 Phys.\ Rev.\  D {\bf 22}, 2227 (1980);
 %%CITATION = PHRVA,D22,2227;%%
%\cite{Lazarides:1980nt}
%\bibitem{Lazarides:1980nt}
  G.~Lazarides, Q.~Shafi, and C.~Wetterich,
  %``Proton Lifetime and Fermion Masses in an SO(10) Model,''
  Nucl.\ Phys.\  B {\bf 181}, 287 (1981);
  %%CITATION = NUPHA,B181,287;%%
  %\cite{Mohapatra:1980yp}
%\bibitem{Mohapatra:1980yp}
  R.~N.~Mohapatra and G.~Senjanovic,
  %``Neutrino Masses and Mixings in Gauge Models with Spontaneous Parity
  %Violation,''
  Phys.\ Rev.\  D {\bf 23}, 165 (1981);
  %%CITATION = PHRVA,D23,165;%%
  %\cite{Magg:1980ut}
%\bibitem{Magg:1980ut}
  M.~Magg and C.~Wetterich,
  %``NEUTRINO MASS PROBLEM AND GAUGE HIERARCHY,''
  Phys.\ Lett.\  B {\bf 94}, 61 (1980).
  %%CITATION = PHLTA,B94,61;%%

%\cite{Arhrib:2011uy}
\bibitem{Arhrib:2011uy}
  A.~Arhrib {\it et al.},
  %``The Higgs Potential in the Type II Seesaw Model,''
  Phys.\ Rev.\ D {\bf 84}, 095005 (2011)
  [arXiv:1105.1925 [hep-ph]].
  %%CITATION = ARXIV:1105.1925;%%
  %40 citations counted in INSPIRE as of 09 May 2014

%\cite{Aoki:2012jj}
\bibitem{Aoki:2012jj}
  M.~Aoki, S.~Kanemura, M.~Kikuchi, and K.~Yagyu,
  %``Radiative corrections to the Higgs boson couplings in the triplet model,''
  Phys.\ Rev.\ D {\bf 87}, 015012 (2013)
  [arXiv:1211.6029 [hep-ph]]; 
  S. Kanemura and K. Yagyu, Phys.\ Rev.\ D {\bf 85}, 115009 (2012) 
  [arXiv:1201.6287 [hep-ph]]. 
  %%CITATION = ARXIV:1211.6029;%%
  %21 citations counted in INSPIRE as of 09 May 2014

%\cite{Aoki:2011pz}
\bibitem{Aoki:2011pz}
  M.~Aoki, S.~Kanemura, and K.~Yagyu,
  %``Testing the Higgs triplet model with the mass difference at the LHC,''
  Phys.\ Rev.\ D {\bf 85}, 055007 (2012)
  [arXiv:1110.4625 [hep-ph]].
  %%CITATION = ARXIV:1110.4625;%%
  %36 citations counted in INSPIRE as of 11 May 2014

  %\cite{Chabab:2014ara}
\bibitem{Chabab:2014ara}
  M.~Chabab, M.~C.~Peyranere, and L.~Rahili,
  %``Degenerate Higgs bosons decays to ${\gamma\gamma}$ and ${Z\gamma}$ in the type II seesaw model,''
  Phys.\ Rev.\ D {\bf 90}, 035026 (2014)
  [arXiv:1407.1797 [hep-ph]].
  %%CITATION = ARXIV:1407.1797;%%

  %\cite{Djouadi:2005gj}
\bibitem{Djouadi:2005gj}
  A.~Djouadi,
  %``The Anatomy of electro-weak symmetry breaking. II. The Higgs bosons in the minimal supersymmetric model,''
  Phys.\ Rep.\  {\bf 459}, 1 (2008)
  [hep-ph/0503173].
  %%CITATION = HEP-PH/0503173;%%
  %735 citations counted in INSPIRE as of 18 Nov 2014

%\cite{Perez:2008ha}
\bibitem{Perez:2008ha}
  P.~Fileviez Perez, T.~Han, G.~-y.~Huang, T.~Li, and K.~Wang,
  %``Neutrino Masses and the CERN LHC: Testing Type II Seesaw,''
  Phys.\ Rev.\ D {\bf 78}, 015018 (2008)
  [arXiv:0805.3536 [hep-ph]].
  %%CITATION = ARXIV:0805.3536;%%
  %164 citations counted in INSPIRE as of 11 May 2014

  %\cite{Melfo:2011nx}
\bibitem{Melfo:2011nx}
  A.~Melfo, M.~Nemevsek, F.~Nesti, G.~Senjanovic, and Y.~Zhang,
  %``Type II Seesaw at LHC: The Roadmap,''
  Phys.\ Rev.\ D {\bf 85}, 055018 (2012)
  [arXiv:1108.4416 [hep-ph]].
  %%CITATION = ARXIV:1108.4416;%%

   %\cite{Chun:2013vma}
\bibitem{Chun:2013vma}
  E.~J.~Chun and P.~Sharma,
  %``Search for a doubly-charged boson in four lepton final states in type II seesaw,''
  Phys.\ Lett.\ B {\bf 728}, 256 (2014)
  [arXiv:1309.6888 [hep-ph]].
  %%CITATION = ARXIV:1309.6888;%%
  %4 citations counted in INSPIRE as of 16 May 2014


%\cite{Akeroyd:2011zza}
\bibitem{Akeroyd:2011zza}
  A.~G.~Akeroyd and H.~Sugiyama,
  %``Production of doubly charged scalars from the decay of singly charged scalars in the Higgs Triplet Model,''
  Phys.\ Rev.\ D {\bf 84}, 035010 (2011)
  [arXiv:1105.2209 [hep-ph]].
  %%CITATION = ARXIV:1105.2209;%%
  %30 citations counted in INSPIRE as of 11 May 2014

 \bibitem{MG5}
%\cite{Alwall:2011uj}
%\bibitem{Alwall:2011uj}
  J.~Alwall, M.~Herquet, F.~Maltoni, O.~Mattelaer, and T.~Stelzer,
  %``MadGraph 5 : Going Beyond,''
  JHEP {\bf 1106}, 128 (2011)
  [arXiv:1106.0522 [hep-ph]];
  %%CITATION = ARXIV:1106.0522;%%
  %972 citations counted in INSPIRE as of 06 Jan 2014
  %\cite{Alwall:2014hca}
%\bibitem{Alwall:2014hca}
  J.~Alwall {\it et al.},
  %``The automated computation of tree-level and next-to-leading order differential cross sections, and their matching to parton shower simulations,''
  JHEP {\bf 1407}, 079 (2014)
  [arXiv:1405.0301 [hep-ph]].
  %%CITATION = ARXIV:1405.0301;%%
  %50 citations counted in INSPIRE as of 13 Aug 2014

%\cite{Sjostrand:2006za}
\bibitem{Sjostrand:2006za}
  T.~Sjostrand, S.~Mrenna, and P.~Z.~Skands,
  %``PYTHIA 6.4 Physics and Manual,''
  JHEP {\bf 0605}, 026 (2006)
  [hep-ph/0603175].
  %%CITATION = HEP-PH/0603175;%%
  %4745 citations counted in INSPIRE as of 06 Jan 2014

\bibitem{Delphes}
  %\cite{Ovyn:2009tx}
%\bibitem{Ovyn:2009tx}
  S.~Ovyn, X.~Rouby, and V.~Lemaitre,
  %``DELPHES, a framework for fast simulation of a generic collider experiment,''
  arXiv:0903.2225 [hep-ph];
  %%CITATION = ARXIV:0903.2225;%%
  %227 citations counted in INSPIRE as of 13 Aug 2014
  %\cite{deFavereau:2013fsa}
%\bibitem{deFavereau:2013fsa}
  J.~de Favereau {\it et al.}  [DELPHES 3 Collaboration],
  %``DELPHES 3, A modular framework for fast simulation of a generic collider experiment,''
  JHEP {\bf 1402}, 057 (2014)
  [arXiv:1307.6346 [hep-ex]].
  %%CITATION = ARXIV:1307.6346;%%
  %114 citations counted in INSPIRE as of 13 Aug 2014

%\cite{Conte:2012fm}
\bibitem{Conte:2012fm}
  E.~Conte, B.~Fuks, and G.~Serret,
  %``MadAnalysis 5, A User-Friendly Framework for Collider Phenomenology,''
  Comput.\ Phys.\ Commun.\  {\bf 184}, 222 (2013)
  [arXiv:1206.1599 [hep-ph]].
  %%CITATION = ARXIV:1206.1599;%%
  %21 citations counted in INSPIRE as of 06 Jan 2014

%\cite{Aad:2009wy}
\bibitem{Aad:2009wy}
  G.~Aad {\it et al.}  [ATLAS Collaboration],
  %``Expected Performance of the ATLAS Experiment - Detector, Trigger and Physics,''
  arXiv:0901.0512 [hep-ex].
  %%CITATION = ARXIV:0901.0512;%%
  %1653 citations counted in INSPIRE as of 18 Jan 2015

%\cite{Dittmaier:2011ti}
\bibitem{Dittmaier:2011ti}
  S.~Dittmaier {\it et al.}  [LHC Higgs Cross Section Working Group Collaboration],
  %``Handbook of LHC Higgs Cross Sections: 1. Inclusive Observables,''
  arXiv:1101.0593 [hep-ph].
  %%CITATION = ARXIV:1101.0593;%%
  %804 citations counted in INSPIRE as of 26 Dec 2014

  %\cite{Cowan:2010js}
\bibitem{Cowan:2010js}
  G.~Cowan, K.~Cranmer, E.~Gross, and O.~Vitells,
  %``Asymptotic formulae for likelihood-based tests of new physics,''
  Eur.\ Phys.\ J.\ C {\bf 71}, 1554 (2011)
  [Erratum-ibid.\ C {\bf 73}, 2501 (2013)]
  [arXiv:1007.1727 [physics.data-an]].
  %%CITATION = ARXIV:1007.1727;%%
  %553 citations counted in INSPIRE as of 24 Dec 2014

  %\cite{Campbell:2000bg}
\bibitem{Campbell:2000bg}
  J.~M.~Campbell and R.~K.~Ellis,
  %``Radiative corrections to Z b anti-b production,''
  Phys.\ Rev.\ D {\bf 62}, 114012 (2000)
  [hep-ph/0006304].
  %%CITATION = HEP-PH/0006304;%%
  %239 citations counted in INSPIRE as of 29 Dec 2014

\bibitem{tt}
  %\cite{Cacciari:2008zb}
%\bibitem{Cacciari:2008zb}
  M.~Cacciari, S.~Frixione, M.~L.~Mangano, P.~Nason, and G.~Ridolfi,
  %``Updated predictions for the total production cross sections of top and of heavier quark pairs at the Tevatron and at the LHC,''
  JHEP {\bf 0809}, 127 (2008)
  [arXiv:0804.2800 [hep-ph]];
  %%CITATION = ARXIV:0804.2800;%%
  %310 citations counted in INSPIRE as of 29 Dec 2014
  %\cite{Baernreuther:2012ws}
%\bibitem{Baernreuther:2012ws}
  P.~B\"{a}rnreuther, M.~Czakon, and A.~Mitov,
  %``Percent Level Precision Physics at the Tevatron: First Genuine NNLO QCD Corrections to $q \bar{q} \to t \bar{t} + X$,''
  Phys.\ Rev.\ Lett.\  {\bf 109}, 132001 (2012)
  [arXiv:1204.5201 [hep-ph]].
  %%CITATION = ARXIV:1204.5201;%%
  %209 citations counted in INSPIRE as of 29 Dec 2014

\end{thebibliography}
\end{document}